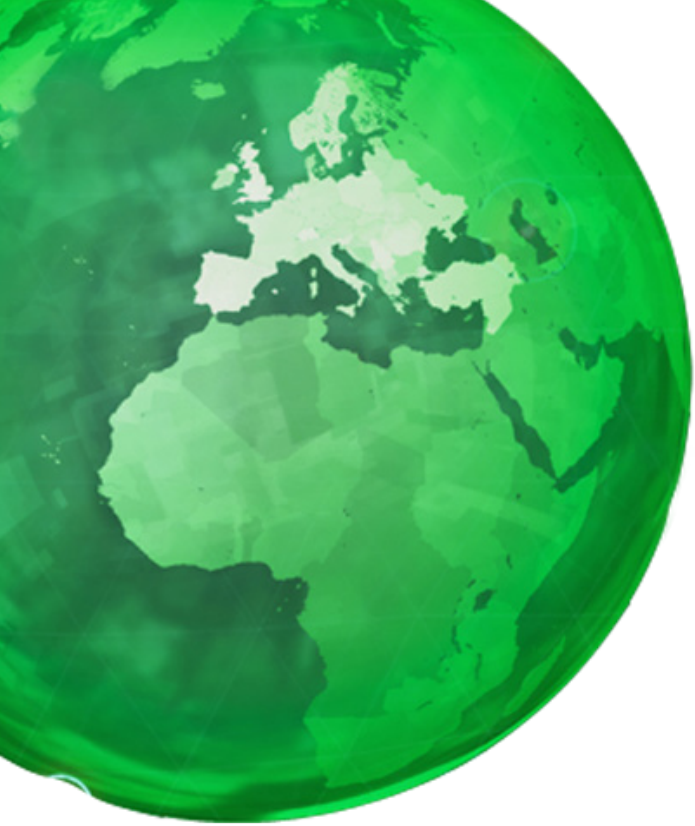

# FuturICT Flagship Proposal Summary

Coordinator: Steven Bishop, s.bishop@ucl.ac.uk
Scientific Coordinator: Dirk Helbing, dirk.helbing@gess.ethz.ch

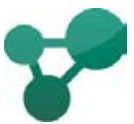

# 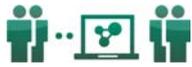 FuturICT Partners

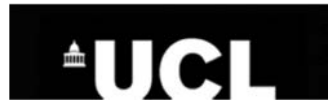 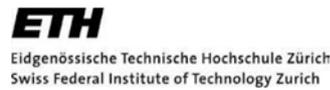

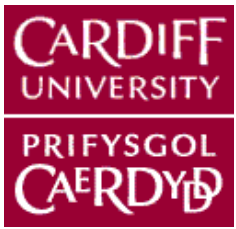 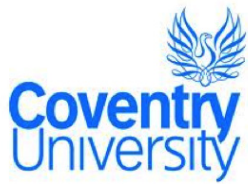 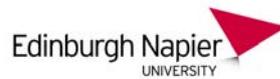 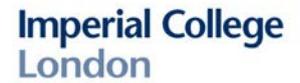

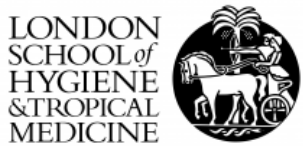 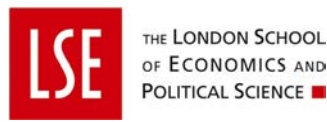 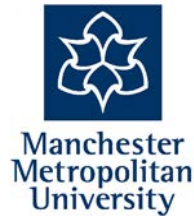 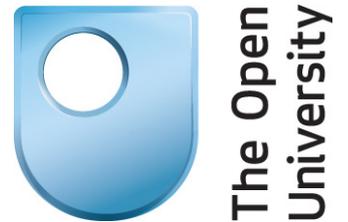

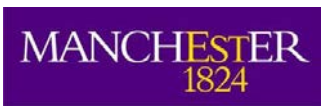 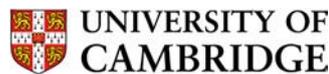 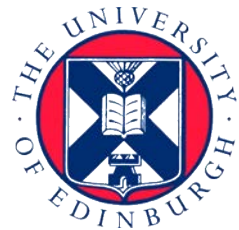 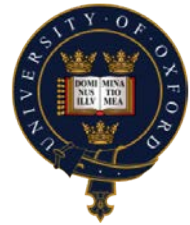

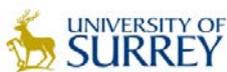 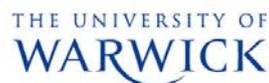 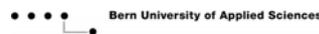 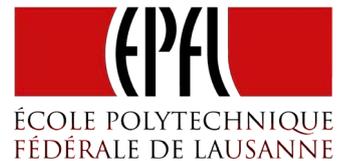

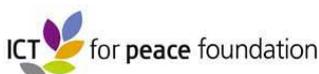 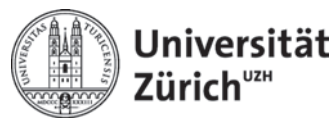 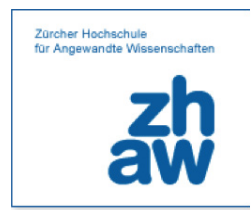 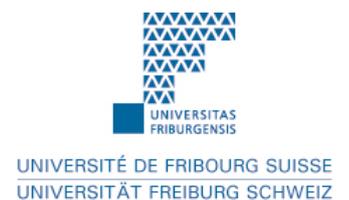

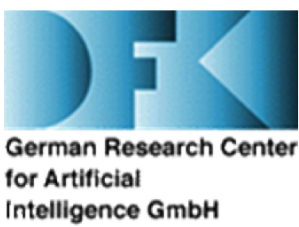 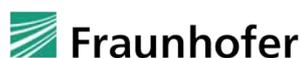

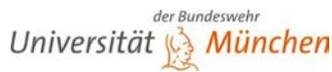 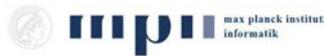 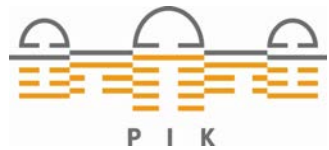 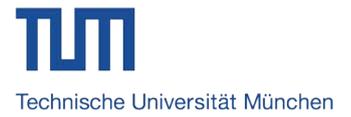
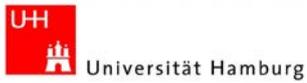 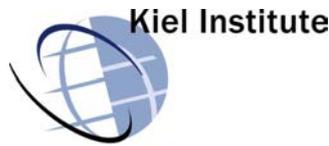 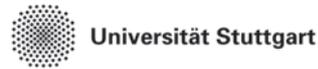 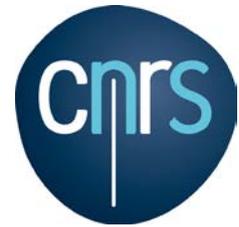
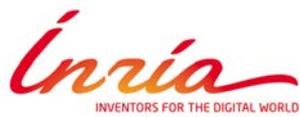 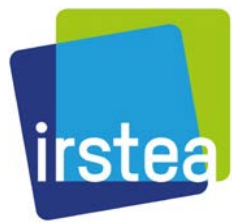 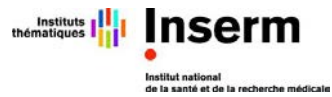 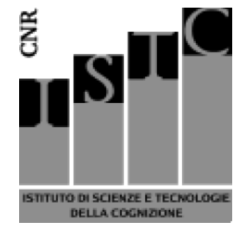
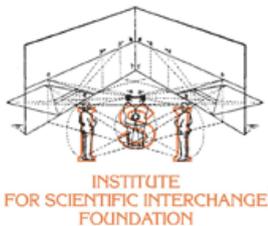 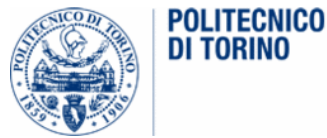 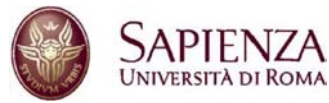 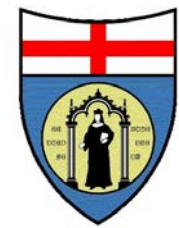
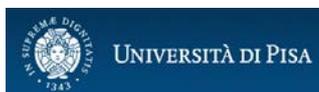 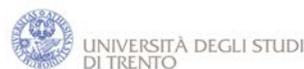 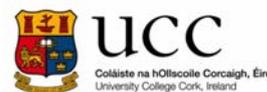 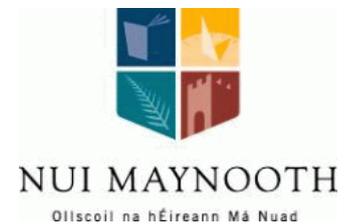
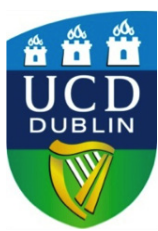 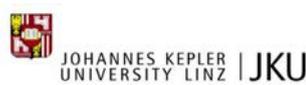 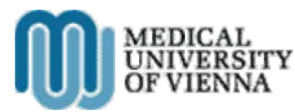 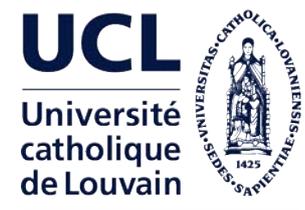
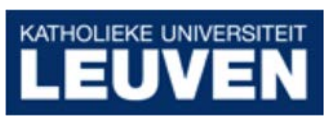 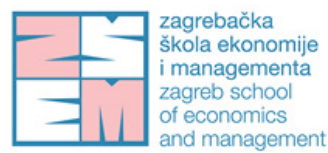 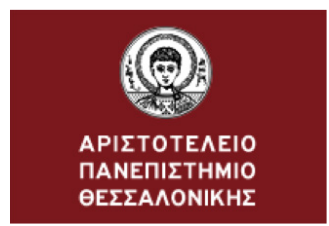 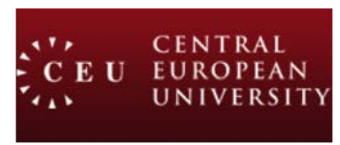
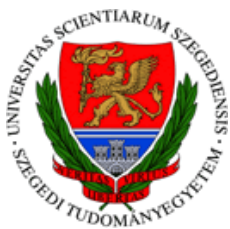 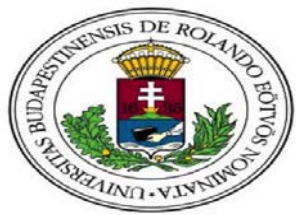 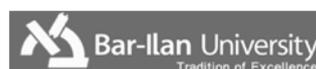 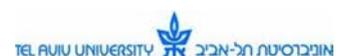

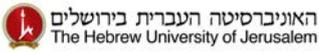 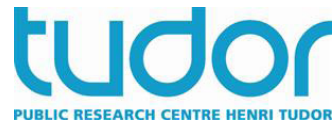 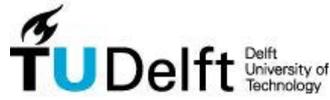 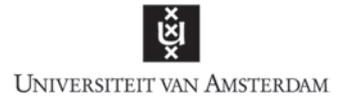
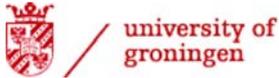 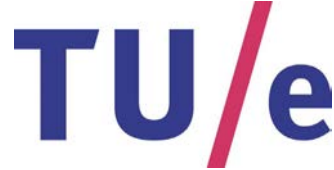 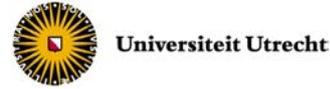 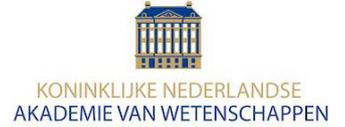
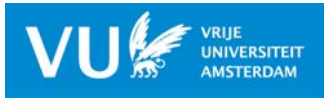 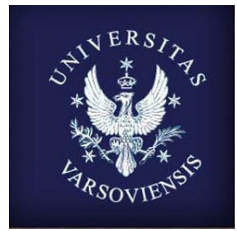 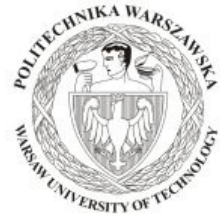 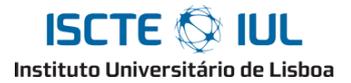
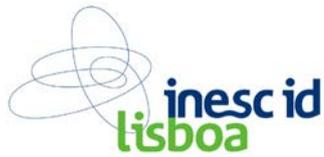 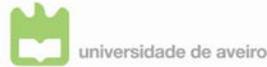 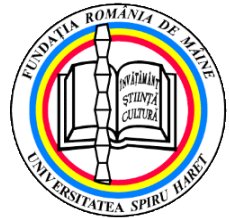 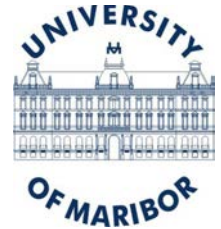
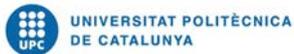 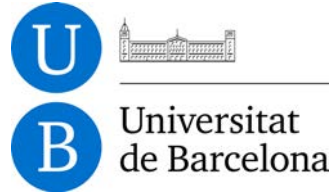 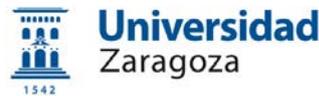 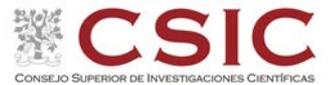
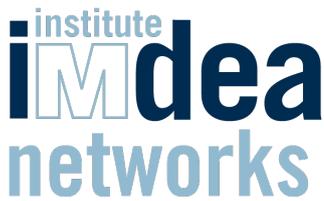 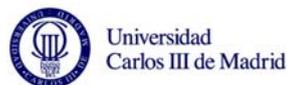 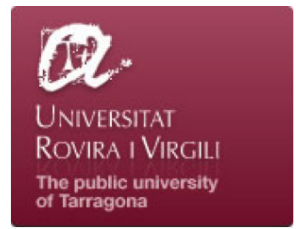 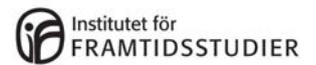
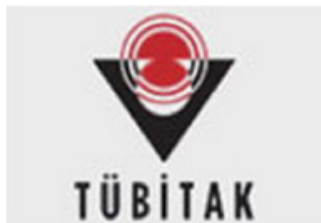 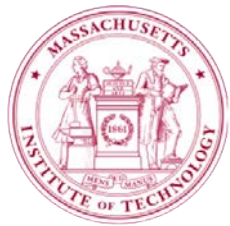 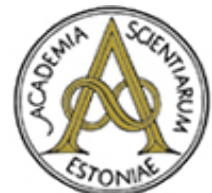 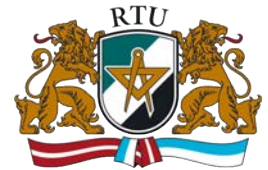

# Proposal Abstract

FuturICT is a FET Flagship project using collective, participatory research, integrated across the fields of ICT, the social sciences and complexity science, to design socio-inspired technology and develop a science of global, socially interactive systems. The project will bring together, on a global level, Big Data, new modelling techniques and new forms of interaction, leading to a new understanding of society and its co-evolution with technology. It will place Europe at the forefront of a major scientific drive to understand, explore and manage our complex, connected world in a more sustainable and resilient manner.

FuturICT is motivated by the fact that ubiquitous communication and sensing blur the boundaries between the physical and digital worlds, creating unparalleled opportunities for understanding the socio-economic fabric of our world, and for empowering humanity to make informed, responsible decisions for its future. The intimate, complex and dynamic relationship between global, networked ICT systems and human society directly influences the complexity and manageability of both. This also opens up the possibility to fundamentally change the way ICT will be designed, built and operated, to reflect the need for socially interactive, ethically sensitive, trustworthy, self-organised and reliable systems.

FuturICT will create a new public resource - value-oriented tools and models to aggregate, access, query and understand vast amounts of data. Information from open sources, real-time devices and mobile sensors will be integrated with multi-scale models of the behaviour of social, technological, environmental and economic systems, which can be interrogated by policy-makers, business people and citizens alike. Together, these will build an eco-system that will lead to new business models, scientific paradigm shifts and more rapid and effective ways to create and disseminate new knowledge and social benefits – thereby forming an innovation accelerator.

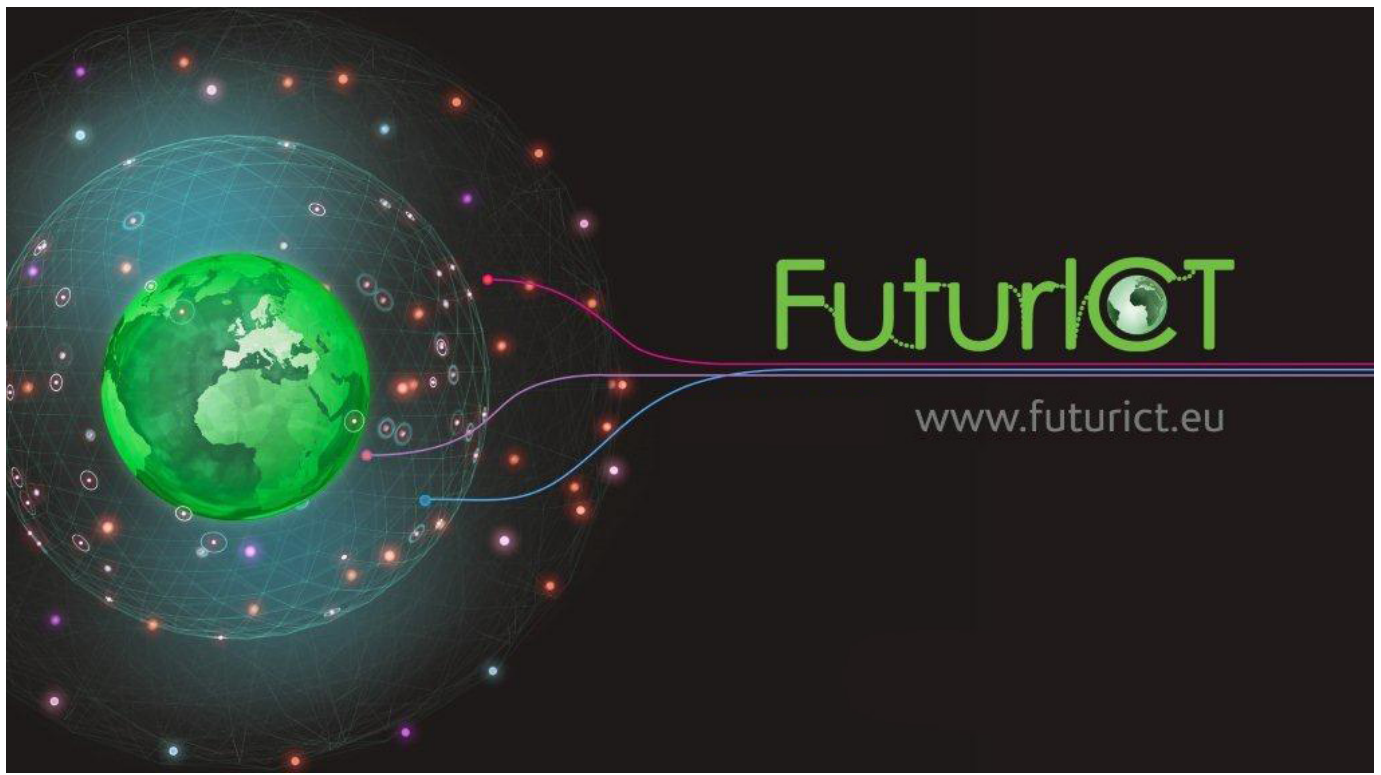

# Executive Summary

**Unifying Goal**

The unifying goal of the FuturICT FET flagship is to integrate the fields of information and communication technologies (ICT), social sciences and complexity science, to develop a new kind of participatory science and technology that will help us to understand, explore and manage the complex, global, socially interactive systems that make up our world today, while at the same time paving the way for a new paradigm of ICT systems that will leverage socio-inspired self-organisation, self-regulation, and collective awareness.

FuturICT will bring together, on a global level, Big Data, new modelling techniques and new forms of interaction, leading to a new understanding of society and its co-evolution with technology. It will place Europe at the forefront of a major scientific drive to understand, explore and manage our complex, connected world in a sustainable and resilient manner.

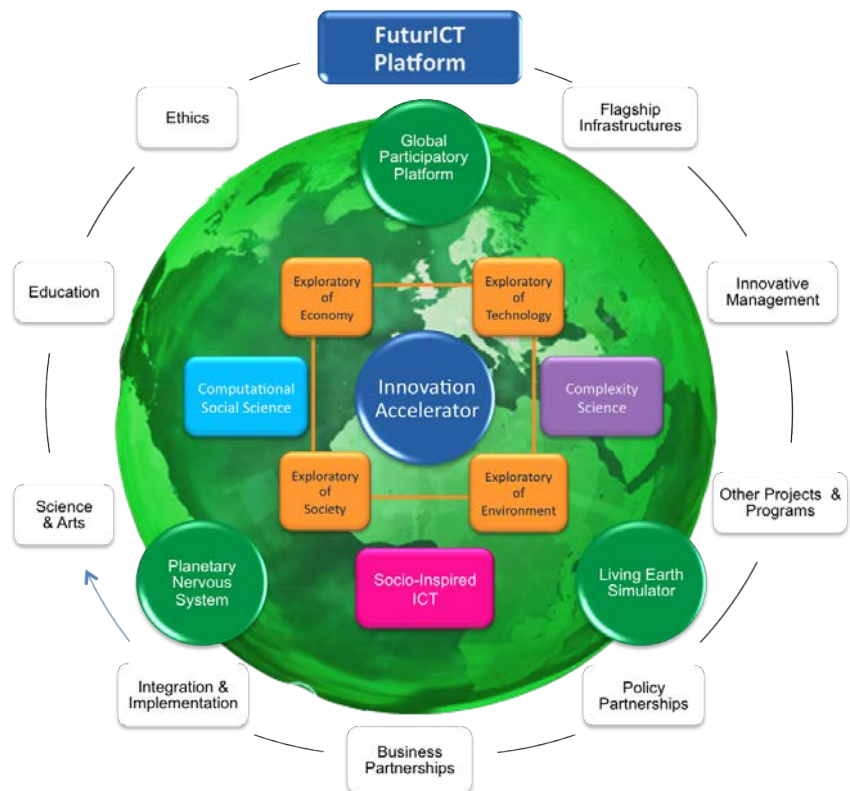

*FuturICT will integrate different scientific areas and activities into a FuturICT platform that will enable participatory science and technology to manage our complex world in a sustainable and resilient manner.*

FuturICT is motivated by the fact that ubiquitous communication and sensing blur the boundaries between the physical and digital worlds, creating unparalleled opportunities for understanding the socio-economic fabric of our world, and for empowering humanity to make informed, responsible decisions for its future. The intimate, complex and dynamic relationship between global, networked ICT systems and human society directly influences the complexity and manageability of both. This also opens up the possibility to fundamentally change the way ICT will be designed, built and operated, to reflect the need for socially interactive, ethically sensitive, trustworthy, self-organised and reliable systems.

FuturICT will create a new public resource - value-oriented tools and models to aggregate, access, query and understand vast amounts of data. Information from open sources, real-time devices and mobile sensors will be integrated with multi-scale models of the behaviour of social, technological, environmental and economic systems, which can be interrogated by policy-makers, business people and citizens alike. Together, these will build an information eco-system that will lead to new business models, scientific paradigm shifts and more rapid and effective ways to create and disseminate new knowledge and social benefits – thereby forming an innovation accelerator.

To realise this vision, FuturICT is organised into a number of closely interacting Focus Areas which will be introduced in Section 2. Given below are a high level overview of the research areas, the motivation behind the research and a plan of action to accomplish the FuturICT goal.

FuturICT will create a "**Planetary Nervous System**" **(PNS)** to orchestrate a high-level, goal driven self-organised, collection and evaluation of Big Data generated from sources such as social media, public infrastructures, smart phones or sensor networks. The aim is to create an increasingly detailed "measurement" and a better understanding of the state of the world. For this, the sensing concept used in the physical and environmental sciences will be combined with machine learning and semantic technologies and extended to social and economic contexts. The information provided by the Planetary Nervous System will fuel the development of more realistic, and eventually, global scale models that bring data and theories together, to form a "**Living Earth Simulator**" **(LES)** enabling the simulation of "what if …" scenarios. The LES will reveal causal interdependencies and visualise possible short-term scenarios, highlight possible side effects and test critical model assumptions. The "**Global Participatory Platform**" **(GPP)** will open up FuturICT's data, models, and methods for everyone. It will also support interactivity, participation, and collaboration, and furthermore provide experimental and educational platforms. The activities to develop a "**Global System Science**" **(GSS)** will lay the theoretical foundations for these platforms, while the focus on socio-inspired ICT will use the insights gained to identify suitable designs for socially interactive systems and the use of mechanism that have proven effective in society as operational principles for ICT systems. FuturICT's "**Exploratories**" will integrate the functionalities of the PNS, LES, and GPP, and produce real-life impacts in areas such as Health, Finance, Future Cities, Smart Energy Systems, and Environment. Furthermore, the "**Innovation Accelerator**" **(IA)** will develop new approaches to accelerate inventions and innovations. A strong focus on ethics will cut across all activities and develop value-sensitive ICT. Targeted integration efforts will push towards the creation of a powerful and integrated ICT platform that puts humans in the centre of attention.

**Fitting the Flagship Call**

FuturICT is a visionary and **ambitious** Big Science project requiring a **large-scale** and collective, **federated** effort of the European academic powerhouses, brought together to provide the necessary expertise and resources to achieve its unifying goal. FuturICT has a work plan of science-driven research ultimately to tackle global **societal challenges**. FuturICT is **interdisciplinary**, bringing together not only the core groups of ICT, social science and complexity science, but also a range of other fields such as economics, engineering and more, leading to innovations of significant **societal** and **economic benefit**.

To achieve European leadership, FuturICT will link Europe's leading academic institutions and businesses to instigate the scientific and technological paradigm shifts needed to achieve FuturICT's goal.

FuturICT capitalises on three developments: the collection of Big Data of techno-socio-economic activities, the availability of previously unseen supercomputing power, and opportunities created by social media and socially interactive technologies. These developments will eventually enable the creation of a "Planetary Nervous System" to measure the state of the world in real-time, a "Living Earth Simulator" to study possible scenarios resulting from causal interdependencies between different processes and systems, and a "Global Participatory Platform" to allow people to solve problems collaboratively that are more complex than any one individual or team can handle. FuturICT will hence develop new science and technology to better manage the opportunities and challenges of our complex, connected world. FuturICT will also create new business opportunities through its new ICT paradigm, characterised by socio-inspired ICT, a co-evolution of ICT with society, platforms for collective awareness, ethical and value-sensitive design, and a whole new information ecosystem fuelled by FuturICT's Innovation Accelerator, Global Systems Science, and Exploratories.

FuturICT is in perfect alignment with Europe's Vision 2020 with its strong focus on innovation and socio-economic-environmental as well as health challenges, and also by supporting European Leadership. Big Data is the "oil of the 21st century". As Europe is scarce of natural resources, it must build on ideas. Big Data and the ability to refine them can become Europe's resource. Big Data is the perfect area of innovation for Europe offering new opportunities for small and medium-sized businesses, for self-employment, for an age of creativity leading to economic, social and cultural prosperity. FuturICT is committed to create an innovation ecosystem jointly with other initiatives such as the Climate KIC project, the Future Internet and the Open Data movement.

# Contents

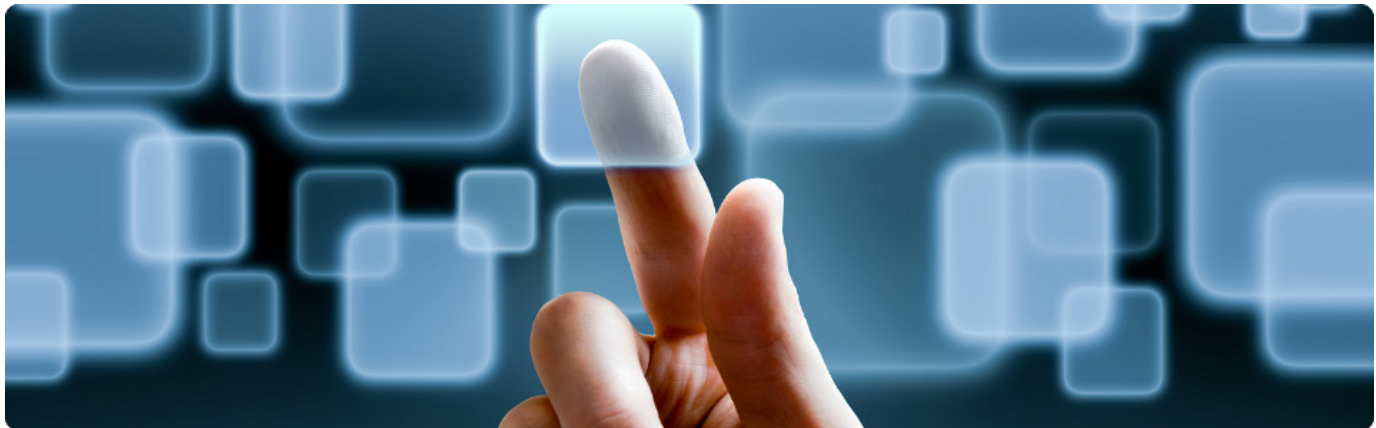

# 1. Flagship Concept & Objectives

## 1.1 Vision

FuturICT aims to extend the horizon of knowledge about our global techno-socio-economic system(s), in particular the effects, consequences and potential benefits of the increasingly complex and intimate relationship between the human society and global, intelligent, networked information processing systems. This will empower humanity to make better informed, responsible decisions for its future. It will also lead to a new generation of socio-inspired and socially interactive information and communication technologies (ICT) that will leverage self-organisation, self-regulation, and collective awareness to foster resilience, stability, trustworthiness, and adherence to basic ethical principles.

The results of accelerating progress in science and technology are increasingly impacting our world. From our earliest roots with language and writing through the invention of the printing press, technology has fundamentally changed society. The invention of communication and transportation technologies in the 19th century (railroads, steam ships, telegraph, telephone) fundamentally changed societies and their relation with nature. The emergence of information technologies in the 20th century is leading to an even more profound transformation in social organisation, as information and communication technologies pervade all aspects of human society. Yet we do not know where this trend is heading, and how it can be harvested to improve human conditions and assure sustainability of the emerging social structure. We lack the tools to understand the emerging system and to anticipate potential consequences of alternative actions.

Hence, a science of society must consider the close interaction with and through ICT systems. Conversely, the scientific study of technological evolution is incomplete, if it is not connected with the scientific study of society. The evolution of the networking infrastructure underlying today's Internet is in many ways much more driven by socio-economic developments than by technological advances. Furthermore, information and communication systems are increasingly becoming "**artificial social systems**". For example, more than 50% of financial transactions previously made by humans are now performed by computers that create an internal picture of their (market) environment, project trends into the future, communicate with other computers, and take autonomous decisions. As recent large-scale ICT failures, the flash crash on May 6, 2010, the explosive spreading of cybercrime and increasing worries about cyber-war demonstrate, such socially interactive systems are increasingly suffering from the same problems that worry societies: coordination failures, instabilities, an inefficient use of resources, crime, and conflict. Hence, our enormous dependence on critical ICT infrastructures calls for novel resilient, adaptive, self-organising systems that go beyond the current ICT design paradigm.

Rooted in the above insights, **FuturICT** is an ambitious and paradigm-changing project that will deliver new science and technology to explore, understand and manage our strongly connected world in a more sustainable and resilient way. It will also fundamentally change the way ICT will be designed, built and operated, responding to the need for systems that are ethically sensitive, trustworthy, adaptive, resilient and able to support not only individuals, but the human society as a whole with its multitude of communities, interests and often contradictory objectives manifested across different spatial and temporal scales.

To address the challenges and opportunities of information society and benefit scientists, decision-makers in business and politics, and citizens alike, FuturICT aims at **integrating computer science with social science and complexity science**. Were any one of these three disciplines neglected, FuturICT could not achieve its unifying goal. Computer Science is needed to collect and evaluate data of our techno-socio-economic world, in order to be able to calibrate, test, and derive more realistic social and complexity science models and generate economic and societal benefits. Complexity science is needed to gain an explanatory understanding of the non-linear and probabilistic feedback, cascading, and side effects in social and ICT systems, and the emergent phenomena occurring in them. Finally, social sciences are needed to understand and properly design the interactions of humans with ICT and **cyber-social** systems, to study the behavioural assumptions that govern the underlying micro-processes, dynamical phenomena, self-organisation, self-regulation, and evolution of socially interactive systems, be it in society, in ICT systems, or cyber-social systems.

In the past years, Europe has carried out a lot of research activities in the areas of complexity science and computational social science, developing, for example, agent-based models of socio-economic systems that go beyond the scope of purely qualitative or purely analytical models. The goal of FuturICT is to integrate these fragmented research activities. Furthermore, FuturICT wants to build an open platform that allows researchers from many other projects in Europe and beyond, businesses, NGOs and citizens to contribute to and benefit from the success of FuturICT's vision and implementation. In fact, scientific communities in the USA, Japan, China, Australia, and South America have voiced a strong interest in contributing to the realisation of FuturICT's vision, and they are continuously interacting with the European core project.

FuturICT will help Europe tackle some of the major issues that we are all facing today, and to find new ways to produce commercial advantages for Europe. During FuturICT's Pilot Phase, the Coordination Action (CA) Consortium made great strides in this direction, building on and amplifying existing trends: ICT is becoming more social, social science is becoming more computational, while complexity science is showing us the way to understand emergent properties of connected systems of systems. The strength of the vision is underscored by the fact that FuturICT has now already about 2000 Supporters and many offers of matching support in excess of €90m over the next two and a half years, and has even been featured among the "Ten World Changing Ideas", a cover story in Scientific American.

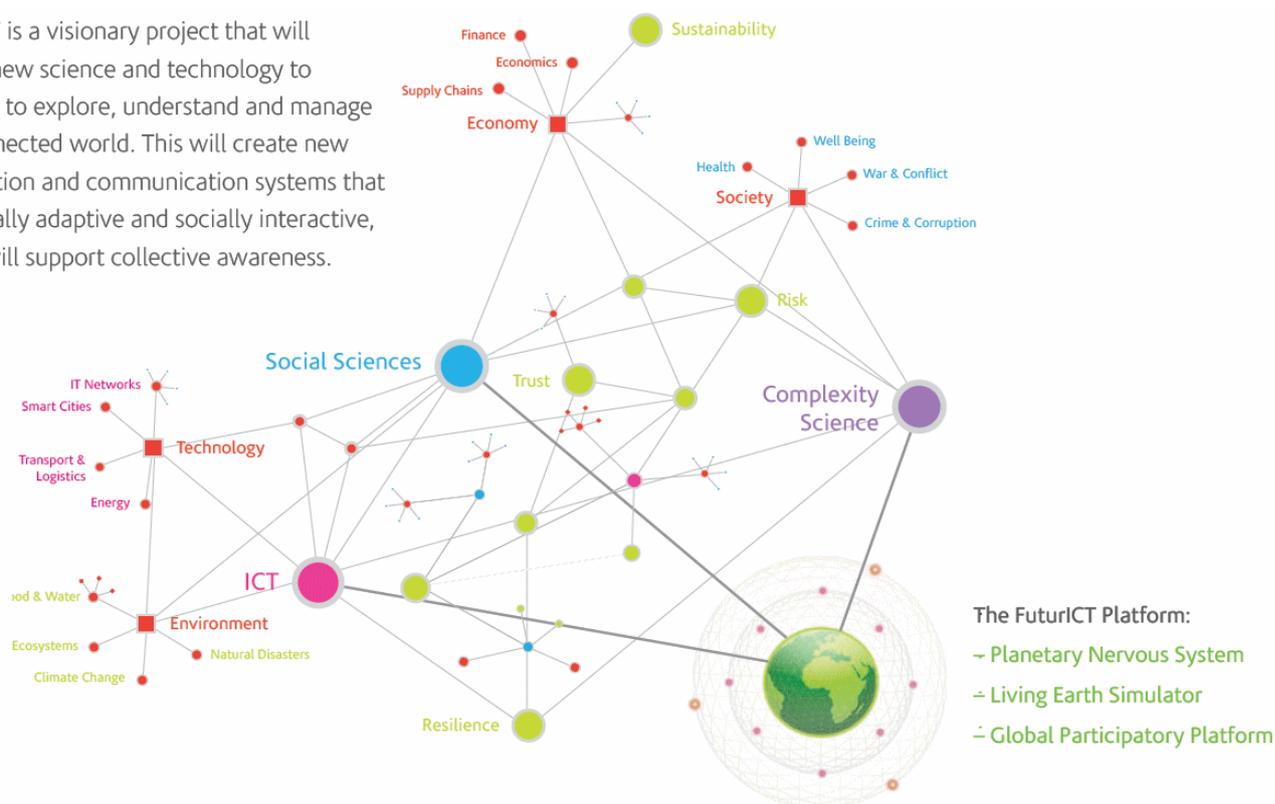

*Figure 1.1*

## 1.2 Aims

To realise the bold vision described in the previous section, FuturICT will develop:

1. Theoretical foundations for a new Global Systems Science and a systemic risk calculus including models to describe how new levels of information spark off new trends and efficient methods to model and simulate real-world complex systems;

2. A new generation of socio-inspired, socially aware, socially interactive ICT that will build on an in-depth understanding of the co-evolution of society and ICT to realise new paradigms for self-organisation, adaptation, trustworthiness and embedding of ethical values and principles of responsible use into its core operational procedures;

3. A new understanding of society that will bring together, on a global level, Big Data provided by major advances in ICT with new modelling techniques.

A **tangible outcome** of the project will be a global but decentralised, democratically controlled information platform to combine online data and real-time measurements together with novel theoretical models and experimental methods to achieve a paradigm shift in our understanding of today's highly inter-dependent and complex world and make our techno-socio-economic systems more flexible, adaptive, resilient and sustainable through a participatory approach. The principal components of this platform will be:

- The "**Planetary Nervous System**" **(PNS)** will bring together existing and new data sources, such as demographic data, mobility and activity patterns, financial and economic data, epidemiological and other data, and enrich them creating higher-level semantic meaning. To this end it will implement a privacy-respecting and user-controlled paradigm of social data mining, featuring question-driven, self-organised, self-optimising, self-regulating, and decentralised measurements, data collection and enrichment. In summary, PNS will turn raw data into semantically meaningful information.

  In summary, PNS will turn raw data into semantically meaningful information.

- The "**Living Earth Simulator**" **(LES)** will combine the information provided by the PNS with models to enable simulations of social, economic, technical and environmental developments, enabling a better understanding of the state of our world, and its possible futures.

  In summary, the LES will turn information into domain-specific knowledge and explore possible impacts of decisions and actions by a pluralistic analysis of "what if …" scenarios, in order to become better aware of possible risks and opportunities.

- A "**Global Participatory Platform**" **(GPP)** will enable citizens, businesspeople, scientists and policy-makers to interact with the Planetary Nervous System, the Living Earth Simulator, and FuturICT's Exploratories. It will display the answers to their questions in new, exciting and engaging ways, using also serious games and immersive technologies. It will feature an Open World of Modelling Platform, a trusted brokerage system to bring together data producers, data consumers, and resource (e.g. computing systems) providers into a novel information ecosystem. The outcome of this work will be a deep embedding of evidence-based decision-making throughout society, enabled by large-scale adoption of the resources mediated by the GPP.

- In summary, the GPP will connect data, models, and knowledge with people, who will produce and consume them at the same time ("prosumers"), to create knowledge and collaboration opportunities that go beyond what any one user or any one team can achieve.

The activities to develop a "Global System Science" (GSS) will lay the theoretical foundations for these platforms, while FuturICT's focus on "socio-inspired ICT" will use the insights gained to identify suitable designs for socially interactive systems. FuturICT will form "Exploratories" at various European institutions, which will integrate the functionality of the PNS, LES, and GPP, and produce real-life impacts in areas such as Health, Finance, Future Cities, Smart Energy Systems, and Environment. The combination of PNS, LES and GPP will empower non-experts to interactively get answers to high-level queries about our complex techno-socio-economic system(s) on a global scale, e.g. about factors relevant for social well-being. This will lower barriers to social, political and economic participation and create an information ecosystem enabling new data- and model-based businesses. By connecting the Exploratories over time, FuturICT will eventually be able to study the interdependencies between the Environment, the Economy, our Society and Technology, and to build an integrated FuturICT platform to study global interdependencies and dynamics.

As a further project component, the **Innovation Accelerator** will speed up the development of new scientific knowledge, and achieve more rapid routes to dissemination and commercial benefit.

As a result, FuturICT will promote an **ICT paradigm** characterised by the following features:

- Participatory platforms as instruments to enable new forms of knowledge (co-)creation, collective awareness, and social, economic, and political participation.
- Socially interactive and socio-inspired ICT as new paradigms for self-organisation, adaptation and trustworthiness.
- A new information ecosystem and a favourable co-evolution of ICT with society.
- Privacy-respecting mining of Big Data and user control of their data as well as ethical, value-sensitive ICT (responsive and responsible).

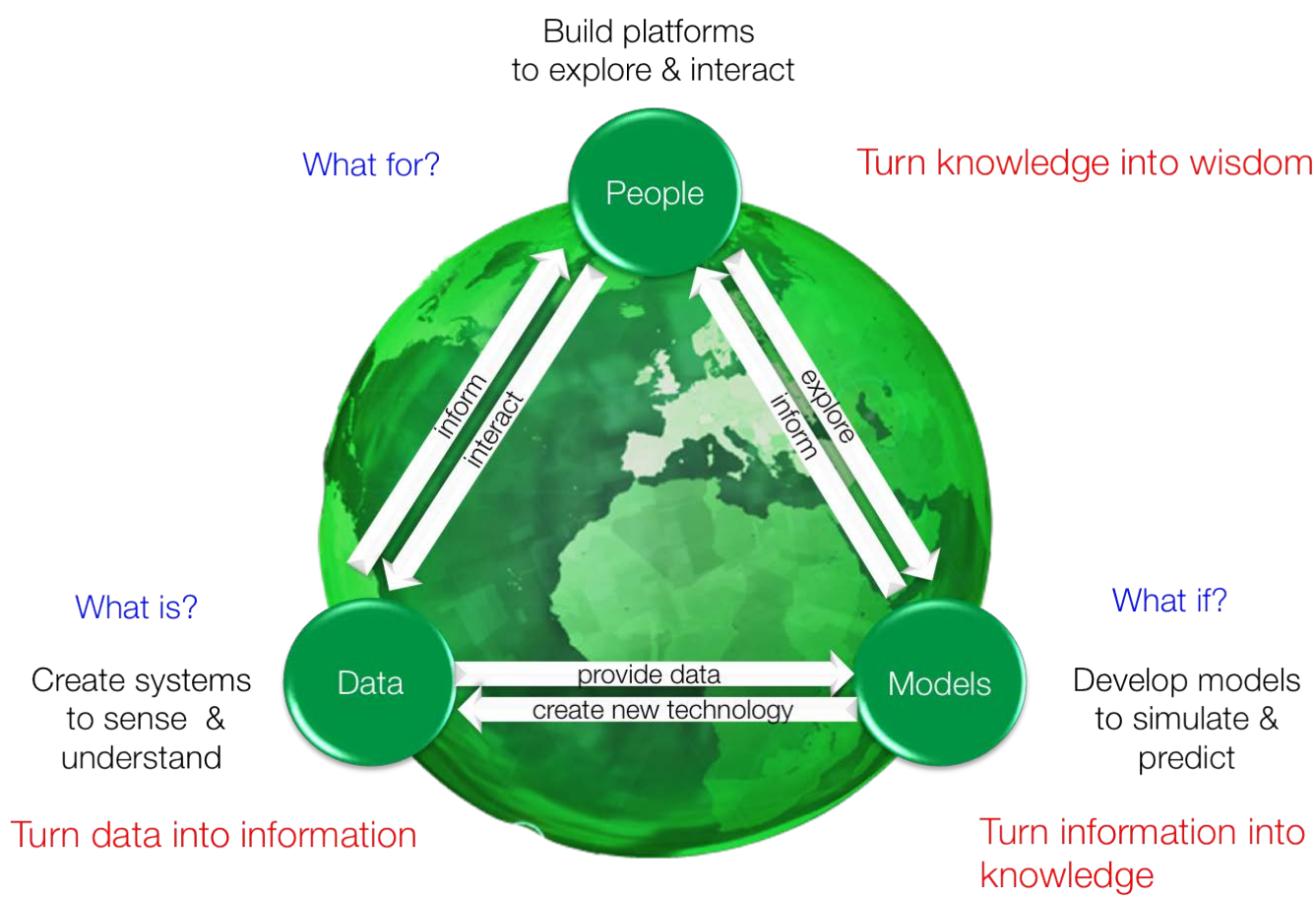

**Figure 1.2** FuturICT engages into bringing together data, models, and people, namely by creating a FuturICT platform that encompasses a Planetary Nervous System, a Living Earth Simulator, and a Global Participatory Platform.

To achieve the above aims, **interdisciplinarity** is at the heart of FuturICT. The world has entered the era of information technology, which can transform society profoundly and dramatically. In order to avoid systemic risks and make use of the opportunities technology affords, it is of fundamental importance to understand the **co-evolution** of ICT and society. To implement features such as trust, reliability, resilience and respect for basic ethical values, a new generation of self-organising and self-regulating, **socially interactive ICT systems** is required.

For this, FuturICT aims to combine approaches from the natural, engineering, and social sciences to a previously unseen extent, moving from multi-disciplinary research towards inter-disciplinary and trans-disciplinary research. Aiming, as it does, at an understanding of techno-socio-economic systems, the FuturICT FET flagship project goes significantly beyond the scope of other large-scale projects. FuturICT's approach is characterised by combining empirically grounded modelling and theoretically founded data mining with advances in quantitative and qualitative social science, paying particular attention to ethical implications. The integration of ICT with the social and complexity sciences creates the opportunity to overcome the limitations of pure data mining and machine learning on the one hand, and of purely qualitative, purely analytical or purely model-based approaches on the other hand. It also creates the opportunity to overcome the basic limitations of todays ICT in the face of the complexity and dynamics of the real world, and the more and more intimately interweaved virtual and physical world.

Within the next 10 years, FuturICT will enable participatory science and computing by providing new methods, instruments and platforms, which will allow general society to interact with data and models of our human society and its possible evolutions.

## 1.3 Scale of Ambition

Given the vision described above, FuturICT by its very nature can only be realised as a FET flagship. It requires a large-scale, federated effort, since no single country alone would have the required expertise and resources. It is **science-driven**, striving to vastly increase our understanding of the behaviour of our global society. It is **transformative,** as it is going to redefine the way we understand, design and manage the modern interconnected world. It is **interdisciplinary**, building on crucial inputs from ICT experts, complexity scientists and social scientists, to push for quick advances in each of these areas. It is **integrative** as the convergence among and across different research areas will define new research challenges, which could not be met by any of these disciplines alone. FuturICT has a **unifying goal**, namely to develop new participatory science and technology to better understand and manage our complex, strongly connected world. Finding new ways how to govern techno-socio-economic systems in a more sustainable and resilient way constitutes an **immense ambition**. This ambitious vision has excited a lot of supporting researchers all over the world, and also the public media and the business world. The power of this unifying vision has been demonstrated repeatedly during the FuturICT Pilot Phase, during which over 160 submissions were sent in to a call for proposals, and FuturICT Hubs were established in more than 25 countries.

FuturICT has a **long-term vision**, focusing on a step change in science and society, driven by novel ICT and the vast quantities of data, which our everyday activities are now generating. Through partnerships with Europe's academic powerhouses, science organisations and companies, FuturICT has built an effective and efficient basis to turn the broad knowledge and skills of its impressive expert network into innovations for the generations to come. FuturICT has an **unprecedented scale** not just in terms of the size of its scientific ambitions but also in the range of areas upon which the project's success would have **immense beneficial impact**, and the vast quantity of scientists, businesspeople, policy makers and citizens, which the project has already succeeded in bringing together. It plays a strong **federating role**, creating a common structure for a broad range of research efforts, which were previously fragmented across disciplines and nations. FuturICT is building a participatory platform of an entirely new scope and power to support and promote Europe's science and innovation.

### What makes FuturICT visionary?

FuturICT aims at defining the **Global Systems Science** needed to address the most pressing social, financial, economic, and political challenges more successfully. FuturICT proposes a paradigm shift in the way we see and plan the world around us. The projects intends to revolutionise our approach to technology and society by an integrated analysis of these two sides of our world as defining a strongly connected techno-socio-economic system, for which we must develop a new science that spans across the current disciplinary silos. Most 21$^{st}$ century challenges cannot be solved by technological innovation alone. They have a significant social component that determines the success or failure of business or policy decisions. FuturICT's has the ambition of a quantitative, scientifically informed approach to the decision making process in the 21$^{st}$ century. In order to fully tap into the power of this new kind of thinking, we need a global infrastructure that responsibly collects, makes sense of, and uses the Big Data being generated today – an endeavour of simply unprecedented scale.

**Why does FuturICT need a large-scale, federated effort?**

FuturICT goes significantly beyond the scope of other large-scale projects, by aiming to gain an explanatory understanding of techno-socio-economic-environmental systems on a global scale, and in much shorter time spans. This requires intellectual and financial resources that go significantly beyond what even big academic institutions or a single European country could set aside. Hence, FuturICT requires a federated approach. It has built a European and even world-wide network of excellence, including currently around 2,000 experts, integrating social sciences, complexity science and ICT expertise in a large number of closely interacting national or regional communities.

In particular, FuturICT has an open platform approach that allows the project to connect with other research projects all over the world. Like an 'app store', the Brokerage Platform (which is part of the Global Participatory Platform) invites contributions from other institutions, businesses, NGOs, and citizens for free or for a fee. The Data and Model Commons that FuturICT wants to create will catalyse activities going far beyond what the EC FET flagship budget itself will be able to cover. In fact, FuturICT has already inspired recent proposals, bringing together novel ICT, Internet data, and social or aspects of complexity science. FuturICT will make sure to stay in close contact with these projects to jointly create greater impact.

Exploratories with a restrictive focus on topics such as Health or Crime or Transportation may produce excellent science and effective guidance for planners and policy-makers, but such topics are obviously interlinked. Without an overarching vision and support for collaboration at the most fundamental levels, the effectiveness of the science of these individual Exploratories could not achieve the ultimate ability to transform the design of Future Cities, the improvement of worldwide Health Systems and deter of Crime. Over time, FuturICT will integrate these and other Exploratories into one FuturICT platform, to yield effective benefits that could not be achieved otherwise.

**Why Europe needs FuturICT as a FET Flagship project?**

As the population grows and ages, environmental conditions change, and ICT becomes more and more pervasive, we are facing systemic challenges and risks that up to now have been impossible to identify, anticipate, and manage. FuturICT will develop a Global Systems Science recognizing that, while each part of a complex system might be aware of its own modes of failure, the impact of these failures on other interdependent systems is often not considered, unknown, or underestimated, so that when these links are exercised, a catastrophic failure may ensue. A FET flagship project that brings together expertise in complex systems, ICT, social sciences and engineering is needed to anticipate dependencies between separate systems, estimate their consequences and help to plan effective strategies for responding to critical situations. FuturICT will be able to provide access to data and demonstrators or case studies (use cases), so that a more resilient and safer infrastructure can be put in place.

Exploiting the science and technology generated by FuturICT will put Europe at the very forefront of a technological revolution that will change the way in which we run our world. Europe will generate the instruments and science that will develop a new integrated approach to better understand and manage systemic risks and opportunities. At the same time, Europe will benefit from leadership in the innovation area of integrating social and technological systems.

In the past, the EU and national funding agencies have already spent hundreds of millions of Euros on projects related to the domains addressed by FuturICT but, because of the fragmented nature of these projects they have not been able to create the visibility and impact that FuturICT will be able to generate. Many research activities and projects scattered all over Europe that are compatible with FuturICT's vision and mission have been identified already. Europe now has the opportunity to pool these resources and expertise to address the grand challenges of techno-socio-economic systems. FuturICT offers an open platform for this. What can be achieved by national projects or EU projects of usual size is too limited in scope and vision, compared to what will be achieved by the FuturICT FET flagship.

## 1.4 Scientific Challenges

For a number of reasons, the increasingly complex techno-socio-economic systems created by humans are becoming less and less manageable with a classical approach, in which systems are analysed in a centralised and top-down way:

- Even managing relatively small systems requires a lot of resources. This difficulty results from the huge number of possible combinations of control parameters and the so-called NP hardness of the optimization problem, which means that the required computational resources "explode" with system size;
- Complex systems are often dominated by self-organised dynamics, which limits possibilities of external control. The interaction of non-linearly interacting system components may cause new, "emergent" system properties, which are often surprising, unexpected, counter-intuitive, or unwanted;
- Complex dynamical systems with many strongly coupled components often show a large variability in time and are hardly predictable. They may sometimes also feature cascading effects that cause undesired extreme events, such as the current financial crisis.

The above issues become apparent even in relatively simple socio-technical systems. For instance, not even traffic light control in big cities can be carried out in real time in an optimal way. Recently, the realization in computer science that even "pure" ICT systems with little or no human interaction (such as high-performance multicore processors) face similar problems, has led to an increasing interest in self-organisation concepts in the ICT community as well. For complex connected dynamical systems, classical planning or optimization approaches do not work well. A flexible, adaptive response is a proper answer to the large variability of the system dynamics and the fact that forecasts are often useless after a short time period and probabilistic. Moreover, a flexible, adaptive response often requires real-time data and decentralised control elements (i.e. a suitable combination of top-down with bottom-up approaches).

Challenges in society imply a *participatory* approach is needed - an aspect covered by FuturICT's **Global Participatory Platform (GPP)**.

Today, most stakeholders still lack the data required for an adaptive real-time management. Developing methods to collect these from the Internet and sensor networks (as they can be established for instance by smartphones) will be the main challenge tackled by the **Planetary Nervous System (PNS)**.

To create an open Data Commons that can benefit everyone, one must address the challenge that collective decision-making processes may be undermined by social influence and that public goods are difficult to establish and maintain. Therefore, in order to create societal benefits by opening up Big Data, one needs to develop **value-sensitive ICT** systems, e.g. to develop methods for privacy-respecting data mining.

More generally, one must find suitable system designs, which promote a beneficial interaction of the system components (including individuals and businesses). For this, FuturICT proposes to develop **socio-inspired ICT**, i.e. to learn from and include principles that make social systems work well. One key concept of socio-inspired ICT is to manage complexity by using the principle of guided self-organisation. The related challenge is to identify and establish the right interactions and institutional settings, as for complex systems there is no simple, general rule how to identify these interactions. This is why FuturICT's **Living Earth Simulator (LES)** is needed. The challenge related to building the LES is to calibrate, validate and develop models with data from the PNS and from experiments (as they may be carried out in the lab, on the web, in multi-player online games, or Interactive Virtual Worlds). Given the bigger and bigger amounts of techno-socio-economic activity data, it becomes possible to do data-driven modelling and perform machine learning. However, gaining an explanatory understanding normally requires the combination with visualisation and advanced theoretical approaches. The latter calls for the development of a **Global Systems Science (GSS)**, which would develop methods to integrate heterogeneous data and heterogeneous models to gain a systemic understanding of global-scale interdependencies.

The **integration** of the above activities is another challenge. The strategy is to connect various **Exploratories** over time, each of which integrates some or all of the above functionalities in an interdisciplinary application area with significant real-life impact. Since all of the above activities will be highly demanding, the production, spreading and practical application of scientific knowledge will have to be significantly accelerated, which is a challenge addressed by the **Innovation Accelerator (IA).**

The ambition of the FuturICT vision goes far beyond anything so far undertaken.

> *The core question is how the intimate, increasingly intractable interweaving of the digital and the social world can be turned from a source of problems into a tool for solving them: both in the social domain and with respect to the next generation ICT systems.*

The main specific challenges resulting from this question fall into four categories

1. Challenges related to the collection, use, and evaluation of Big Data generated at the interface of society and global ICT systems. In particular, emphasis will be on: (a) the Data Orchestration Challenge and (b) the Challenge of Privacy Respecting Social Sensing and Mining;
2. Challenges related to the knowledge and fundamental scientific insights needed for, and generated by the leveraging of such Big Data. In particular, emphasis will be on: (a) the Modelling Challenge and (b) the Computational Social Science Challenge and (c) the Challenge of the co-evolution of ICT and Society;
3. Challenges related to the basic advances in Computer Science and ICT needed for, and facilitated by such data, knowledge and insights. In particular, emphasis will be on: (a) the ICT Systems Challenge, (b) the Ethical and Value Sensitive ICT Challenge and (c) the Challenge of Collective Awareness and Socially Interactive ICT and (d) the Challenge of Socially Inspired Self-Organisation.
4. Challenges related to facilitating real-world impact of the above advances. In particular, emphasis will be on: (a) the Challenge of Democratising Big Data and Models, (b) the Challenge of Building Interactive Exploratories and (c) the Integration Challenge.

**The Data Orchestration Challenge**

More information has been created in the last few years than in the entire history of humanity, and the data creation rate is rapidly increasing. Much of this new information relates not to archival facts, but to real-life events and trends. Such data range from government data banks released into the public domain in the wake of open government initiatives through online digital media, social media, blogs, and online videos to volunteered sensor data (e.g. location and travel patterns) from users' mobile phones. The stated mission of FuturICT is to leverage such data as real-time "measurement" of the socio-economic state of the world. A core difference to previous projects in the area (see State of the Art) is that FuturICT aims to lay the foundations for broad, interactive use of such data by non-experts. Thus, while previous work concerned systems that were carefully designed and implemented for a specific task and a specific configuration of the information sources, FuturICT needs to facilitate the automatic transition from high-level queries that can be formulated by a non-expert user to the self-organised configuration of the required information collection and processing system. Thus, in essence, FuturICT must provide products with functionality similar to today's Internet search engines. However, where search engines deal with string matching on largely static, indexed, and organised information (within HMTL pages), the **Planetary Nervous System (PNS)** must (1) interpret semantically complex queries, (2) automatically identify (and acquire) suitable sources of information within the myriads of dynamically changing data streams, and (3) autonomously select appropriate complex pattern recognition and reasoning algorithms to be applied to those streams. Given the complexity of the task, a core aspect will be the ability of the system to learn from the success or failure of past queries and autonomously evolve new, increasingly complex and subtle strategies for the organisation of the data collection and processing for certain query classes.

**The Privacy-Respecting Social Sensing and Mining Challenge**

Much previous research in Reality Mining and Participatory Sensing concentrates on few, mostly statically selected information sources (e.g. mobile phone mobility data) and spatially and temporally constrained questions (e.g. daily traffic patterns within a city). Todays sensed data also fail to describe human activities at an adequate, multi-faceted semantic level, which makes it difficult to observe human behaviour. We thus need to develop a new paradigm of Social Sensing to overcome these limitations by novel semantic-based techniques that can map human behaviour by automated annotation and enrichment methods, which extract multi-dimensional social semantics from raw digital breadcrumbs. Social Sensing needs to facilitate the analysis of data over different temporal and spatial scales, including truly global questions and leverage any data source available: from YouTube videos through Tweets to mobile phone sensor data from volunteers. This poses fundamental challenges with respect to interoperability, data fusion, semantic enrichment and reasoning at different levels of abstraction and spatio-temporal granularity.

Towards further interpretation of the data we need to move towards Social Mining to provide the analytical methods and associated algorithms needed to understand human behaviour by means of automated discovery of the underlying patterns, rules and profiles from the massive datasets of human activity records produced by social sensing. Social mining needs novel concepts for multi-dimensional pattern discovery and machine learning, and for multi-dimensional social network analysis whose ultimate goal is to understand human behaviour through: the discovery of individual social profiles, the analysis of collective behaviours, the analysis and monitoring of information diffusion and social contagion phenomena, the analysis of sentiment and opinion evolution. Social sensing and social mining must also be able to exploit domain knowledge and models of the underlying processes. To this end it will have to integrate the macro vision of complexity and social sciences with the micro vision of data mining and machine learning within a unifying theoretical framework.

Social Sensing and Social Mining need to take into account ethical considerations through informed consent and privacy-respecting algorithms, and by enabling user control. For privacy and practicability reasons, the PNS will also try to refrain from storing excessive amounts of raw data. Instead, the algorithms will be applied "on the fly" to the data as it will be aggregated and, where possible, mostly high-level conclusions will be propagated to the user. Significant advances in pattern analysis and reasoning will be facilitated by this vision.

**The Modelling Challenge**

Human problems that emerge from physical or social problems are becoming increasingly complex as new technologies provide more ways to communicate and interact. It is known that many key societal problems evolve bottom up, but while trying to solve these problems, macroscopic top-down approaches as used, for example, in economic policy making, transportation planning, the organisation of health care etc. often create other problems. Therefore, new kinds of models that combine micro- and macro-perspectives and incorporate the way systems evolve are sorely needed.

FuturICT will pioneer models of social self-organisation, built on elementary units that are treated as agents, which interact in ways that capture the essence of the richness and complexity of socially interactive systems. However, self-organising systems do not necessarily imply sustainable or optimal states. To understand how major societal problems come about, we require new kinds of models that enable us to understand bottom-up self-organisation, while at the same time, these systems are designed and managed partly top down.

The decisive step forward in modelling is to address the challenges of multi-level dynamics, internal complexity, and learning. Complex social systems are systems in which many different levels interact and mutually affect one another. In particular, while the elementary units of complex socially interactive systems adapt to one another and thereby give birth to new, macroscopic system properties, these may produce micro-level effects, behaviours, and properties that need to be identified and considered as well.

To model such a complex dynamics is of vital scientific and practical interest. It is a critical requirement for a **Global Systems Science (GSS)** to make a serious step ahead in the understanding, design or management of such highly non-trivial systems. In particular, we need to model how humans perceive, interpret and respond to emerging collective phenomena or critical situations in order to better understand how agents might react to new opportunities and challenges. To enable acceptable and effective management solutions, any such modelling attempts must be closely connected with meaningful data mining activities as well as suitable and efficient experimental approaches.

Therefore, a related core challenge is to calibrate, validate and develop models with newly available data from experiments (as they might be carried out in the lab, on the web, in multi-player online games, or Interactive Virtual Worlds, for example). Given the bigger and bigger amounts of techno-socio-economic activity data, it now becomes possible to do data-driven modelling and to use powerful machine learning approaches. However, in order to gain an intuitive understanding, such data mining and machine learning activities must be combined with advanced theoretical and visualisation approaches. To eventually achieve a solid foundation of a Global Systems Science and a systemic overview of global-scale interdependencies, one must also develop methods to integrate heterogeneous data and heterogeneous models, and to assess the transferability of modelling approaches from one problem area or situational context to another.

### The Challenge of Democratising Big Data and Models

A core requirement towards the FuturICT vision is to provide a platform and support for the growing communities of "prosumers", who both consume and create information, models, analyses and other resources. Towards this aim, the **Global Participatory Platform (GPP)** faces two challenges. On one hand, it needs to make the power of the Planetary Nervous System and the Living Earth Simulator available to a spectrum of people, from professionals to motivated members of the public. The goal is to embed the capacity for evidence-based decision- making into our society by empowering individuals to go beyond their bounded rationality to embrace the availability of data and models to interpret the data. On the other hand, the GPP will need to facilitate and broker contributions from stakeholders including the public, scientists, computing centres, government agencies. Such contributions can be data, models, software, time, participation in serious games (or the right to observe gaming behaviour), and viewpoints in debates about policy implications. Thus, a key component of the GPP will be a trustworthy, transparent, privacy-respecting brokerage platform. Designing and implementing such a platform on the envisaged scale will be a major challenge; so will be making the data, methodologies, models, algorithms, and libraries available to third party developers through appropriate high-level toolkits that empower a far wider user base.

In short, it can be said that the challenge of the GPP is to make Big Data and the associated models a pubic good ("Data Commons"). However, public goods are difficult to establish and maintain. Some of the possible problems are: collective delusions, privacy intrusion, data pollution, cybercrime, or misuse. Related key challenges are incentives (how motivate people to contribute) and trust (how to convince them and ensure that their contributions and data will only be utilised towards a specified aim).

As is often the case with major scientific and technological developments, Big Data and the ICT infrastructures for analysing it, represent an opportunity for further developments while creating a new set of challenges and new problems to manage. In particular, a major question is posed to the study and management of sociotechnical systems, i.e. what is the effect of participatory data provision and treatment on the organisation and structure of society? What is the expected impact on different levels of social structures, on the one-to-one *versus* one-to-many relationships between individuals, and between individuals and the different segments of society, be they aggregates, groups, communities, or institutions? What will be the expected impact of participatory data provision on truth-certifying institutions? Intellectual property is only one aspect of a more general problem that invests the role of authorities (including scientific ones) in assessing and publicising facts. In short, we need to make scientific, technical, and methodological instruments available to observe, explore, and study the emergence of new forms of social regulation, considering growing awareness at the population level and making use of it.

> **The Fitness Universe game.** The Fitness Universe Game utilises the GPP to bring together a wide range of stake- holders in a research-driven approach to adaptive problem solving. The connectivity of the GPP is leveraged to allow game developers to implement a wide range of different assets sourced semantically from the web within a game. In turn, these com- ponents allow for ethical data capture from players, and its subsequent analysis. This data is then used to refine the game, and inform policymakers of its impact. Where this differs from other adaptive gaming platforms is the power leveraged by the big data and complexity modelling techniques at the heart of FuturICT: adaptation is dynamic, flexible, and informed fully by an understanding of the data generated by not only the user base of the game, but also its contextual backdrop and links to other chains of cause and effect.
>
> What kind of platform would need to be in place to deliver such scenarios? We use the concept of a "platform" to refer not only to digital technology, but more holistically, to include the motivations and skillsets that different stakeholders in society bring, and the practices they evolve as they appropriate technologies into their daily lives, as a means to many different ends. As we will see, when the ambition is to develop a participatory platform, the societal engagement issues are even more acute.

### The Challenge of Computational Social Science

The increasing integration of technology into our lives has created unprecedented volumes of data on society's everyday behaviour. Such data opens up exciting new opportunities to work towards a quantitative understanding of our complex social systems, within the realms of a new discipline known as Computational Social Science. Against a background of demographic explosion, financial crises, riots, increasing crime and worldwide epidemics, there is an urgent need for a greater comprehension of the complexity of our interconnected global society and an ability to apply such insights to policy decisions.

In order to tackle the Big Problems of our time, humans are developing valuable instruments and techniques for generating, gathering and analysing data related to the 21st century challenges; but we also need novel **theories** matching these challenges.

For this reason, the aim of FuturICT is to develop complexity science and Computational Social Science further towards a **Global System Science (GSS)**. The new approach will connect the understanding of the generative, cognitive and behavioural social mechanisms and the knowledge of complexity science regarding self-organisation, emergent properties and systemic risks with novel experimental techniques and the mining of Big Data of techno-socio-economic activities.

Simulations of societal processes with the **Living Earth Simulator (LES)** will provide integrated analytical functionality, enabling to base decisions on the consideration of possible consequences. The models simulated in the LES will be used to discover new opportunities and applied to critical social, economic and environmental problems. The problem-centred **Exploratories** will provide critical and effective policy information and advice to decision-makers, facing problems such as financial regulation, health, crime, conflict, security, urban and environmental sustainability, environmental-friendly transport or smart energy systems, all of which involve a social dimension.

In this context, the Computational Social Science pursued by FuturICT project will meet the challenge to generate novel knowledge and theoretical understanding not only by processing previously acquired data, but also by triggering further data acquisition through new theory-driven questions and real-time measurements.

This will allow one to understand the dynamics of complex social systems: not only helping to analyse Big Data for addressing Big Problems, but also providing new instruments for Big Thinking.

### The ICT-Systems Challenge

Today, many large-scale computational experiments are carried out in a way that does not facilitate the reuse for other experiments (e.g., from different domains), so that a great deal of resources is wasted to rebuild them from scratch. Ad-hoc experiments carried out by individual research groups with innovative ideas are often not possible because of a lack of resources. To mitigate such problems within FuturICT, the **Living Earth Simulator (LES)** will develop a framework with suitable programming paradigms and abstractions for a wide spectrum of analytical and simulation experiments, thereby enabling the reuse of important components such as data, models, code, and computational resources (i.e., clusters of machines). The LES needs to support also scientists who are not IT experts and thus must be based on easy-to-understand abstractions that automate mundane IT tasks and provide a platform to carry out simulations and what-if explorations, using models of heterogeneous forms and data from heterogeneous sources. To this end it needs to address four main challenges. First, it needs to provide scalable robust, uniform IT infrastructure, to automatically map models and simulations onto code and (modern) hardware, including the degree of parallelism and scheduling as part of the mapping, support data- and computationally-intensive tasks in a highly performing and energy-efficient way, and scale in many dimensions. Second, it needs develop new visualisation and interaction techniques that empower scientists to understand complex relationships between an input dataset and possible input conditions, input and resultant simulation output parameters of a model, and simulation accuracy as well as between different models and prediction qualities. Third, it needs to provide the right abstractions and tools for scientists to carry out mega-modelling. That is, to define new models from scratch and compose new, more complex models from existing models, to link models to other models, data sources, and to experiments, including what-if analytics model validation, to define meta-models and rules that specify the assumptions of a model and how it can be composed, to provide a unifying language for mapping available data and models into new knowledge, and to compose mega-modules enabling a rich repertoire of possible compositions, in order to form complex explorations with simple means. Finally, using the above components, it needs to provide tools for multi-scale modelling, causality discovery and uncertainty quantification. This involves the development of new mathematical and computational methods to reason about the quality of results achieved with computational experiments, the development of new methods for multi-scale simulation that will integrate spatio-temporal model descriptions of different granularity, and the integration of multi-scale modelling with uncertainty quantification techniques and scalable computing infrastructures, which are linked to the applications developed across all areas of FuturICT's research.

### The Challenge of the Co-Evolution of ICT with Society

The realization that systems where complex information technology closely interacts with humans cannot be understood and modelled just with tools developed for classical ICT systems, has lead to many research initiatives devoted to "techno-social systems" (e.g. the FET COSI-ICT initiative in which many of the FuturICT concepts have its roots) and "human computer confluence". So far, most models consider constrained isolated aspects, such as for example the spread of opinions in social networks or the dynamics of flash mobs. On the other hand, striving to increase the understanding of the global ICT-society systems in all its facets, we need to consider the interaction of different aspects over different temporal, spatial and social scales. The challenges that we face when striving to adapt the models developed in these and similar efforts to describe the systems considered by FuturICT are manifold. First, most efforts to date assume a clear system boundary between the ICT and social structures, implying an interaction between two separate systems over a well-defined interface. In contrast, FuturICT recognises the increasingly blurred boundaries between the digital and the social world, which contradicts the existence of a well-defined system boundary and of a clean interface. Second, as any other complex system, techno-social systems are multi-level entities, characterised by a multidirectional dynamics among their levels and components (see the above section on the modelling challenge), so that emergent properties of techno-social systems retroact on the lower-level, be it human or artificial, social or behavioural, and determine new properties on that level. Third, techno-social systems undergo evolutionary dynamics based on different forms of mutual adaptation among their technical, social, cognitive and behavioural components. More precisely, they show co-evolving processes, in which each of the components presents the others with problems of adaptation, and providing at the same time an environment in which emergent properties and solutions can compete and be positively or negatively selected. Hence, each component modifies all others and undergoes their adaptation pressures. In substance, such co-evolving processes lead humans to develop properties, mechanisms, and operating modalities that are selected for surviving and finding their way in a techno-social system; on the other hand, technological development takes place in an evolutionary way by alternative designs that compete for better adaptations to human users in specific contexts.

### The Challenge of Ethical and Value-Sensitive ICT

Some of the main ethical concerns addressed by FuturICT are privacy, trust, and responsible use. Thus, advanced computing for the social sciences requires among other things, conceptual clarity and technical translations of the concept of (i) informed consent and control over own data, identity and identification, (ii) the right of information to be forgotten, (iii) personal data, (iv) technological ways to enable the implementation of ethical principles in open platforms that are usable for everyone. The breakthrough required is a framework for systematic, sustained and seamless collaboration between software engineers, information architects, hardware and database specialists, ethicists, legal scholars and social scientists on the articulation of ethical principles of an information society on the grounds of social values and our constitution and their effective implementation. It should result in system architectures that incorporate ethical considerations and values as a core operational principle, not as "add-ons". Note that such architectures must address questions such as to what extent ethical decisions about system use can be "outsourced" to the system and who can specify values and principles of system use. FuturICT is, therefore, in favour of architectures that support pluralism and diversity.

### The Challenge of Collective Awareness and Socially Interactive ICT

When humans interact, their actions are influenced by the awareness of their respective social roles, and their anticipation of possible consequences of their actions. Such social awareness provides the foundation for a stable environment in which dense populations of individuals can thrive and grow. The challenge of socially interactive ICT is to adapt the principles of social organisation in order to provide models that can be used to organise and observe planetary-scale systems composed of both, humans and ICT components.

Socially aware ICT will require techniques to construct systems that are aware of the social roles of partners, are able to behave in a manner that complies with social expectations, and able to anticipate the consequences of actions. This will require technologies to perceive and understand human contexts that go far beyond current ICT capabilities. Perception of human context will require abilities to perceive the actions of people, to learn models for social roles, and to understand social behaviours. Understanding human context well require the ability to use social models to explain and predict likely human actions and interactions. Building on the capabilities of the PNS, the LES and the GPP, such abilities will make a new form of pervasive computing possible, in which ICT systems are fully a part of human society.

Collective awareness will extend concepts from classical Human Computer Interaction from individuals to groups, facilitating socially interactive ICT. Thus, interaction concepts will be driven by the understanding of the impact that the system has on the community, rather then on a single user, and directed at optimising the system's usability, utility and benefit for the society rather then on the classical notion of user experience.

To this end, the challenge for collective awareness is to provide tools by which teams of human and artificial systems can share understanding, including the ability to observe their current situation and to anticipate consequences of trends and actions. In such a system, no single individual is able to apprehend or understand everything. It is the collective, the working together, that can provide a more complete understanding. Our challenge is develop models by which such organisations can be enabled, as well as metrics by which they can be assessed and controlled.

### The Challenge of Socially Inspired, Self-Organised ICT Systems

How to ensure stability, reliability, and trustworthiness in systems where billions of complex, autonomous devices interact in dynamic, often unpredictable ways, is a core challenge faced by computer science today. This challenge will be even harder within systems that incorporate social awareness, social adaptiveness and the complex composition and orchestration of humans and ICT components.

It has long been accepted within the ICT community that adapting mechanisms found in nature is a good approach for dealing with complexity in computational systems. This has lead to the rise of the field of bio-inspired systems. FuturICT will go a step further towards developing **socio-inspired ICT**. This means to learn from mechanism that make social systems work well and use them as operational principles for socially interactive ICT systems. One key concept of socio-inspired ICT is to manage complexity by using the principle of self-organisation. The related challenge is to identify and establish the right interactions and institutional settings, as for complex systems there is no simple, general rule how to identify these interactions. A further challenge is to move from the classical ICT design process to the co-design and co-development of ICT systems with society. This requires (1) formal models of the envisaged forms of human machine, machine-machine and human interaction, (2) methodologies, where the design of ICT systems and of their interdisciplinary consequences on individuals and society are evaluated at the same time, and (3) a new generation of architectures which allow for the seamless orchestrations of ICT components' actions with humans and human communities towards shared collective goals. Finally, an interesting question is how far collective social awareness can be leveraged to allow systems to autonomously evolve new socio-inspired operational principles.

### Challenge of Implementing Interactive Exploratories

The Exploratories will face their separate and distinct technical challenges, as well as the challenges of integrating, as far as possible, the data, models, simulation codes and visualisation tools. However, they will all face in some way the challenge of modelling and integrating agent and social models, at many scales and with sophisticated models of agents' cognitive behaviour and responses. For example, the Health and Epidemics Exploratory will be challenged by the social response to its predictions, i.e. will need to define, test and validate behaviour-contagion models, able to close the feedback loop between behavioural changes in the population in response to the actual disease impact. This would require a breakthrough in novel experimental behavioural and cognitive science models, as well as the exploitation of novel experimental capabilities through a sensing and monitoring infrastructure, along with secondary and proxy data harvesting. The Social Challenges Exploratory will also face the issue of social feedback, since this Exploratory will need to model people's reactions to policies and other measures of prevention, in the struggles against critical social phenomena. The Future Cities Exploratory is facing the challenge of developing a city simulation that integrates multiple systems and their data, while meeting appropriate confidentiality and privacy standards, and enabling citizens to participate in routine as well as strategic ways, using these tools to enhance their quality of life.

**The Integration Challenge**

The **integration** of the above activities and functionalities under the umbrella of one FuturICT platform will be another challenge. The challenge is how to connect available data with available models that cover different policy domains and scientific disciplines, especially to real-time data input, which is typically required in response to a crisis situation such as an earthquake or tsunami. Since all of the above activities will be highly demanding, the production, spreading and practical application of scientific knowledge will have to be significantly accelerated. This challenge requires a new approach to both scientific and business innovation, and it is jointly addressed by FuturICT's **Innovation Accelerator** and Focus Areas targeted at Integration and Coordination.

# 1.5 State of the Art: The Big Picture

The idea of managing the complexity of our world with scientific models and/or ICT systems is not itself new. It was a driving force behind cybernetics (Wiener 1948), systems theory (Bertalanffy 1968), system dynamics (Meadows 1972), synergetics (Haken 1983), operations research, computer science and artificial intelligence, or also Fuller's World Game (see http://www.bfi.org/about-bucky/buckys-big-ideas/world-game), to mention just few examples. On the engineering side, the increasing complexity and need for adaptiveness in ICT systems has been studied in research areas such as autonomic computing (Kephart and Chess 2003) and organic computing (Muller-Schloer 2004). There has also been significant interest in the implications of the increasing blurring of the boundaries between the digital and the physical worlds, which is demonstrated, for example, by concepts such as the Internet of Things (Atzori et al. 2010) and Reality Mining (Eagle and Pentland 2006).

There have also been some large-scale projects that relate to the overall goals of FuturICT, in particular the notion of leveraging large-scale electronic data collections for a better understanding of social, economic and environmental phenomena. One certainly needs to mention the Digital Earth project suggested by Al Gore and some similar initiatives of the European Union (e.g. run by JRC). Moreover, Japan's Earth Simulator aimed at large-scale environmental simulation. CISCO and NASA have joined forces to launch the Planetary Skin Institute. This institute uses data collected from sensors across the globe to monitor global environmental change. The climate science community is engaged in large-scale simulations of the Earth system and gets interested in both, the impact of environmental change on society and economy as well as the impact of human behaviour on environmental change. Moreover, a large-scale „Agent-Based Laboratory for Policy Analysis" (N-ABLE) was built for scenario and policy analysis in response to the challenges after 9/11.

> **Resilience** relates to the systemic capability to anticipate, absorb, adapt to, and/or rapidly recover from a potentially disruptive event. Resilience has been subject to a number of studies in complex networks (Albert et al., 2000) and social-ecological systems (Berkes & Folke, 2003).

With a similar motivation as FuturICT, the UN has recently launched an initiative called Global Pulse. This project aims to exploit the vast amount of data generated as a by-product of human activities, and uses this information to improve the overall wellbeing of society and its resilience to crises. The initiative is developing an international network of projects working towards these aims. Google.org also features a number of projects to counter global crises and to address some of the challenges that FuturICT is also interested in. However, both initiatives do not offer the same level of integration of ICT expertise with social and com-plexity science, the integration of computational approaches with theoretical and experimental ones, and the unified vision and strategy that FuturICT has developed over the past 3 years.

Another fundamental difference between all of the above projects and FuturICT is the feedback into ICT and computer science. Thus, while all of the above initiatives use state-of-the-art ICT as a tool, FuturICT emphasises the synergy between Computer Science, Social Science and Complexity Science to develop a new paradigm of ICT inspired by and oriented towards social phenomena and needs. Thus, capabilities of gathering and evaluating information about the society will become part of the operational principles of the system, allowing it to build an internal representation of the world (collective awareness), driving adaptation, self-regulation and the way in which the systems interacts with users, communities and the society as a whole. Social scientists will not be just using the system, they will be providing crucial input to system architectures on how to leverage principles that have proven successful in regulating social processes, in order to address issues of resilience, reliability and trustworthiness faced by today's ICT systems. The above interdisciplinary synergy sets FuturICT apart from many current ICT-driven initiatives that aim at managing complex computing systems and dealing with the convergence of the digital and the physical world, such as the above-mentioned autonomic computing or organic computing efforts or the EU FET Human Computer Confluence, PERADA and Fundamentals of Collective Adaptive Systems initiatives. The FuturICT vision has evolved from the FET proactive COSI-ICT cluster, from which the core ideas of the project originated and in which many of the FuturICT core partners play a leading role.

Perhaps the most important difference between FuturICT and all related large-scale projects lies in our ethical and value-oriented approach. The trend towards gathering and evaluating more and more data generated at the interface of the digital and the physical world is currently driven by commercial interests and often by narrowly defined national security needs. Privacy, benefit to the individual, and ethics in general are a boundary condition at best, often conflicting with commercial interests. In contrast, privacy and ethical aspects are a core research topic of FuturICT and will be embedded into the core operational and design principles of the underlying ICT technology. The project is built around the notion of participation, ensuring that data "producers" will also be the ones benefiting from them, thereby turning Big Data into a public good.

## 1.6 State of the Art: The Detailed Picture

To achieve the high-level advances summarised in the previous sections, the project will build on and extend the state of the art in all of the areas related to the specific scientific challenges outlined in section 1.4.

**Goal Driven Orchestration of Data Collection and Processing**

The notion of collecting large amounts of data from the interface between the digital and the physical world has generated a lot of interest and applications over the last few years. Early origins of the concept can be traced back to research on the use of ubiquitous computing technology for group collaboration and the study of group dynamics (e.g. Grudin 2002). Among the first explicit descriptions and implementations of large-scale sensing with personal devices was the Reality Mining concept put forward by Pentland at MIT (Eagle and Pentland 2006). It focused on recording and analysing social interactions and relationships. Other examples are using car- and mobile-phone-integrated sensors to monitor things like traffic congestion, commuting patterns and WiFi Hotspot density (Hull et al. 2006) or collaborative monitoring of road state (Ericsson et al. 2008). A broader and well-publicised effort is the MetroSense project that envisages "urban sensing at the edge of the Internet, at very large scale" and refers to the concept as "People Centric Sensing" (Campbell et al. 2008). There are also a number of other projects that are going on all over the world (e.g. CenceMe by Miluzzo et al. 2008), AnonySense (Cornelius et al. 2008), or CitySence (Murty et al. 2008), and others). In a recent study, data coming from millions of mobile phones was used to track the spread of malaria in Kenya (Wesolowski et al. 2012).

What all of those efforts have in common is that they are based on systems specifically designed and built for a concrete sensor configuration and application. In contrast, FuturICT proposes to automatically configure the data collection and evaluation process based on high-level, human-understandable queries. The evaluation of such queries relates to Natural Language Processing (NLP, Joshi et al. 1991), which is a huge research field that produced such well-known successes as IBM's Jeopardy-winning Watson computer (Ferrucci et al. 2010) or the iPhone Siri interface. Another large related area is the Semantic Web, which aims to simply speaking and add semantic meaning to the lexical data stored on the Internet (Berners-Lee et al. 2001). FuturICT will build on such developments and extend them (top NLP researchers are part of the Consortium). However, parsing a query is just the first step. How to transform it into a specification that can be used to acquire relevant information sources, select appropriate processing algorithms and orchestrate the collection and analysis process in the face of dynamic changes, depending on the available information sources, computing and communication resources, is an open problem. The problem is made more difficult by the need to combine heterogeneous sources ranging from volunteered sensor data by mobile phones over social media contents to public infrastructure information, which all have different access and privacy issues, data rates, formats etc. The closest previous work comes from FET's OPPORTUNITY project, which was dedicated to human activity recognition in dynamic ubiquitous sensor environments (Kurz et al. 2011) (several FuturICT partners were part of that project). There is also some relevant related work in ad-hoc and self-organised sensor networks (e.g. Gupta et al. 2006, Sohrabi et al. 2000). While interesting, the above are far more limited in scope (in terms of query complexity, system size, heterogeneity and dynamism) than what FuturICT is trying to accomplish. Existing work has also little to offer with respect to the envisaged goal of incremental evolution of orchestration strategies.

**Social Sensing and Mining**

Social Mining, Human Data Mining, Reality Mining, and Mining Human Behaviour are the new keywords from the emergence of novel data mining, statistical learning and network science methods built around the digital footprints of human activities. The availability of such new ICT methods and data is reflected by thousands of papers published in the last ten years, and it enables the convergence of a number of fields such as data mining, knowledge management, pervasive computing, complex systems, network science, and social sciences.

- Social network analysis refers to the study of interpersonal relationships, with the purpose of understanding the structure and the dynamics of the fabric of human society (Kleinberg 2010). Roughly, speaking, two major lines may be distinguished: the statistical laws regulating statics and dynamics of complex networks (Watts 98, Barabási 99, Caldarelli 2007, Newman 2010), and the methods aimed at discovering patterns, evolutionary rules, community structure and the dynamics in large social networks: i) community discovery (Fortunato 2010, Coscia 2011); ii) information propagation (Kleinberg 2000, Kempe 2003, Satorras 2001); iii) link prediction (Kleinberg 2003, Kashima 2009, Leskovec 2010); iv) temporal evolution (Leskovec 2005, Holme 2012, (Holme 2012, Mucha, 2010, Berlingerio 2010, Bringmann 2010); v) multilevel networks (Jianxi 2012, Berlingerio 2012, Tang 2009).
- Social media mining refers to methods for data mining from different online sources (tweets, mails, blogs, web pages, link structures etc.): i) public opinion/sentiment mining, (Pang 2008); ii) automated tagging; iii) ranking (Page 1998, Kleinberg 1999).
- Mobility data analysis leveraging the spatio-temporal dimensions of Big Data (such as mobile phone call records and GPS tracks, generated by current mobile communication technologies and wireless networks) with the purpose of understanding human mobility behaviour (Brockmann 2006, Gonzales 2008, Song 2010, Giannotti 2008, Moussaid 2011), daily activity patterns (Jiang 2012, Ferrari 2012), or geographic patterns (Trasarti 2011, Giannotti 2007, Monreale 2009).

A comprehensive discussion can be found in Giannotti et. al. (2012). The key point of social mining is to combine the macro and micro laws of human interactions within a uniform multi-dimensional analytical framework. Two main complementary research directions are envisaged for social mining, focusing on the invention of methods for the extraction of high-level models and patterns of human behaviour that combine the different dimensions of human activities, for example, statistical of social ties or mobility, collective opinions and preferences, socio-economic behaviour etc.: What is the interplay between the social and mobile network (Wang 2011)? How can we automatically identify strong and weak ties in organisational networks by analysing human interactions patterns? How can we use the social and mobile networks to answer question about the socio-economic structure of a city? Another focus is the engineering of such methods in complex scenarios of techno-social systems at global scale.

## Modelling and Global Systems Science

Supercomputing has been used everywhere, from car to plane and drug development, process control, weather forecasting and climate modelling. Compared to this, modelling social and economic systems has been particularly difficult. However, computer modelling has persistently moved on to address more and more complex challenges. Since the invention of cybernetics (Wiener 1948), systems theory (Bertalanffy 1968), catastrophe theory (Zeeman 1977), and the "Limits to Growth" study (Meadows 1972) using a systems dynamics approach (Sterman 2000), much progress has been made. Scientists have learned that techno-socio-economic systems are complex dynamical systems with many mutually adaptive system components, which come along with a number of particular challenges:

- Complex systems (Nicholis 2012, Strogatz 2001, Bar-Yam 1997, Holland 1998) can have multiple equilibria or rich dynamics (including "chaos" or "turbulence"). They may show resistance to external control attempts or systemic shifts, instabilities and cascading effects. In case of probabilistic network interdependencies, the same causes may have different effects, or the same effects may have different causes.
- Spatial and network interdependencies, or the heterogeneity of system components can make a big difference. Details of history or context may matter.
- Forecasts for complex dynamical systems are often probabilistic and useful for short time periods only.
- Complex dynamical systems may resist to classical control approaches. For example, modelling attempts which are oriented at commercial or political interests tend to underestimate that individuals can easily undermine governance or manipulation attempts, when they do not share the same interests.
- The internal complexity of socially interactive agents (Haykin 2008) complicates the system dynamics. It typically implies "subjective" worldviews, learning, and a feedback of the macro-level on the micro-level behaviour.

The theory of self-organisation and complex systems (Prigogine 1982, Strogatz 2001, Nicolis 2012), agent-based modelling (Epstein 1996, Gilbert et al. 1995, 2005, 2008, Luke 2005, Miller 2007, Helbing 2012), network theory (Barabási 1999, Huberman 2001, Watts 2004, Newman 2010, Christakis 2011), statistical physics (Wolfram 1983, and computational social science (Lazer 2009) have all made significant contributions to the computer modelling of techno-socio-economic systems (Vespignani 2009). While these approaches were limited in scope and often theoretical in the beginning, these methods have matured over time (Helbing and Balietti 2011), and they are increasingly used to address real-life problems such as cities (Allen 1997), traffic (Helbing 2001), economic systems (Anderson 1988, Arthur 1997, Farmer 1999, Blume 2006), innovation (Bettencourt 2007), or social dynamics (Weidlich 1991, 2000; Epstein 1996, Gilbert et al. 1995, 2005, 2008, Luke 2005, Miller 2007, Castellano 2009, Buchanan 2012, Helbing 2012).

Now, Microsoft runs a "Modelling the World" lab (http://www.modellingtheworld.com/). Also the US Homeland Security invests a lot into the computational modelling of social challenges. For example, the Sentient World project launched by the US Department of Defense combined models of the geographic and physical resources of a location with information about the cognitive state of the inhabitants. There are also recent IARPA and NSF projects amounting to about 200 Mio. dollars, targeted to unleash the power of Big Data. In comparison, FuturICT pays particular attention to an ethical, value-sensitive and privacy-respecting approach.

There is no doubt that the modelling of techno-socio-economic systems is now a thriving research field. In the complex systems arena, there are various multi-agent simulation platforms for specific domains like traffic (e.g. Transims, Matsim), economics (e.g. Eurace), finance (see, e.g., the Santa-Fe Market Simulator, Japan's U-mart, and the outcomes of the European Crisis and FOC projects), and policy (N-ABLE). Also environmental scientists get increasingly interested in the impacts of environmental change on behaviour and in the question, how behavioural change can be reached to protect environment. However, they are often lacking crucial functionalities, such as the integration of real-time data streams or the interdisciplinary integration of expertise from the natural, social and engineering sciences. FuturICT's experts are involved in Matsim, Eurace, Crisis, FOC and other comparable initiatives and are aiming at the integration of this knowledge, and of theoretical approaches with experimental and real-time data through FuturICT's PNS and GPP (Bainbridge 2007, Slganik 2006, Szell 2010, Helbing 2010). Moreover, FuturICT learned from the challenges that past projects have been facing:

- FuturICT is aware of the limits of forecasting techno-socio-economic systems and, therefore, focuses on studying parameter dependencies (e.g. conditions for systemic stability) and causal interdependencies (implying possible cascading effects), and it tries to identify suitable feedback mechanisms to benefit from probabilistic short-term forecasts, where these are possible.
- FuturICT addresses generic problems (such as systemic instabilities) and emerging problems (such as implications of new information technologies). FuturICT's goal is to reach improvements in system performance and enhance the ability to absorb shocks.
- FuturICT goes beyond mean-field and representative agent models. It takes into account spatial and network dependencies, heterogeneity, randomness, correlations and co-evolutionary effects.
- FuturICT uses and combines network theory, complex systems theory, social science models, multi-agent simulations, multi-level models, experiments, and interactive elements (e.g. to use the wisdom-of-crowds effect).
- FuturICT develops entirely new methods of investigation (such as the Living Earth Simulator, the Planetary Nervous System, or the Global Participatory Platform).
- FuturICT will use Big Data, reality mining and machine learning (Goldberg 1989) to calibrate, validate, and develop models, and it will also use insights into underlying principles of socially interactive systems to identify better institutional designs and develop innovative approaches to manage complexity.

**Computational Social Science**

In the Big Data era, the gathering, formatting, analysing and manipulation of massive amounts of digital information has deeply transformed such fields as physics and biology. In comparison, the emergence of computational social science has been slower, but the increasing complexity of techno-social-systems and the capacity to collect and analyse data with an unprecedented breadth, depth and scale calls for complementary social science research using a quantitative, data-driven approach (Lazer et al. 2009).

However, computational social science has a more complex history, which is not only quantitative: computational models were first introduced to make experiments, thus exploring and understanding the dynamics of complex social systems. In particular, agent-based models (ABM) and agent-based social simulations (ABSS), complementing classical models of rational decision-making, have been so far been mainly used as a qualitative tool, e.g., to provide plausible explanations of social phenomena (Epstein 2007). One important topic in this line of research is the assessment of the sensitivity and validity of simulation models (Gilbert 2007).

This led to the *generative* method: a generative explanation of an observed social phenomenon consists of a description of the external (environmental and social) and internal (behavioural) mechanisms generating them (Epstein 2008). In this way, ABM explains macro-level behaviours in a bottom-up way, in terms of local mechanisms that have generated them, i.e., micro-motives on the agent level when behaving in one way or another (Schelling, 1978).

This approach showed the way forward to a multiplicity of agent models, sometimes characterised by a certain arbitrariness. For this reason, ABM should be put on an empirical and a theoretical ground (Conte 2009). It is a challenge that meets new needs and opportunities of the modern age. On the one hand, the needs are mainly related to the necessity of facing crises: this is translated into the observation of more complex social dynamics, e.g. multi-level dynamics, and requires sophisticated models to address questions such as "how do individuals react to the knowledge of the macro-phenomena they produce?" (Arthur 2005). On the other hand, a new impulse to computational social science is coming from complexity science, data mining and Big Data science, enabling the extraction of knowledge from huge datasets.

**Co-Evolution of ICT and Society**

Since its very start, the design of ICT systems for human non-expert users posed interesting scientific and application-oriented challenges. System designers soon realised that ICT systems are hybrid constellations of functionalities, in which not only human activity is computer-supported, but users and ICT systems synergistically contribute to execute common activities (Nardi 1995). Early developments in the field of Human-Computer Interaction (HCI) (Dix, 2004) stimulated an interdisciplinary and highly innovative perspective in the study and design of user-friendly or enabling technologies, where scientists from AI, cognitive science and psychology, linguistics, and formal logics cooperated to model interactive systems (Miller and Page, 2007). On the other hand, developments in HCI contributed to launching the field of intelligent systems (Albus, 2001), at the same time favouring the application of computational modelling to the study of abstract categories of entities, which subsume both natural and artificial systems (Portugali, 2005). Later on, the application of computational instruments to the study of human intelligence had profound implications not only for the development of the behavioural sciences (Kautz et al. 2002), but also for the modelling and design of ICT systems (Carroll, 1991). Since the early nineties, the notion of agents (Conte 2009), i.e. computational units interacting with other agents and with the environment, appeared in computer science, and soon proved highly influential. Its relevance became still more evident when one of the critical features of agency, namely autonomy, was brought to the stage and applied to automated coordination, cooperation, coalition formation, etc. (see the Proceedings of several series of events, like Modelling Autonomous Agents in a Multi Agent World (MAAMAW), Agent Theories, Architectures and Languages (ATAL), International Conference on Multi-Agent Systems (ICMAS), Autonomous Agent and Multi- Agent Systems (AAMAS) from the early 90s onwards). Far from a user-system dyad, the critical interaction unit for useful technological applications became a larger and more complex system, in which a set of agents, natural and artificial, and the associated artefacts shape and get adapted to one another, at the same time carving their social, cultural, and institutional environment. The study and design of electronic institutions (Noriega and Sierra, 2002) was good evidence supporting the necessity of a new way of looking at and modelling ICT tools as systems inspired by the society itself (Conte 2009), with its micro- and macrostructures, functionalities, systems of relationships and artefacts. In turn, a new way to conceptualise and model societies proved necessary, working out the conceptual, methodological, and technical instruments accounting for the co-evolving dynamics (Spangenberg, 2005), responding to adaptive pressures of technological, cultural, demographic and socio-economic developments. To account for such co-evolving processes is an on-going necessary condition for a successful management of social complexity, and it is pivotal for any significant advance in the science of global systems.

> **Social Computing, human computation, crowdsourcing.** The term social computing is used to denote broadly the intersection of social behavioural sciences and computational systems. In its weaker sense, it pertains to supporting any sort of social behaviour in applications where groups of people interact socially. In its stronger sense, it deals with supporting and orchestrating computations that are (partially) carried out by people (e.g. the US NSF funding program on social-computational systems). Concrete examples include: the Darpa Balloon challenge, the Human Flesh Search engine and FoldIt. The term human computation usually denotes approaches where only a single person is engaged in a given task. Examples include the reCAPTCHA Project and work on ESP games (Von Ahn, 2006). They build on the notion that there exist tasks where humans are much better than machines. In crowdsourcing, such as the Amazon Mechanical Turk, people voluntarily accept to have their work scheduled and coordinated by a machine. In this context a core research issue is the development of computational models of trust and reputation.

## ICT Systems Support for Complex Modelling

Computational social science has made important advances in the theoretical modelling of complex social phenomena. But the development of tools that allow an integration of large, heterogeneous multi-scale computational models with similarly large, multi-scale and heterogeneous data still lies ahead (Oden et. al. 2006). Such tools are needed before the challenge of better understanding and possibly predicting real-life social phenomena can be addressed, both in social sciences (Conte 2012; Lazer 2009) as well as natural sciences (Emott et. al. 2006). Current simulation platforms (e.g. MASON, MASS, GLEAMviz, SimExplorer) are still lacking crucial functionalities for effective integration of heterogeneous models, automated experimentation, validation and model uncertainty quantifications, visual analytics of complex structures in model and data, and efficient mapping of algorithms to available computer architectures. Research in computer science has focused on providing general-purpose tools. Imielisnki and Mannila (1996) advanced the idea of a data mining query language supporting the knowledge discovery process. This idea has recently been implemented as part of the M-Atlas system for human mobility data mining and visual analytics (Giannotti et al., 2011). Another framework for data analysis is the Map Reduce paradigm (Dean, Ghemawat, 2004), which is used by most Web companies such as Facebook, Google, Microsoft, Twitter, or Yahoo. One of the reasons for the popularity of the Map Reduce framework is the development of key-value stores, which enable the efficient storage and routing of key/value pairs in a distributed system of thousands of machines. Two early examples of such a key/value store are Google's BigTable (Chang et al., 2006) and Amazon's Dynamo system (DeCandia et al., 2007). Distributed file systems such as the Google File System (GFS) and the Hadoop File System (HDFS) can also be seen as incarnations of this technology. The underlying principle of key/value stores is a concept called "distributed hashing" (or Distributed Hash Table, DHT), which was pioneered in the projects Chord (Stoica et al., 2003) and Pastry (Rowstron, Druschel, 2001). Most of these efforts to build such large-scale computational systems in order to support new, data-driven science go back to Jim Gray's approach of a 4th scientific paradigm (Hey et al., 2009) and the first Big Data initiatives postulated in Emmott et al. (2006) and Bryant et al. (2008). While all these Big Data technologies have paved the way for projects like FuturICT, today's generation of Big Data systems still lack a comprehensive integration of data mining and simulation techniques and what-if capabilities. A first step to address the latter has been taken in Haas et al. (2011). Another deficiency of most modern tools is their missing integration with the models of complexity science.

### Ethical and Value-Sensitive ICT

One approach to characterise the development of ICT is to note that it has become more sensitive to its environment: to users, to organisational and social context, and to society at large. In the first phase, in the sixties and seventies, ICT was largely the outcome of a technology-push focused on core computational functionality. The social and psychological and organisational context, the users and usability were of no special concern. Digital computers were fascinating technology, solutions looking for problems. In the second stage of the development in the eighties and nineties – after many failures – developers gradually started to realise that there were human users out there with specific needs and requirements, and that real organisations and society at large were involved. It therefore became clear that it would be wise to accommodate user requirements and conditions on the work floor, already in the early stages of the development of applications. But this was still a very small step towards taking the needs and wishes of users, organisations and society into account, namely as mere constraints for a successful implementation of systems. We are now entering a third phase in ICT development, where the needs of human users, the values of citizens, and the big societal questions are in part driving the research and developments in the ICT arena. The social and moral values are no longer seen simply as risk factors and constraints, but also as drivers of innovation and R&D. It brings within reach the idea that applications are developed in order to serve and support values and serve the interests of society. The idea of making social and moral values central to the development of ICT and to embed values in ICT artefacts (architecture, code, interface, integrity constraints, ontologies, authorization matrices, identity management tools) first originated in Computer Science at Stanford in the seventies. This approach is now referred to as "Value Sensitive Design" (see the later description on ethics and Bibliography in Appendix 3, Friedman 1997 -2007) or "Values in Design" (Friedman and Nissenbaum, 1996) or "Design for Values" (van den Hoven, 2005; 2010). In this approach, the focus is on incorporating human values – as a type of non-functional requirements – into the design of technical artefacts and systems. In doing so, one takes a look at software engineering from an ethical perspective. One can intentionally and explicitly design for values such as accountability, safety, inclusion, privacy, trust, or sustainability. A number of groups in human computer interaction and software engineering around the world have experimented with the VSD approach and the results are very encouraging. A number of studies have been conducted, which concretely show how values can be incorporated in ICT products and services (Friedman et al, van den Hoven et al 2005, 2011).

### Collective Awareness and Socially Interactive Computing

As the functionality of ICT systems becomes increasingly complex, the ability of users to interact with the system is a critical factor. Ben Shneiderman argues in his book "Leonardo's Laptop" that the new computing is not about what computer can do, but what users can do with computers (Shneiderman, 2002). The interaction concepts and the overall "user experience" will increasingly become the central issue in the design and will impact the overall success of the system. While "hard" technical performance parameters remain relevant, they are increasingly taken for granted. User experience has become a key factor by which consumers discriminate products. Thus, today's human computer interaction (HCI) concepts go beyond reactive systems driven by explicit user input towards implicit interaction (Schmidt,2000), and towards context and situational awareness (Schilit 1994, Abowd 1999, Schmidt 1999, Crowley 2002 and Gellersen 2002). Such systems are designed to anticipate the needs and expectations of users based on machine perception and user models. Such proactive systems adjust their capabilities, and functionalities and interface in accordance with the physical and social context. By reducing the complexity of interactions, such systems become easier to use and create a more satisfying user experience.

By awareness, we refer to the ability to perceive phenomena, project trends and anticipate consequences. This definition is drawn from the cognitive sciences (Johnson-Laird 1983, 1998, 2006) and in particular from the domain of human factors (Endsley 1995a, 1995b). These theories are consistent with understanding human awareness from cognitive neuroscience (Damasio, 94, 99) and provide a framework for building artificial systems that exhibit awareness.

For both biological and machine awareness, the combinatorial nature of the interactions between phenomena and consequences imply significant limits to the awareness of individuals. A challenge for FuturICT is to extend these limits. As with Distributed Cognition (Norman 1993, Hutchins 1995), Collective Awareness does not imply that all individuals are directly aware of all phenomena, but rather that trends and consequences of phenomena are shared. The core challenges in this area concern the management of attention. Moving from individual to collective awareness requires sharing the knowledge of phenomena and consequences, including the underlying ontologies. This constitutes an entirely new research problem that is not yet sufficiently addressed by the computing or cognitive sciences.

Context-aware computing has so far focused on physical context, in particular on location. FuturICT will extend the notion of awareness to the social domain. With socially aware systems, we expect that we can reduce the interaction complexity and create systems that become more adapted to the social context and hence feel smarter and more appropriate. Over the last 30 years, we have seen a transition for a user interacting with a computer to groups of people interacting with a set of computing systems and services. FuturICT envisages to increase the scale of interaction by several orders of magnitude, both with regard to the user and with regard to the systems and data. We can view a navigation system that is used by 1 million people, receiving output on their screen, and providing input through their mobility, as a single large interactive system. The input by all people in this system impacts the system state (e.g. the business off the roads) and the output, in particular the suggested routes, which is then again reflected in the sensed data. Current ways for modelling interactive systems are not fully adequate to deal with such questions, and we expect that, following the paradigm of socially interactive systems, we can tackle this class of applications. The benefit for the society will then go way beyond the classical notion of user experience.

### Democratising Access to Big Data and Models

Easing access to computing has been a driving force in the design and development of new computing technologies. For the last 50 years, an increase in usability and accessibility was always followed by an increase in uptake of computing technologies, initially in the workplace and later for personal use. In the beginning of the computing age, computer operators where required to use the systems, and very few people made use of computing. Later on, with interactive textual terminals, the user base was increased to many professional users in science and engineering, and with the advent of graphical user interfaces a widespread adoption was reached. This trend is continued till today, and with smart phones, the use of computers has become ubiquitous. In analogy to this development, we see that, in the next 50 years, access to Big Data and models, now requiring experts, has the potential that computing had 50 years ago. The major challenge we face now, is how to empower large numbers of professionals as well as end-users to make use of Big Data and models. While the access to computing is not the limiting factor anymore, we see that making use and sense of Big Data and models still is.

Historically, Vannever Bush and his description of the Memex in the article "As we may think" (Bush, 1945) can be considered as a source of inspiration and motivation. In his article, he foresaw the concepts that 50 years later emerged on the WWW. FuturICT has a vision with similar power, and pursuing this vision will open many new research directions. FuturICT can be seen as being in the tradition of Licklider's idea of a man-machine symbiosis (Licklider, 1960), but foresee it as a mankind-data symbiosis for the future. From a technical perspective, FuturICT will advance Doug Engelbart's concept of "Augmenting the Human Intellect" (Engelbard, 1962) and see us on the way of inventing systems that allow individuals to interact with data, very much like Ivan Sutherland pushed forward the idea of individuals interacting with computing (Sutherland, 1969). This historically perspective is to highlights that the ideas proposed by FuturICT are not just a vision, but a logical continuation of cutting edge ideas towards a new ICT paradigm. Note that Alan Kay's motivation for the Dyna-Book (Kay, 1972), the first concept of a personal notebook computer, and much other work in the 60ties and 70ies was motivated by moving computing from the corporations and governments into the hands of everyone, ranging from school children to professionals. In computing, a trend towards democratisation has shaped the world we live in, and FuturICT believes that in the next step, democratising data can have an even bigger effect on the way we live.

> **Diversity.** Diversity, especially manifested in language and knowledge, is a function of local goals, needs, competences, beliefs, culture, opinions and personal experience (Giunchiglia, 2006) (Giunchiglia et al., 2012a). It is an unavoidable and intrinsic property of the world, is at the root of the resilience of cultures and pervades human discourse (e.g., synonymy and polysemy). In the LivingKnowledge EU FET project diversity has been formalized in terms of diversity dimensions (i.e. the dimensions by which knowledge is framed). In the faceted approach (Ranganathan, 1967) in library science they are topic, space and time. These, together with the notions of domain and context, are at the basis of diversity-aware knowledge bases (Giunchiglia et al., 2012b).

In the last decade we have seen that companies, governments (UK Government, 2012) and organisations have released Big Data to the general public. Participatory sensing platforms have been explored in different contexts that have created large data sets for people to use and experiment with (Campbell et al., 2008). Some of these data sets have become openly available. One early example is a reality mining experiment conducted at the MIT (Eagle and Pentland, 2006), and this can be taken further towards community sensing (Krause *et al.*, 2008, Aberer *et al.*, 2010). A further direction that is widely explored is the implicit collection of data from usage of digital systems, ranging from the WWW to sensors on mobile devices.

The great challenge now is to allow all people to use the data that is publicly available, in principle. Linking the data is one crucial challenge and there are initial efforts using web technologies to address this (Berners-Lee 2005, Heath and Bizer 2011). There is a clear need for working with Big Data in many areas (Linch 2008), and we require forms to organise and exchange Big Data (Tin-Lap 2008). Working with the data requires different skills, ranging from algorithms for processing and analysing raw data to understanding the meaning of data, which is often related to social phenomena, political questions, economic issues, or mobility. Over the last years, special workshops ("hack days") have brought together people with different skills to make use of publicly available data. In order to turn larger and larger amounts of Open Data into useful knowledge in many different domains, FuturICT's Global Participatory Platform will enable Citizen Science and scale up this concept considerably.

**Socio-Inspired ICT**

*"The network paradigm is shifting. Today's open, distributed and dynamic networks are no longer artefacts that we construct, but phenomena that we study."* (Gkantsidis et al., 2003). This quote illustrates the shift in attitudes within computer science that has taken place over the last decade. Self-organisation and adaptation as ways of harnessing the dynamics and complexity of modern, networked, environment-aware ICT have become central research topics leading to a number of concepts such as autonomic computing (Kephart et al., 2003) or organic computing (C. Muller-Schloer et al.,, 2004). Existing work on collective adaptive systems is another example considering features such as self-similarity, complexity, emergence, self-organisation (Holland et al., 2006), and recent advances in the study of collective intelligence (Brown00). Collective awareness is related to the notion of resilience, which means the systemic capability to anticipate, absorb, adapt to, and/or rapidly recover from a potentially disruptive event. Resilience has been subject to a number of studies in complex networks (Albert et al., 2000) and social-ecological systems (Berkes et al., 2003).

Great innovations in computer science and artificial intelligence are linked to bio-inspired systems. Such systems take functional principles of biological systems (e.g. swarms, genetics, etc.), and design algorithms and computer systems after these principles (e.g. [BDT00, GH88]). In many cases, the increase of efficiency or reduction of complexity that can be achieved by mimicking biological systems is significant. In the domain of artificial neural networks, this approach was pioneered by Rosenblatt (1958), and since then this concept is widely applied. The interesting observation is that, on one hand, algorithms are inspired by biological systems (ICT learning from biology), but at the same time experimenting with the algorithms can help to better understand the biological systems (ICT providing insights for biology). FuturICT envisages a similar symbiosis between social systems and ICT. By creating algorithms and computer systems that are modelled based on social principles, FuturICT will find better ways of tackling complexity, while experimenting with these algorithms may generate new insights into social systems.

In computer science and related subjects, people have started to explore socially inspired systems, e.g. in P2P networks (Hales and Edmonds, 2005), in robotics (Fong et al, 2003), in neuroscience (Todorov et al., 2006), and in the area of agent systems (e.g. Di Marzo Serugendo et al., 2006; Do et al., 2003; . V azquez-Salceda et al., 2003).

Despite the above initial work, the overall field of socio-inspired system is still in an early stage of development. It will be one of FuturICT's ambitious goals to demonstrate the great potential of social principles for operating large-scale complex ICT systems.

### Interactive Exploratories

In the last years we have witnessed tremendous scientific and technological progress that has caused a qualitative change in the ways we model socio-technical systems (Vespignani, 2009). Now, Big Data and computational infrastructures put us into the position to envisage Exploratories, providing capabilities for quantitative scenario analyses of phenomena as they occur, ranging from the spreading of diseases to urban mobility to the analysis of conflict and crime. Although crisis rooms around the world are already using some of those novel capabilities, it has become clear that any new and significant scientific progress to gain knowledge on human, social and economical systems requires a novel interdisciplinary integration of computational and data-intensive science with theoretical and computational modelling concepts, experimental methods, and an innovative use of Information and Communication Technologies (ICT) (Conte *et al.*, 2012). Many scientific projects especially in the FET program have pushed the frontier of scientific research in this area (Bishop *et al.*, 2011). However most of them take only one problem area into account, without creating appropriate integration between different domain knowledge, between various disciplinary perspectives, and between modelling, simulation, data mining, and experimental approaches (Helbing and Balietti, 2011). There are no integrated projects of the kind outlined here, which would allow one to understand highly interconnected techo-socio-economic-environmental systems (Crane and Sornette, 2008). Until we develop the Global Systems Science required for this, we will not be able to frame proper and appropriate policies. However, the experts assembled by FuturICT have the abilities to forge this new science, thanks to their diverse skill sets and the experience collected in previous projects.

## 1.7 Innovation, Advances over the State of the Art

As described in the previous section, FuturICT builds on a vast body of research in a broad range of areas of computer science, social science and complexity science. Within each of those areas, sketched out below are the significant advances over the state of the art and paradigm shifts that will be needed to be achieved by FuturICT over a 10 year period.

Note, however, that the innovation of FuturICT goes far beyond breakthroughs in individual research areas. It lies in the broad, interdisciplinary research approach promoted by its unifying vision, namely to understand, explore and manage complex, global, socially interactive systems that make up our world today, in particular the increasingly interweaved digital, physical, and social worlds. It boldly goes beyond the state of the art both, in **the ambition** of what FuturICT aims to achieve and in **the approach** that the project will be taking.

**Ambition**

The chain of ground breaking scientific and technological innovations facilitated by the interdisciplinary, integrative approach of FuturICT can be summarised as follows:

1. As outlined in the previous sections, the notion of collecting data from the interface between the physical and the digital world to "measure" ecological, economic or social phenomena is currently a hot research topic. While the Planetary Nervous System will build on such work, it will take it to an entirely new level. Today, systems are hand-crafted to a narrowly defined goal and a well-defined narrow configuration of information sources. The evaluation is often constrained in terms of spatial and temporal extent. Given the need of a certain application, it takes a team of engineers to build a corresponding system. Sometimes, systems feed information into social or economic models, but there is typically no meaningful integration between the models and the data collection. Making an analogy to Internet search, we are today in the "pre-search-engine era", when finding information on the web would have required one to program crawlers based on hand-coded IP addresses and search terms, with no rankings to build on, and no possibility of leveraging theoretical models of relevance with connectivity.

   - The Planetary Nervous System (PNS) will change that. It will allow non-experts to simply formulate human-understandable, high-level queries. From such queries, the system will automatically identify, configure, acquire, and manage the required information sources and algorithms. Methodologies to integrate the macro vision of complexity and social sciences with the micro vision of data mining and machine learning within a unifying theoretical framework will allow an optimal integration of domain knowledge and a close, iterative coupling with the models. In summary, FuturICT will make the evaluation of complex information coming from the interface between the digital and the real world as simple to use as standard web search is today.

2. Significant steps ahead in the area of modelling concern

   - Operational models of multi-directional dynamics in multilevel systems, in particular models of human reactions to the growing awareness of macroscopic phenomena and political interventions;

   - The evolutionary and co-evolutionary dynamics of techno-socio-economic systems, paying particular attention to the mutual influence of ICT and society. In general, computational models of evolutionary processes are still poor, focused on the distal causes (adaptation problems) rather than on the proximate ones (local mechanisms and properties of evolving systems). Thanks to the PNS, FuturICT will extract possible models from data, exploring them further by the Living Earth Simulator (LES), where the predictions of competing models can be compared under various conditions;

   - What-if analyses of the effects of certain interventions, with a special focus on governance measures and policies. FuturICT is fully aware of the difficulties of forecasting techno-socio-economic systems, where multiple heterogeneous paths may lead to the same macroscopic effect, and various macroscopic effects may result from the same local rules in combination with hidden variables (e.g. history dependencies). Hence, FuturICT does not commit to provide exact long-term predictions. However, different paths need to be explored, hidden variables must be hypothesised, different local rules need to be implemented and checked through all the way up to resulting effects, in order to get a meaningful range of possible outcomes of "what if" analyses. This will be enabled by a combined application of FuturICT's main infrastructures, the PNS and the LES, with the novel modelling techniques emerging from advances in FuturICT's scientific framework, the Global Systems Science (GSS). FuturICT is not restricting itself to a few modelling approaches, but welcomes and combines all methods that help to extend the horizon of knowledge, which implies a pluralistic modelling approach. Far from a pastiche, this will be based on an evolutionary process of collection, competition, selection-by-results, and recombination of diverse modelling techniques, implemented on shared infrastructures. In particular, the following features of the project are expected to yield innovation over the state of the art:

a. FuturICT will engage in the development of standard methods to test the sensitivity of theoretical or computational results to the underlying model assumptions (not just to the parameter choices).

b. FuturICT will collect the Big Data needed to calibrate, validate and develop plausible anticipatory interpretations of trends observed in its Exploratories, with a special consideration of indicators for crises, by combining data-driven modelling with explanatory theories and experimental approaches.

c. FuturICT will take into account effects of randomness and heterogeneity, non-linearity and dynamics, network and context effects, parameter and history dependencies.

d. FuturICT will profit from a scientific task force able identify, adapt, revise and, if needed, provide new agent models and architectures with different levels of cognitive complexity and abstraction for large-scale agent-based simulations of phenomena that are of interest for the project. The role of the different levels of cognitive complexity and of specific features of agency – such as individuality, subjectivity, emotional life, motivational drives, self-perception, etc. – will be carefully studied. The respective efficacy of such agent requirements in dealing with cultural traits, such as norms or values, will be worked out.

3. Participation and uptake are key to the impact any technology. The Internet would have little value without the content that users put online. The content would have little meaning, if there were not millions of people accessing it. The majority of applications that have transformed our lives were not thought of by the designers of the Internet, but were produced by people who leveraged the fact that it is an open platform that anyone can use or program. Computing had dramatically smaller impact in the time of mainframes than it has now, with all the smart phones.

- While even today, the notion of participation is central to research work on gathering information from the interface between the digital and the real world, the use of such Big Data in complex techno-socio-economic models is at a stage that is comparable with where computing was in the era of mainframes: accessible only to big organisations and research groups. Through the Global Participatory Platform (GPP), FuturICT will open up data-driven models for information society: Everyone will be able to easily pose complex questions, explore the answers, organise their own data gathering and analysis communities, and create value from new applications. Putting together a small-scale Exploratory will be as simple as programming an App. Using the system to analyse, say, the consequence of somebody's life style on the environmental footprint will be as simple as starting an app today. Everyone will be able to become a contributor, with self-control of what his/her data (or models or software) will be used for, and what sort of compensation will be offered. A core innovation is that the project will not just provide the technological infrastructure needed to facilitate such a vision. FuturICT will also perform research on questions related to incentives, trust, privacy, and embedding in social structures and community processes. As a consequence, the technology will contribute to novel ways of social, economic and political participation and collaboration.

4. As shown above, the increasing blurring of the boundaries between the real and digital world creates unparalleled opportunities. However, they also come at a cost. Even today, existing design methodologies and operational principles are increasingly failing in the face of the complexity, dynamism and intractable interactions with humans and their environment, which characterises modern ICT systems. Therefore, FuturICT will improve our understanding of society and how it interacts with the global ICT system will be used to develop a theory of the co-evolution of ICT with society, which in turn will build the basis for using principles inspired by society to drive adaptation, self-organisation and self-regulation within ICT systems as means of fostering reliability, resilience and trustworthiness.

The FET Flagship vision has often been described by the European Commission as aiming at a "man too the moon-like" program. Staying within this analogy, it can be said that, with respect to things such as Reality Mining, modelling global techno-socio-economic systems, participatory computing and self-organisation in ICT systems, FuturICT wants to go from the early day of rocket science in the 1940ies (which had proven that building rockets is feasible, in principle) to actually flying to the moon (where the technology has been taken to a stage where it could really make a meaningful difference that humanity can be proud of). The science and technology gap that needs to be bridged is certainly comparable, while the required investments into data science and information technology are certainly much cheaper.

However there is a big difference between FuturICT and the "man on the moon" program. FuturICT does not aim at a single spectacular, headline-grabbing success that fits into a "sound bite". Instead, the project is aimed at creating a string of sustainable, transformative breakthroughs with impacts on science, technology and society that are comparable to that of the transition from the mainframe to personal computer and with the raise of the Internet.

## Approach

To facilitate the above ambition FuturICT is fully committed to the integration of theoretical, computational, experimental, and data mining approaches in a true synergy. Thus, none of the three disciplines is at the service of the other, but rather all disciplines **fundamentally profit from the collaboration.**

Other key difference and advantages of FuturICT over currently existing projects are:

- FuturICT combines theoretical with experimental approaches and large-scale data mining, creating platforms for efficient social experiments.
- FuturICT will create an open participatory approach and platform, which will bring together a much larger and more diverse brain power than it could be recruited by a single company or state, combining youth and experience, scientific and managerial excellence, and experts from ICT, complexity science and social science.
- FuturICT engages in collaborations with many other projects, for example, Climate KIC, The Future Internet, CRISIS, and many others (see Appendix 5 for an overview). In this way, FuturICT stands on the shoulders of giants. In particular, the project extends experiences collected in weather and environmental modelling towards measuring and simulating techno-socio-economic systems.
- FuturICT is performing open research and will, therefore, benefit from external participation and critique. It will unleash the power of open participatory platforms, crowd sourcing, and co-creation.
- As a publicly-funded project, FuturICT will be conducted openly and with a particular attention to ethical issues, so that its outcomes will receive scrutiny and debate from scientists, who will contribute to the development and dissemination of the science, but also from professional ethics experts, businesspeople, and citizens, who will ensure that the outcomes of the project are produced for the common good, and will be available for exploitation and utilisation to the widest possible range of users.

Therefore, FuturICT is different from previous efforts in the USA to simulate the behaviour of real individuals, using personal activity data, known under the name Sentient World. These were terminated primarily due to a lack of transparency and public control. This is a problem that does not apply to the FuturICT FET flagship due to its transparent, privacy-respecting and publicly controlled approach. Moreover, FuturICT does not want to simulate an exact copy of each individual, or track individual data. From complexity science it is known that a complex, multi-component system often shows universal behaviours close to a so-called tipping point, at which the system changes its properties. In fact, FuturICT is not interested in tracking the private behaviour of individuals. Instead, it wants to understand the interdependencies between social, economic, technological and environmental systems on a macroscopic scale, with the goal of identifying the driving factors underlying fundamental changes in these systems. FuturICT's technologies and platforms will furthermore be designed to promote privacy-respecting, ethical, value-oriented, and responsible use.

1. Flagship Concept and Objectives> **Example of FuturICT Participatory Scenario: The High School H1N1 Observatory**
> Alessandro Vespignani (one of FuturICTs partners) was able to model accurately the spread of H1N1 through mathematical models of infection combined with global travel data (http://www.gleamviz.org). Inspired by this, Ms. Teacher in Little Village challenges her pupils to set up an observatory to predict how soon H1N1 would reach Little Village, given outbreaks in the nearest city 10 miles away, and several locations around the world, and to demonstrate their understanding of why they reach the conclusions they do. The students build their H1N1 portal using the GPP web toolkit to drag and drop a set of widgets together to interrogate public datasets and global travel patterns, mash them up using rules defined in a simple visual language, and then render the results using a range of visualisation widgets. They also devise a sensor network game in which villagers "infect" each other via their phones when they meet under certain conditions, allowing them to study the spread of the disease within their own school and local streets, which really drives home the seriousness of the illness. The conclusions are not definitive, so they summarise policy recommendations to their Minister for Health, using argument maps to distill on a single page the key issues for deliberation, the tradeoffs between different options, and the evidence-base underpinning of each one. Hyperlinks in the maps reveal more detail on request, showing different states in the simulation models and visualisations, at which key turning points are judged to be seen, with automatically generated textual narratives, summarizing the key assumptions, variables and interdependencies.

## 1.8 Benefits

### General benefits of the FET flagship response

The benefits of creating a large-scale, federated effort across Europe have been described earlier in Section 1.1 including: scientific justification; creation of European scientific leadership; societal benefits; and potential new innovations. The recent economic crisis vividly illustrates why we need better tools to understand, design and control socio-technical systems. Even more compelling however, are the fantastic opportunities for business, the public sector, policy and academia to use the fruits of FuturICT's research to bring about transformational change in our economy, in academic excellence, and in civic value. The scale and necessary integration of these benefits justify and require a FET flagship. The leaders of the universities that joined the FuturICT meeting at the Royal Society in London all agreed on this. Although research teams around the world are working on small-scale initiatives, Europe has the breadth, depth of talent, diversity of interest and culture, and the capacity to reach for goals at scale. Only a Europe-wide flagship can raise and integrate the funding and human resources, which go far beyond what could be achieved by any one nation or enterprise.

### Tangible benefits

FuturICT will deliver primary benefits following directly from a better understanding of how socially interactive systems work. The Pilot Phase developed the argument for three broad categories of benefit: technology, science, and society. See Figure below.

FuturICT    27

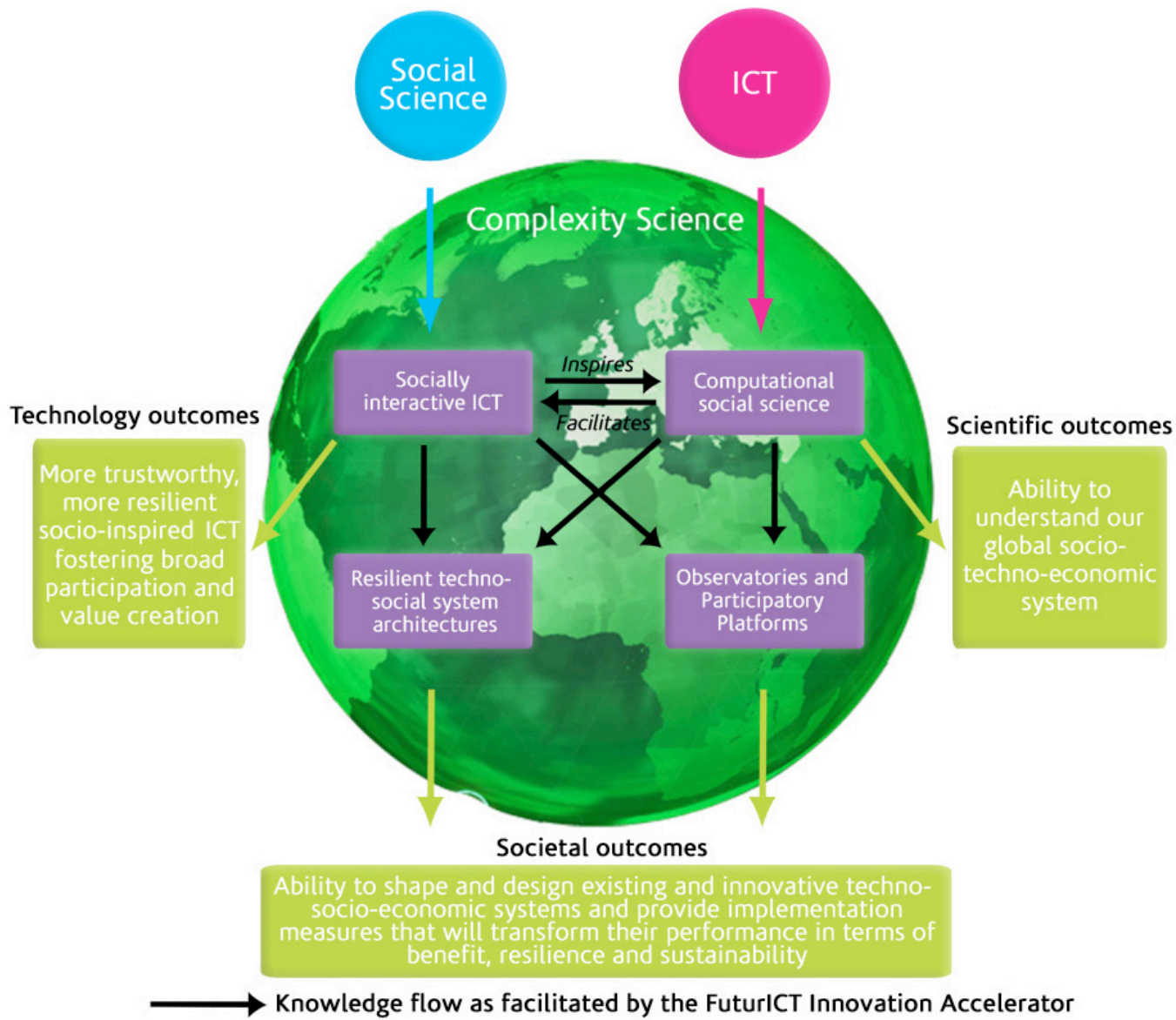

*Figure 1.8 Outcomes*

**Secondary and serendipitous benefits**

There are many other secondary benefits that are likely to create long-term benefits. The FuturICT research environment will act as a magnet, attracting the best young talent worldwide, helping retain a talent pool in an area that will define the future of our economies and societies. Eminent researchers such as László Barabási and Alex Vespignani will be able to extend and empower their European research groups, creating a European excellence cluster that will drive the future vision of science, work and economies. There is also enormous synergy with many other groups such as the Santa Fe Institute who would bring their skills in dealing with the future of cities and economies to the project

The alliances that will be forged between FuturICT and academic institutions will add value in different, surprising ways. It will drive universities to fund new posts and create new educational programs at both graduate and undergraduate levels (e.g. on Big Data, Integrated Risk Management, Global Systems Science, Sustainable Cities), so that Europe can build the human capital required to successfully deal with challenges of our complex, strongly connected techno-socio-economic systems. A recent Wall Street Journal article estimated a 1.9m person shortfall in big data skills in the US Economy over the next 10 years. FuturICT will stimulate effort in Europe to close the corresponding gap in European skills. The integration of ICT expertise with knowledge in the social and complexity sciences will provide emerging key skills area for the 21st century. For example, the move towards a personalised education could potentially have a massive social benefit both in terms of equality and in terms of cost savings.

In the public sphere, FuturICT benefits include transparency and participation. The Global Participatory Platform (GPP) will empower and enable cooperation across society. The GPP will provide access to sound evidence and exploratory models for policy-makers, businesspeople, and entrepreneurs. It will be engaged with a wide range of active citizens groups. This broader and deeper engagement will unleash fresh insights on pressing problems at local, regional, national and global scales. The step change that is needed in the policy-making infrastructure, ethical responses to Big Data and other scientific breakthroughs, can be best achieved by an open, public process.

### Societal benefits

It is difficult to attribute changes in society due to a single issue but in the fields of ICT it is quite clear that these have life changing capabilities. Key aspects of the FuturICT are of participation and transparency. The whole purpose of the GPP is to muster the cooperation of many more sectors of society than those who usually engage with scientific research, such as policy-makers, businesspeople, entrepreneurs and citizens, who will participate in the development of fresh insights on the world's most pressing problems. A step-change is needed in the policy-making infrastructure, ethical response to Big Data and scientific breakthroughs. This can be best achieved by an open process engaging with stakeholders, supported by a major initiative of public and private funding.

Potential benefits will generate handsome dividends, both by mitigating problems that are currently generating tremendous societal costs, and by developing new socio-inspired technologies and a participatory information ecosystem that will create new business opportunities – not only for companies, but also for SMEs and individuals, particularly along the lines of the paradigm of co-producing consumers ('prosumers').

### FuturICT: All aboard!

Working in interdisciplinary fields is not easy but the project will create an atmosphere of collegiate action that helps to remove barriers elsewhere in our societies. During the Pilot Phase certain domains were perhaps seen as easier than others to integrate and so these form the thrust of research for the Ramp-Up Phase. But there are other areas which, once research begins, could lead to significant advances and financial and societal gain. For instance healthcare, is one of the biggest ongoing costs in most countries. A new approach to sections within healthcare activities, taking into account the complexity of the problems and the broader picture (for example the fact that advertising, travel, schools, policy all have implications for obesity, and the location of hospitals, clinics and related medical services with implications for access, hence quality of care) may prove to be extremely beneficial. For business, the same is true for logistics and planning for global businesses. Therefore, taking a lead from climate issues, we put off until tomorrow what we should really do today, at our peril. It can only lead to false economy. FuturICT simply is a project that we cannot afford not to do.

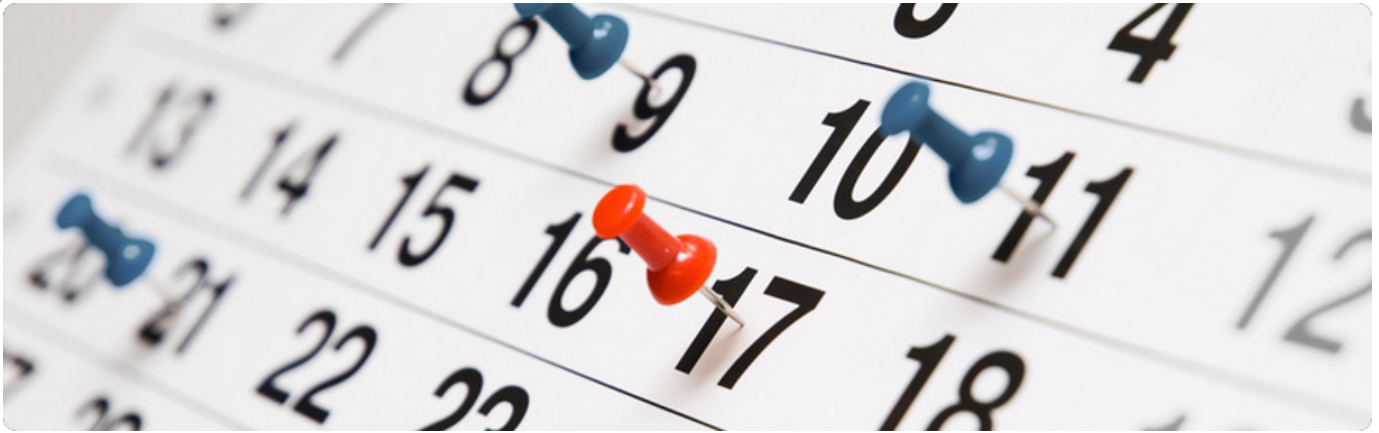

# 2. S&T Methodology for the Flagship

FuturICT's vision is to extend the horizon of knowledge and understanding of global techno-socio-economic systems. FuturICT aims at defining a transformative change in the scientific approach to studying human society and its interaction with global, networked information systems. It will also lead to a new generation of socio-inspired and socially interactive information and communication technologies (ICT), which will leverage self-organisation, self-regulation, and collective awareness to foster resilience, stability, trustworthiness, and adherence to basic ethical principles.

## 2.1 Overview of Roadmap

The vision and objectives of FuturICT lay out a roadmap that, starting from multi-disciplinary research through inter-disciplinary and trans-disciplinary approaches, defines a new integrative research area that build the Global Systems Science needed to target the scientific and technological challenges of today and in the future. The path to accomplish the FuturICT goal and to realise the FuturICT vision, will proceed in four phases:

1. Ramp-Up Phase. In this 30 month phase of FuturICT, we will pursue three main stages:
   - Define: Identification and definition of key methodological and architectural foundations needed to realise the FuturICT vision. Establishing key processes needed for effective interdisciplinary collaboration towards the envisaged goals. Open calls for new Partners.
   - Refine: Based on the results of the initial work and experiments, refine the approach to produce solutions of some selected problems as exemplified in small-scale prototype proof-of-concept case studies underscoring the value of the FuturICT interdisciplinary approach and the overall plausibility of the vision.
   - Align: Demonstrators of success and continued community building to align all stakeholders behind a detailed work plan for the next FuturICT phase under Horizon 2020.

2. Core Development Phase (2.5 years). Given that FET flagships are expected to promote basic research leading to ground-breaking discoveries in many areas, a classical waterfall model for the linear development of systems and solutions is not suitable for FuturICT. Instead, FuturICT will follow an iterative, exploratory approach driven by continuous interaction and discussion and frequent empirical validation through interdisciplinary case studies that are increasing over time in scale, accuracy, ambition, and user base. An important aspect of the development will be regular public releases of new methods, algorithms and models used and feedback from a broader community as well as incremental contributions to the envisaged FuturICT integrated System-of-Systems.

3. Consolidation Phase (2.5 years) with a focus on improving, evaluating and applying the most promising technologies and approaches from the Core Development Phase, finalising the envisaged

FuturICT System-of-Systems to the public and fostering commercial uptake. Developing advances that have arisen serendipitously giving the FuturICT opportunities to explore concepts not imagined (so not formulated in a work package structure).

4. Demonstrations of success. Working models and techniques which realise the FuturICT vision and successfully accomplish the FuturICT goal.

The roadmap will be characterised by the following processes:

- Building Blocks: FuturICT will use from multidisciplinary research endeavours that involve several researchers from different academic disciplines, with different research paradigms, working together on a theme or problem but with multiple disciplinary goals. In a multidisciplinary framework participants exchange knowledge, but do not aim to cross boundaries between disciplines to create new shared knowledge and theory.
- Engineering Inter-disciplinarity: FuturICT is therefore aims to dissolve scientific silos and domains within its community to generate interdisciplinary research that goes beyond multi-disciplinary research by involving researchers from several unrelated academic disciplines and leading them to cross subject boundaries to create new knowledge and theory and solve a common research challenge.
- Integrative process: FuturICT will move toward an integration of efforts and building trans-disciplinary research that combines interdisciplinary research with a participatory approach to involve both unrelated disciplines and non-academic participants to create new knowledge and theory.

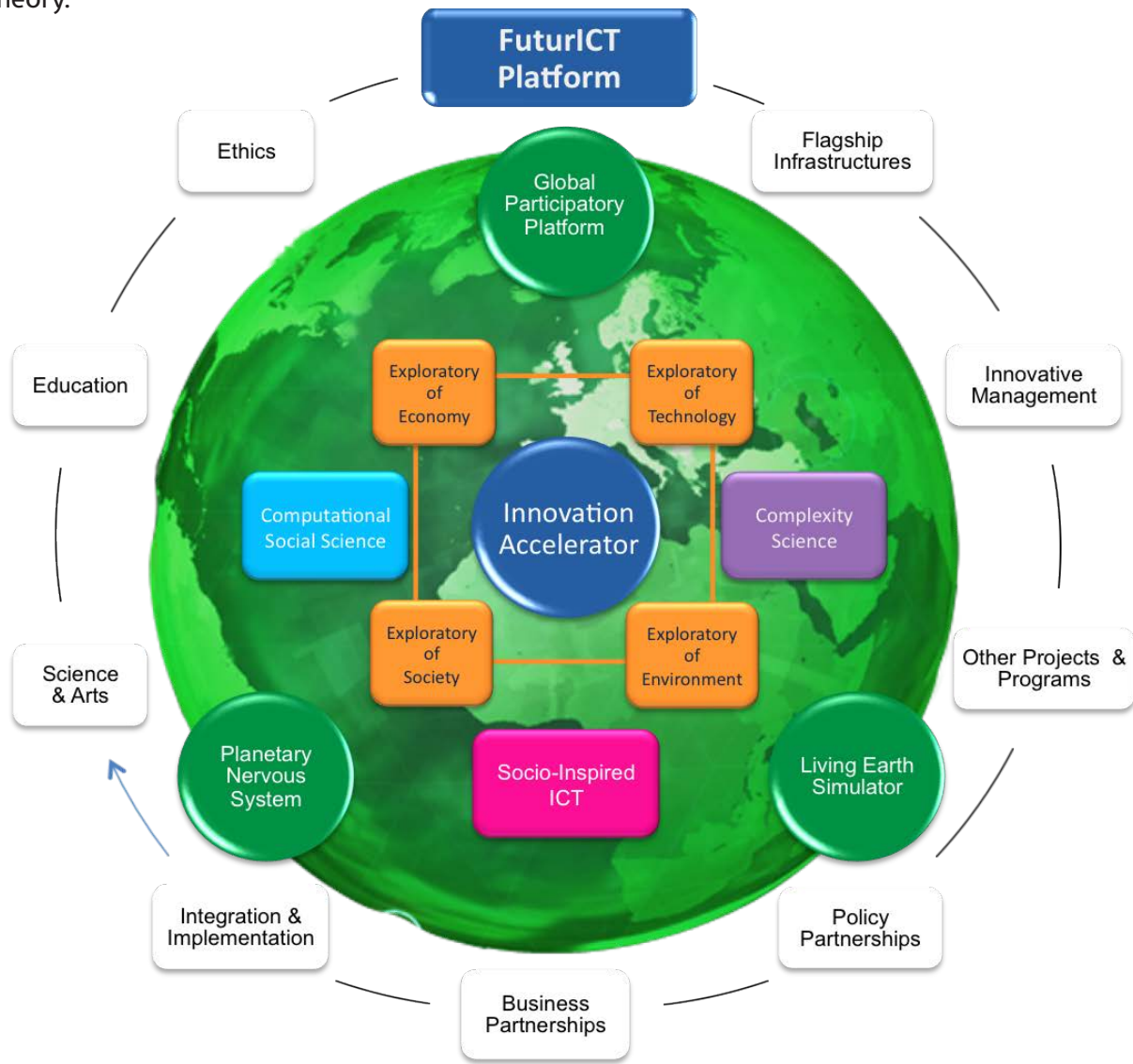

**Figure 2.1** *The FuturICT FET flagship will integrate many different scientific areas and activities into one FuturICT platform that will enable participatory science and technology to manage our complex world in a more sustainable and resilient way.*

## 2.2 Focus Areas for the Flagship

In order to accomplish its goal, FuturICT will develop new scientific methods and technologies by integrating key elements from the fields of ICT, the social sciences and complexity science using a collective, participatory approach and integrated research. During the Pilot Phase a strategy was formulated to accomplish this unique goal and a Work Plan was devised based on research activities integrated across these topic areas. Our work is strongly collaborative across 9 scientific areas, termed Focus Areas (FA), which form the Core Project (the CP). These research oriented Focus Areas, covering the core areas of Data, Models and People, follow the description of FuturICT established during the Pilot Phase, and consist of:

- FA1: PNS - Planetary Nervous System - creating systems to measure and organise information.
- FA2: LES - Living Earth Simulator – models to simulate and forecast.
- FA3: GPP - Global Participatory Platform – a platform to interact and explore.

These will be supplemented by two key Focus Areas dealing with fundamental advances in ICT and science, namely:

- FA4: SocioICT - Socio-Inspired ICT & Principles of Systems Design
- FA5: GSS - Global Systems Science

Exploratories form the next Focus Area, which combine heterogeneous data and models in specific application areas to explore interdependencies between the Economy, Environment, Society and Technology:

- FA6: Exploratories

Furthermore, the CP-CSA will have four cross-cutting Focus Areas:

- FA7: IA – The Innovation Accelerator
- FA8: Ethics - Ethical and Value-Sensitive ICT
- FA9: Integration - Integrated Technological and Scientific Research
- FA10: Coordination and Dissemination

The first 9 of these Focus Areas are research-oriented Focus Areas forming the Core Project (the CP of the CP-CSA). To bring new Partners into the FuturICT, there will be a Focus Area that covers a series of Open Calls. These will offer opportunities for new academic institutions and businesses, but also a staircase to excellence for young researchers:

- FA11: Open Calls and Staircase to Excellence

FA10 and FA11 also support the Core Project, but are more formally part of the Coordination and Support Action (the CSA bit of the CP-CSA). Finally, there is an administrative Focus Area:

- FA12: Project Management and Flagship Framework.

The Focus Areas are diagrammatically represented in Figure 2.2.a below:

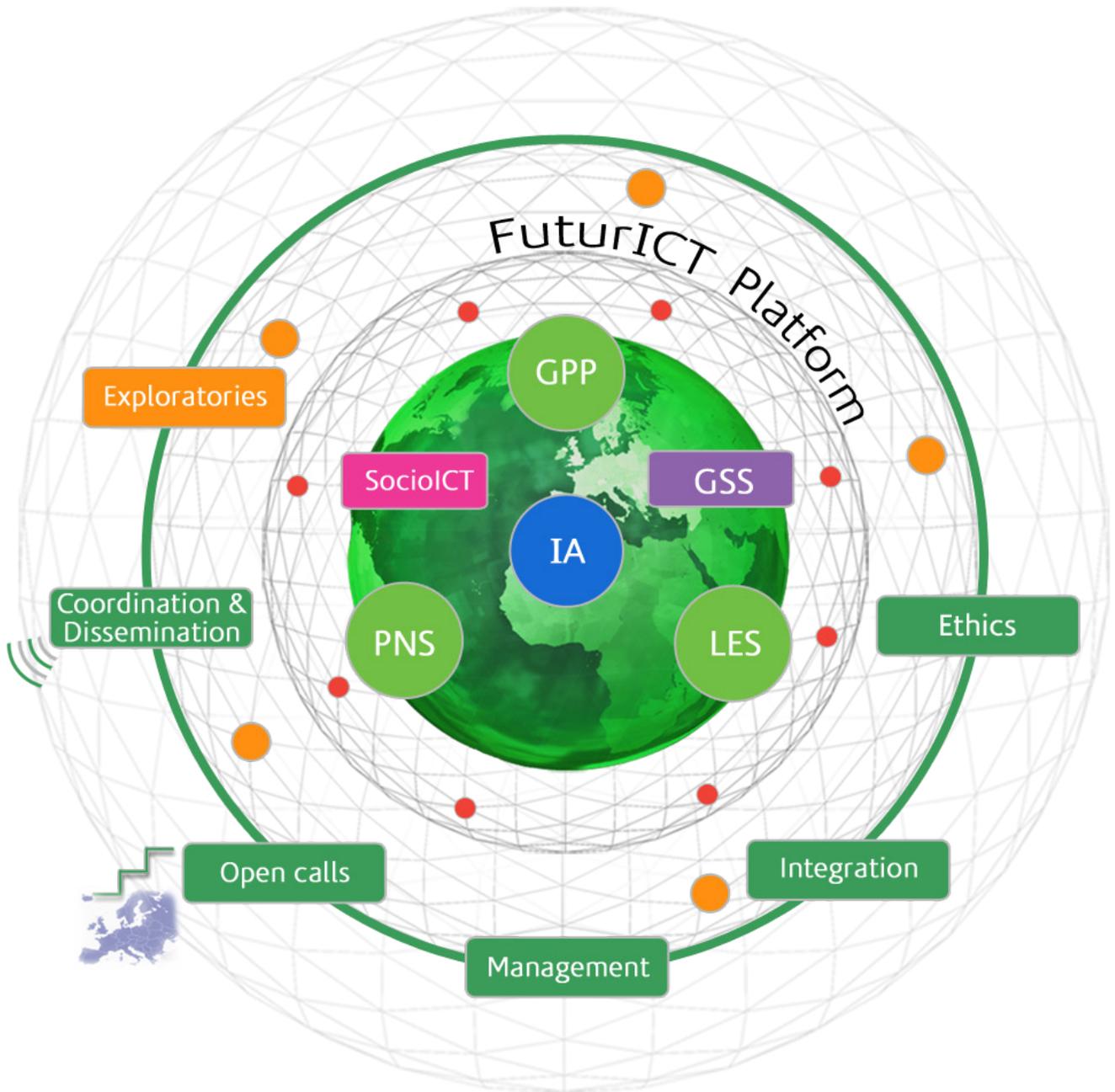

**Figure 2.2.a** *Illustration of the organisation of FuturICT research activities grouped into Focus Areas.*

The Focus Areas have been designed to enable interdisciplinary working in reasonably sized, effective teams. There will be support within each FA for travel and exchanges between FA members, including staff employed or engaged in FuturICT, in order to promote scientific integration. During the Pilot Phase the FuturICT team demonstrated the usefulness of National and international FuturICT events to foster collaboration with a wide community and so these experiences will also be taken forward during the Ramp-Up Phase.

### FA1: The Planetary Nervous System (PNS)

will be a real-time "measurement" instrument for our global techno-socio-economic system – what might in perspective be considered a "CERN for the social sciences and global economy". The PNS will be an interactive, self-organizing measurements system facilitating the transition from high-level human-understandable questions to the global-scale collection and evaluation of highly heterogeneous sensor and Internet data to produce an ever-changing picture of the current state of our world. Providing a fundamental evidence base, it will support a collective sense-making of causal interdependencies and hence a collective awareness of the implications of human decisions and actions ('social footprint'). This will improve the ability of our society, as a whole, to be aware of and make sense of developing issues at an early stage, allowing decision-makers to effectively identify problems, "steer" policies around them and/or initiate efforts to solve them (where necessary). This would, for the first time, create the real-time data and a substantial capacity to adaptively manage our complex world, achieving improved sustainability, and providing the resources for Exploratories of financial instabilities, social challenges, health and epidemics etc. (FA6). At the same time the PNS will foster fundamental advances in distributed systems, machine learning, knowledge discovery, distributed reasoning participatory and self-organised sensing, and give rise to a new research field of social sensing and privacy-aware social data mining.

### FA2: The Living Earth Simulator (LES)

will perform what-if scenarios to reveal the opportunities and challenges as well as the possibilities and limitations of decisions and policy interventions. It will allow one to identify the hitherto unanticipated consequences (including unintended side-effects) of decisions and natural events, such as planning decisions concerning major urban development projects (e.g. future cities), evacuation scenarios, the spread of epidemics etc. Modelling tools will be designed and developed for use by non-experts through new scalable computing capabilities that will support the interactive integration of heterogeneous models (such as multi-agent and multi-level models) with Big Data provided by the PNS. Visualization tools scaling from large domes to mobile devices will make the data and models easy to interact with and understand. This will enable a broad range of users to gain new insights into our complex socio-economic world and support a truly informed policy-making process. Creating the Living Earth Simulator will lead to groundbreaking advances, including algorithms and methods, in large-scale visualization, simulation, cloud computing, data management and visual analytics.

### FA3: The Global Participatory Platform (GPP)

will connect data, models and projections with people, and provide the access to PNS and LES functionality to a range of users, ranging from school children to highly skilled domain experts. It will create a new ICT paradigm focused on enabling bottom-up participation in social, economic and political affairs, creating opportunities to be involved in social, economic and political matters on all levels. At the core of this new paradigm will be a trusted, scalable brokerage platform designed to promote communication and cooperation between data collectors and data users, incentivizing community data collection, provision and curation, and providing an abstraction layer to allow the use of data as a resource. The GPP will facilitate personalised education and customised services; create non-expert systems (accessible to everyone without prior training); empower citizen science; and develop principles for open data and modelling that will promote their responsible and ethical use. The GPP will offer a layer of abstraction and application programmer interfaces that will empower software developers to create and distribute applications on top of the LES and PNS. The power of this approach is that the creation of such applications is bottom-up and participatory, allowing a broad use of the envisaged technologies. Additionally the platform aims at creating interactive multi-player online games and Interactive Worlds as platforms for quick, large-scale decision experiments, which allow one to explore and test various decision rules, urban designs, financial architectures, and possible futures. This social "in vitro approach" will accelerate the ability to gain insights into social interaction processes by providing almost-real data and experience of proposed systems, in a way that might be compared with the way gene sequencing has revolutionised genetics. In short, the GPP will provide the interface "glue", which will allow a new fusion of participatory discussion, complex scientific networking and collaboration, and a virtual learning/experiencing of proposals. Such a fusion would allow new social institutions to emerge, breaking down the "crisp" barriers between professional researchers, stakeholders, students and politicians. In computer science, the GPP will create a new paradigm of participatory computing where the notion of resources is extended to data and activities in the physical world and where the physical world becomes part of the input and output.

### FA4: Socio-Inspired ICT and Principles of Systems Design

will develop a new stream of technologies that are inspired by social systems, thereby leveraging some of the principles underlying social organisation (such as coordination, cooperation, self-regulation, conflict resolution, resilience, reputation, trust, social norms, culture, social capital, values and ethics), understanding and distilling them and then applying them in distributed ICT systems. In particularly, this Focus Area will aim at developing a kind of socially interactive ICT that is characterised by self-organization, social adaptability and collective awareness, seeking to enhance human and cyber confluence, system adaptation and a beneficial co-evolution of ICT with society. Underpinning this new kind of ICT will be a transition from the currently dominating "single user – single system" perspective and the associated notion of Human Computer Interaction towards a "system of systems – human community" perspective and the resulting notion of System Society Interaction. Socio-inspired, socially interactive systems promise ICT systems that fit more "naturally" within human society, and offer increased flexibility and robustness. This will not only produce direct benefits by the provision of better and more accessible ICT, thereby improving the lives of European citizens, but it will also help with the design of the major FuturICT components (FA1, FA2 and FA3), enabling large-scale participation far beyond the pool of FuturICT experts.

### FA5: Global Systems Science (GSS)

will develop a deeper and better founded understanding of strongly coupled, complex, global techno-socio-economic systems – made up of components with internal complexity and cognitive ability – including the analysis of causal interdependencies, parameter-dependent system behaviours, and probabilistic/'possibilistic' short-term predictions. Part of this activity is the further integration of agent-based models of socially interactive systems with network theory and complexity science. The establishment of a new data science, which studies how the interaction with and between information impacts the dynamics of techno-socio-economic systems, will enable an empirically powered global systems science and thereby boosts complexity science from the world of abstract models and analogies towards having real-live impacts. One such impact will be a new theory of the co-evolution of ICT and society as theoretical underpinning of the envisaged new wave of socially-inspired, socially interactive ICT technologies. GSS will have breakthrough consequences for the modelling of social policies as well as many other areas. The explanatory understanding gained by Global Systems Science will also inform and empower other parts of FuturICT (FA2, FA3, FA4 and FA6).

### FA6: Exploratories (EXP)

are cross-sectional, application-oriented activities that bring together data, models and people. Hence, the main breakthrough will be the integration of PNS, LES, and GPP functionality (i.e. data mining, computer simulation, and interactivity) and the application to practical challenges in concrete exemplar case studies, namely in the areas of (1) sustainable financial systems, (2) health protection from epidemics, (3) measuring and mitigating social challenges, (4) measuring systemic risks and increasing systemic resilience, (5) finding principles for sustainable future cities, (6) developing smart energy systems, and (7) reaching sustainable environmental systems.

In particular, the Exploratories will implement massive data mining, innovative (agent-based) simulation platforms, and suitable filtering techniques to detect forth-coming or possible crises, e.g. bubbles or crashes in financial or housing markets ("market monitoring"); develop and integrate sophisticated theory-driven models, allowing to explore testable hypotheses and interpretations of the complex; identify advance warning signs for social, financial and economic instabilities, for shortages in supply (e.g. energy, water, food), wars and social unrests, epidemics, environmental change, etc.; extract laws of systemic instabilities; identify the social and behavioural mechanisms on which the phenomena under observation are based, also find the interdependencies, feedback loops, and causality chains that may lead to different levels and types of spreading effects (such as cascading effects, contagion spreading, or emerging collective behaviours).

This will require new behavioural and cognitive models of socio-economic learning and reasoning, as well as new economic models that track instabilities and spreading effects. One direction of special interest in the monitoring of instabilities lies in finding correlations with spatial outcomes, possibly down to the city level in its most significant aspects, such as the housing markets and the migratory movements. Generating economic data and predictions, which are spatial and local is a main priority, linking various themes in this Focus Area across Work Packages, involving financial systems, health and epidemics, social challenges, crime, and energy systems.

As a very specific result, the Exploratories will for example develop techniques to provide, whenever possible, early warnings of impending crises, such as:

- possible bubbles or crashes housing and other markets ("market monitoring");
- economic instabilities or shortages in supply (e.g. energy, water, food),
- wars and social unrest,
- emerging epidemics,
- environmental change.

Thus the Exploratories are the applied end of FuturICT, delivering concrete advances in these various areas that will complement and go beyond our currently available understanding and techniques. The experience from these Exploratories will feedback into the other FAs in order to inform their design and improve their eventual contribution to the Exploratories.

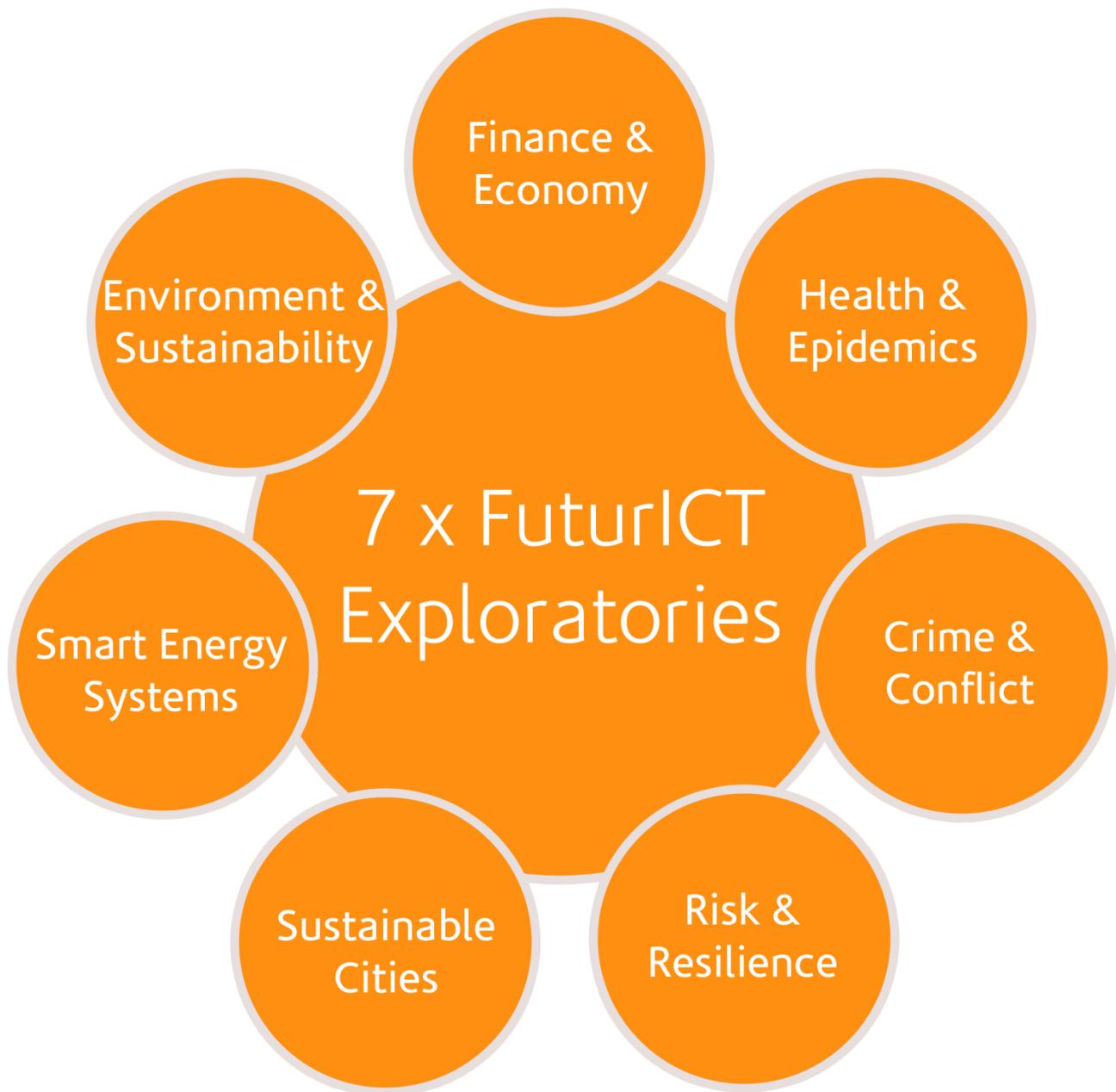

**Figure 2.2.b** *FuturICT Exploratories*

### FA7: The Innovation Accelerator (IA)

will develop an integrated and participatory approach to employ the 'wisdom of crowds' and the 'power of networking' for purposes of learning, with a view to establishing new trends, inventions and innovations. This will include learning how to measure science and innovation in a multi-criteria way; enabling personalised education; developing new IPR approaches and principles for differential, collective innovation (including suitable incentive schemes); facilitating crowd funding; eventually leading to the creation of an integrated communication, coordination, and collaboration platform for large-scale projects. Another aspect of the IA is to fuel businesses and spin-off companies. The impact of the IA will be spread across the sciences, improving their effectiveness as well as the effectiveness of large-scale projects such as FuturICT itself.

### FA8: Ethical and Value-sensitive ICT (EVI)

will be fundamental to all FuturICT activities, ensuring the integrity at all levels, including the technical level. It will help the FuturICT platform to become a proper public good, and create appropriate value for the citizens of Europe and beyond. EVI will identify the dangers and potentials of Big Data, develop privacy-respecting data mining; learn and demonstrate how to promote the responsible use of open platforms; understand the downsides of innovation; and create a culture of ethical awareness and respect. The understanding of how to achieve EVI within the distributed and complex FuturICT platforms will also have a wider impact by promoting a paradigm of ethics and value-orientation by design, considering ethics and values from the very first moment, when new technology is born. It will also have an impact by establishing and spreading an ICT paradigm characterised by openness and transparency, participation and democratic control, tailored to the culture of the users.

### FA9: Convergence towards Technology and Systems Integration

will consider the breadth and depth of the technical basis of the Fut$\sum$urICT project, enabling standards to be agreed upon between the Planetary Nervous System (PNS), Living Earth Simulator (LES), Global Participatory Platform (GPP), Exploratories (EXP) and the Innovation Accelerator (IA). This will enable the FuturICT consortium to continue to grow as new technologies can be included and new components, such as tools and sensors, can be added to the FuturICT infrastructure in a seamless manner. Such standards and interfaces may well be useful outside FuturICT, enabling an "ecosystem" of other projects and spin-offs that, whilst not part of FuturICT, are facilitated by its open standards. On a scientific level, FA9 will develop fundamental new principles of integration and interoperability in the spirit of a "federated system of systems" with a focus on bringing together dynamically evolving, autonomous, highly heterogeneous components by an approach based on the principles of convergence and co-evolution.

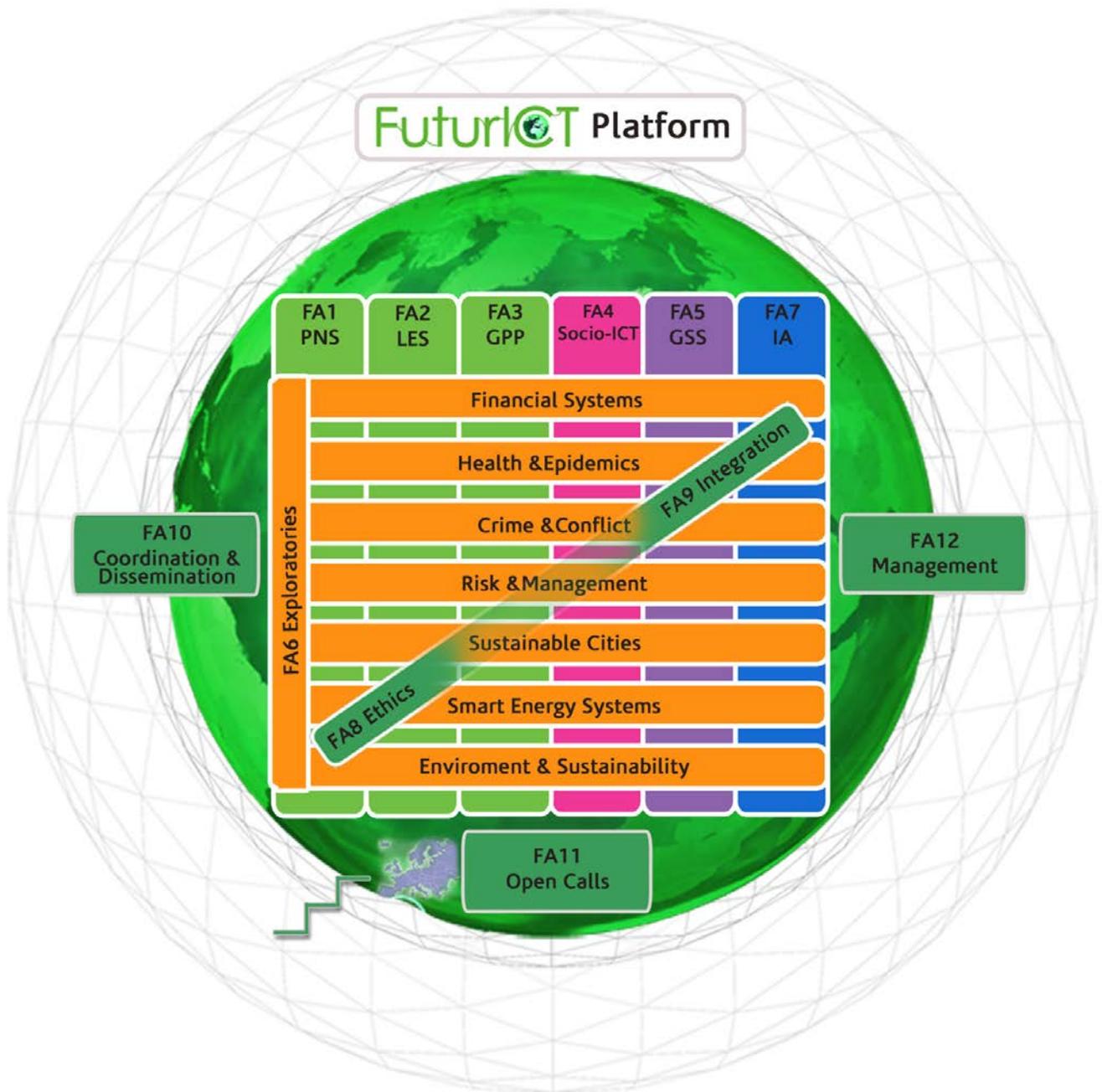

*Figure 2.2.c* Illustration of the organisation of the FuturICT Flagship activities. FuturICT is organised in 12 closely interacting Focus Areas (FA). Vertical Focus Areas have a strong focus on fundamental aspects of research and ICT development. They include the development of the 3 main ICT components of the FuturICT Platform, namely the Planetary Nervous System (PNS), the Living Earth Simulator (LES), and the Global Participatory Platform (GPP), furthermore, the development of Socio-Inspired ICT (SocioICT), and of a Global Systems Science (GSS). The Exploratories (Focus Area 6) and the Innovation Accelerator (IA, Focus Area 7) will be the outcomes of horizontal and more application-oriented activities. They integrate the elements of the vertical activities, in particular data mining, scenario simulation, and interactivity. Ethical research activities (Focus Areas 8) take place across all Focus Areas. The same applies to Integration activities (Focus Area 9). FuturICT is supported by Network Coordination and Dissemination activities (Focus Area 10), Open Calls for new Partners - also offering a staircase to excellence for young researchers (Focus Area 11), and finally the Management of the FuturICT Flagship (Focus Area 12).

# 2.3 Integration on a Technological Level through Technology Convergence

FuturICT's goals of technological integration can be summarised as follows:

5. The core scientific and technological innovations need to be made available to a broader community in a form in which they are useful, usable and will be used.
   a. As new algorithms are developed in the course of the project, they must be made available for use by a broad community
   b. The Exploratories must be usable by the relevant stakeholders.
   c. Eventually, a version of the GPP able to use core PNS and LES functionality and able to access the most relevant FuturICT models, should be publicly downloadable.
   d. The methodologies developed to enable collective awareness and socio-inspired self-organisation need to be made available for take up by other scientists and system developers and moved towards the integration in complex real-life systems, also to create spin-off companies.

6. The project must facilitate easy take up of the core project innovations by industry. This in particular implies that
   a. appropriate standards for data interchange, protocols for interaction between different system components and reference architectures must be defined and demonstrated (even if they are not necessarily the basis for all collaboration and cooperation within the project itself);
   b. concepts/roadmaps must exists for the integration of the methods and tools developed by FuturICT with existing standards and systems.

The difficulty of technical integration stems from the fact that the project has a focus on basic science and involves an exceptional breadth of different research approaches and cultures. Thus, integration in the classical sense of defining and in particular enforcing a common software development process, architecture, and an official, exclusive set of tools and standards is neither desirable nor practicable. Instead, FuturICT relies on an entirely novel approach to technology integration, a new paradigm that is tailored for large-scale, heterogeneous, quickly changing science and technology environments, in which creativity must be unleashed rather than obstructed by too many operational and frequently short-sighted constraints. This is the paradigm of co-evolution and convergence, rather than the classic top-down integration that defines and enforces common research and development processes, architectures derived from waterfall methodology, and exclusive sets of tools and standards.

FuturICT will achieve this by engaging into different forms of integration, combined with a coherent collaborative approach of co-evolution. The fundamental concept, and FuturICT's ultimate ambition, is that of a "System-of-Systems". Such a FuturICT system emerges naturally, with individual research groups being the origins of high-quality scientific contributions, each and any evolving in its domain and thus parallel and concurrent to its peers. The integration effort will complement this approach by being facilitator that encourages overall convergence without touching scientific agendas and the work process of the individual groups. To achieve this ambitions, the following top-level objectives are defined:

- Building a mutual understanding between "problem solvers" (e.g. computer scientists and complexity scientists, whose aim is to deliver innovative components and algorithms to solve problems and create new opportunities) and "problem users" (e.g. social scientists, domain experts and the broader public, whose aim is to better understand social structures or to project the effects of decisions onto future states).
- Establishment of data interchange standards for the ever-growing body of qualitative and quantitative data to address the FuturICT vision of supporting policy makers, domain experts and the public in engaging with Big Data to support sense-making and decision-making.
- Encouraging and supporting the convergence and interoperability of technical infrastructures, platforms, components, and systems developed by the individual research groups.

In summary, innovation is achieved through permitting diversity: through allowing researchers to investigate novel possibilities, unconstrained by established standards or requirements to conform. At the same time, to reach maximum impact, new opportunities for integration will be created by weaving supportive technologies into the overall fabric of FuturICT's System-of-Systems through appropriate interfaces and middleware components.

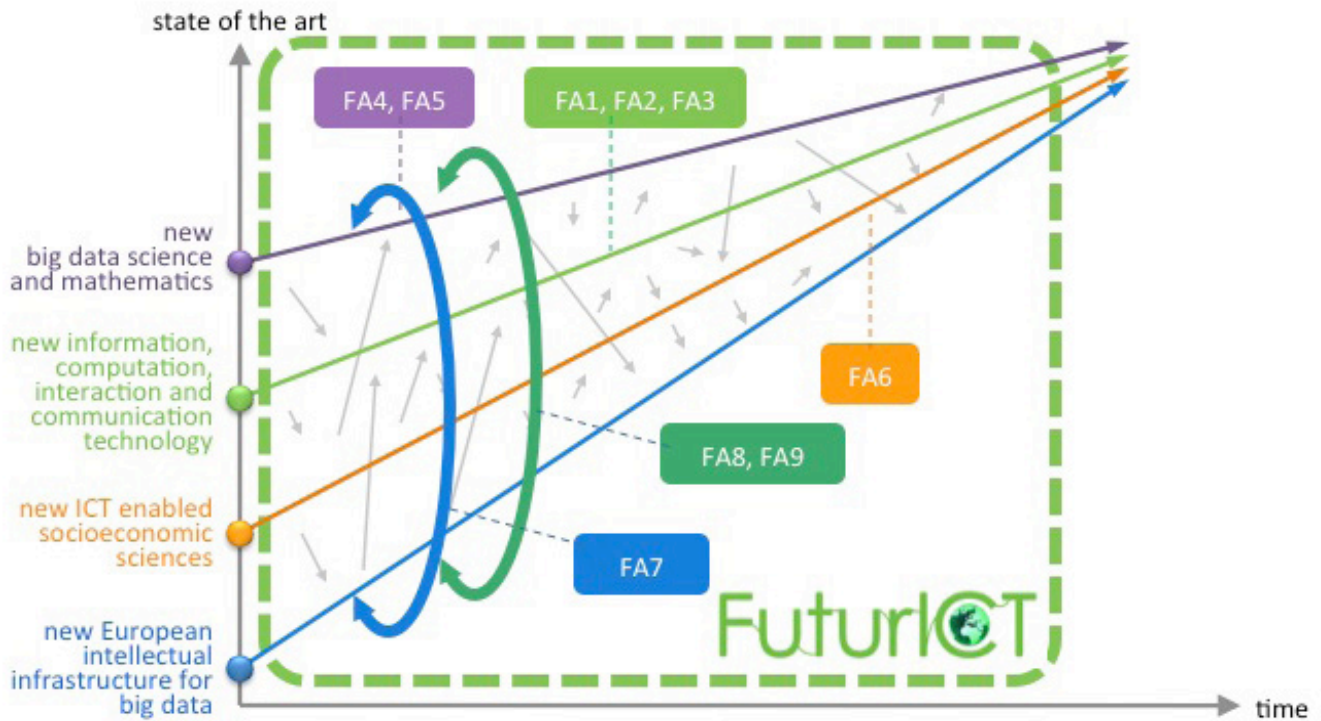

**Figure 2.3** Co-evolution of the main FuturICT objectives

## 2.4 Feasibility & Roadmap for the Flagship

### a) Financial Feasibility

The FuturICT FET flagship asks for support by the European Commission of under 60m EUR during the Ramp-Up phase and about 500m EUR altogether for a 10-year project period. There are substantial commitments to match this budget. For the Ramp-Up phase they add up to 92m EUR from academic institutions, funding agencies, and ministries, as documented by letters of support in the appendix of this proposal. We therefore expect that we can raise matching funds during the Horizon 2020 phase, which equals the EC contribution. By the end of the 10 years time period, the commercialisation of FuturICT outcomes through licenses and services of the FuturICT Platform will have progressed to an extend that allows to continue FuturICT in a self-sustained way via the FuturICT Foundation.

In addition to matching funds, the FuturICT's partners will bring in considerable complementary resources and are involved in many related research projects,as listed below. These resources will allow FuturICT to create synergy effects, added value, and an impact much higher than the FET flagship funds alone would allow. This high impact will be enabled by FuturICT's open platform approach that connects with other projects and promotes increasing levels of integration with them. The paragraphs below give a representative, but by far not complete picture of the complementary resources and projects that contribute to the feasibility of FuturICT's grand unifying vision.

### b) Scientific Feasibility

FuturICT does not start from scratch – it can build on extensive previous work that impressively demonstrates the feasibility of the modelling approach and of the technological concepts needed. A number of these FuturICT-relevant projects coordinated or conducted by the Partners of the FuturICT Consortium are listed below. Thanks to FuturICT's open platform approach, they will create substantial added value. The feasibility to organise synergy effects and stimulate convergence of these projects along FuturICT's grand vision in a fleet-like manner was already demonstrated during the Pilot Phase.

In recent years, DFKI, INRIA, CNR, Fraunhofer and many other Partners have been working on the various dimensions of social sensing and data mining, which provide basic methods, algorithms, systems and demonstrators for FuturICT's kind of research. Several award-winning systems and platforms for social and reality mining have been developed within European and international projects, and deployed in large-scale analytical studies: the Reality Mining project developed at the MIT Media Lab by Sandy Pentland and his group uses sensor data to extract subtle patterns that predict future human behaviour; the M-Atlas analytical system, a platform for mining Big Data of human mobility and reasoning over the discovered patterns (Giannotti et al., 2011), achieved as a result of the GeoPKDD FET-Open project (www.geopkdd.eu), was recognised by the EC as one of the top 10 results of all FET projects, and showcased at the European parliament in Strasbourg in April 2010.

A partial list of further relevant projects is given in the following: FP6 FET-Open GeoPKDD (2004-2008), FET-Open LIFT (2010-2013), FET-Open DATASIM (2011-2015), FET-Open ICON (2012-2016), FET-SAC Haggle (2006-2010), FET-SAC Bionets (2006-2010), FET-SOCIONICAL(2009-2012), FET-OPPORTUNITY (2008-2011), FET-PERADA SOCIALNETS (2008-2011), FET-Awareness RECOGNITION (2010-2013), FP6 FET-Proactive DELIS (2004-2008), FET-Open FRONTS (2008-2011, FET-Open FOC (2010-2012), FET-Open Bison (2008-2011), ICTeCollective (2009-2012). Marie-Curie SEEK (2012-2015); relevant machine learning networks and projects include FP7 PASCAL (2003-2008) and PASCAL2 (2008-2013), SMART (2006-2009), PinVIEW (2008-2011) and CompLACS (2011-2015). Some partners (MIT, CNR, YAHOO, Univ Helsinky, Univ. Tartu) have established Big Data labs, which will provide in-situ access for large-scale experiment.

To add further value to the FuturICT project, there are many Big Data sets, which will be available for the FuturICT partners, including (anonimised) call graphs from mobile phone call data, and networks of many online social networks, including Facebook, Twitter and Flickr; transaction micro-data from diverse retailers, Skype datasets, consisting of 600 million users and over 8 billion connections tracked across more than five years, query-logs both from Yahoo! search engine and e-commerce, GPS tracks from personal navigation devices and data from location-aware social networks. We also have memes extracted from 170+ million news articles and blog posts from 1 million online sources over a period of one year from September 1, 2008 till August 31, 2009. In addition, there will be projects and campaigns to collect data with smartphones and pervasive sensor networks jointly with MIT's Media Lab. The Consortium also has access to (among others) mendeley.com, Google Scholar, datacite.org, Scopus, Microsoft Academic Search, Arnedminer and Sciencewise. Last but not least, FuturICT has teamed up with OpenData.ch and other international Open Data communities. And finally, QLective's www.LivingArchive.eu has just been launched to make hundreds of OpenData from various sources and archives more easily accessible, and the archive grows every week.

In the modelling arena, FuturICT will build on the previous results of a variety of simulators in various application areas of social and complexity science. This includes computational modelling environments such as MASON, MASS, SimExplorer, MatSim or Gleamviz, and Big Data analytics systems such as M-Atlas. The FuturICT experts are involved in building large-scale computational modelling, visualisation and data analysis systems in social sciences and other fields such as evolutionary biology, ecology, and weather forecasting. FA members have proven experience and deep knowledge of the required computational and statistical methods as well as of the relevant areas of complexity sciences and social science such as models of activity patterns in large cities or dynamics of world-wide epidemic spreading, to mention just a few examples. This world-leading expertise is impressively demonstrated by the lists of Science, Nature, and PNAS papers, which have been published by FuturICT consortium partners.

FuturICT's research activities can also build on previous results of a variety of simulators and EC funded projects such as Eurace (economic modelling), FOC (financial modelling), Crisis (systemic risks), Epiworks (epidemic modelling), Cyberemotions (sentiment analysis), Socionical (crisis response), and QLectives (quality-oriented collaboration), to mention just a few. The experts within Focus Area 5 (FA5) and the related work in FA6 on Exploratories have a deep knowledge of models of social coordination, cooperation, norms, conflict, crime, war, opinion formation, crowds, traffic, production, logistics, innovation, economics, and other fields. Their expertise encompasses a wide range of methods and disciplines.

Furthermore, when investigating routes to innovation, the FuturICT project will build on previous results in building platforms such ScienceWise, VIVO, and QScience: ScienceWise shows how to use a crowd sourcing model in order to obtain semantic annotation of a large body of literature; VIVO shows how to use the Linked Open Data model (RDF, OWL) to build a network of researchers that can be used for expert finding and monitoring research activities in groups and fields; QScience is a platform for experimenting with self-organised collective behavior as basis for peer collaboration. The concept of nanopublications (http://nanopub.org/) is an early vision of new models for publishing and exchanging scientific information. A crucial enabling factor for innovation and many of the above platforms is the rapid growth of Linked Open Data over the past few years, which enables the sharing of semantically annotated data in machine-processable formats, where 35 billion triples with hundreds of millions of interlinks have been created in just a few years. The team of Focus Area 7 has been closely involved in all of these earlier developments, and were in most cases Principle Investigators of the key projects in the area. Other projects that FuturICT will build upon are the Workflow4Ever platform, agent-based simulation platforms, peer-to-peer technology, semantic web data formats and platforms, data visualisation suites, and Etoile's educational technology – technology for automated marking and educational feedback.

The FuturICT consortium also represents some cutting edge expertise in the fields of ethics, IT, complexity and social science that are all highly relevant to address ethical issues related to FuturICT and information society in general. There is

a) work on generic ethical categories such as pro-social behaviour, reciprocity, altruism (Helbing et al.), but also on preferences, utility, rational choice, and welfare among computational economics and rational choice theorists (Hommes);
b) work on specific values, such as privacy, accountability, trust, security (Conte, Pedreschi, Giannotti, Domingo Ferrer, Stauffacher, van den Hoven);
c) considerable experience and knowledge about ethical and fundamental methodology on computer ethics and value sensitive design (van den Hoven, Friedman, Nissenbaum); and
d) expertise on governance and policy analysis from a complex systems angle (Mitleton-Kelly).

### c) Resources

The work in the FuturICT project will be carried out, using extensive hardware and software resources provided by a large number of supporters. In the ramp-up phase, it is important to have a large variety of different kinds of hardware and software resources to meet the specific requirements of the work packages as they get started. In the main project phase (Years 3 to 10) and with an increasing convergence, integration and standardisation of the ICT services across the Focus Areas, there will be a concentration to about 10 data centers and a few key software platforms. Furthermore, at ETH Zurich, there are negotiations underway regarding the possible use of the Cupola of the ETH main building for an interactive visualisation dome to demonstrate the power of the Exploratories led or co-led by ETH Zurich, and finally the FuturICT Platform. The choice of this symbolic place underlines the value of the project to ETH Zurich and will achieve a great visibility of the FuturICT FET flagship in the scientific and general public.

To help achieve its goal many institutions have offered FuturICT resources in terms of hardware, software, data, and labs. In addition, many business partnerships also exist, as documented in the letters of support at http://www.futurict.eu/the-project/whos-involved#Business-and- Industrial-Partners.

Structure, Alignment and Support of European and National Communities and Programs

FuturICT is a forward-looking FET flagship that addresses many priorities of the agenda "Europe 2020", including the following areas:-

- Health and social wellbeing
- Sustainable economy
- Smart energy systems
- Resilient supply and logistics
- Inclusive, innovative, and secure societies

Moreover, FuturICT is also aligned with Europe's commitment to innovation and scientific excellence, to making industry more competitive, and to improving society. Leaders of other major EU projects will advise us, either via a membership in FuturICT's Boards, or by aligning their projects with FuturICT and becoming involved at various levels. This concerns, for example, the Climate KIC (led by ETH Zurich) and also The Future Internet (which some FuturICT experts are simultaneously involved in). Besides, there is a large connectivity with hundreds of smaller-scale research projects through FuturICT's experts (see Appendices 5a and 5b for lists of EC-funded and national projects they are or were involved in). Furthermore, a number of research organisations related to existing EU projects (such as Networks of Excellence or Coordination Actions) have signed letters of support indicating their willingness to engage with FuturICT. This allows FuturICT to effectively reach out to big research communities, even beyond the 2,000 experts large pool of FuturICT supporters. The enormous number of invited keynote talks by FuturICT's core experts demonstrates the phenomenal interest of many scientific communities in the inspiring vision and goals of FuturICT.

## 2.5 FuturICT at the Heart of European Research

The fundamental approach of the flagship is to coordinate research on FuturICT themes across Europe. The FuturICT CP-CSA project will drive integration with researchers in many universities, businesses and organisations across Europe. The aim is to form a collaborative agreement with other projects, individuals and businesses, so that the FuturICT project will drive forward, and scientifically lead a whole movement in connected social systems. This will create a fleet of integrated projects on a scale that Europe has not seen before.

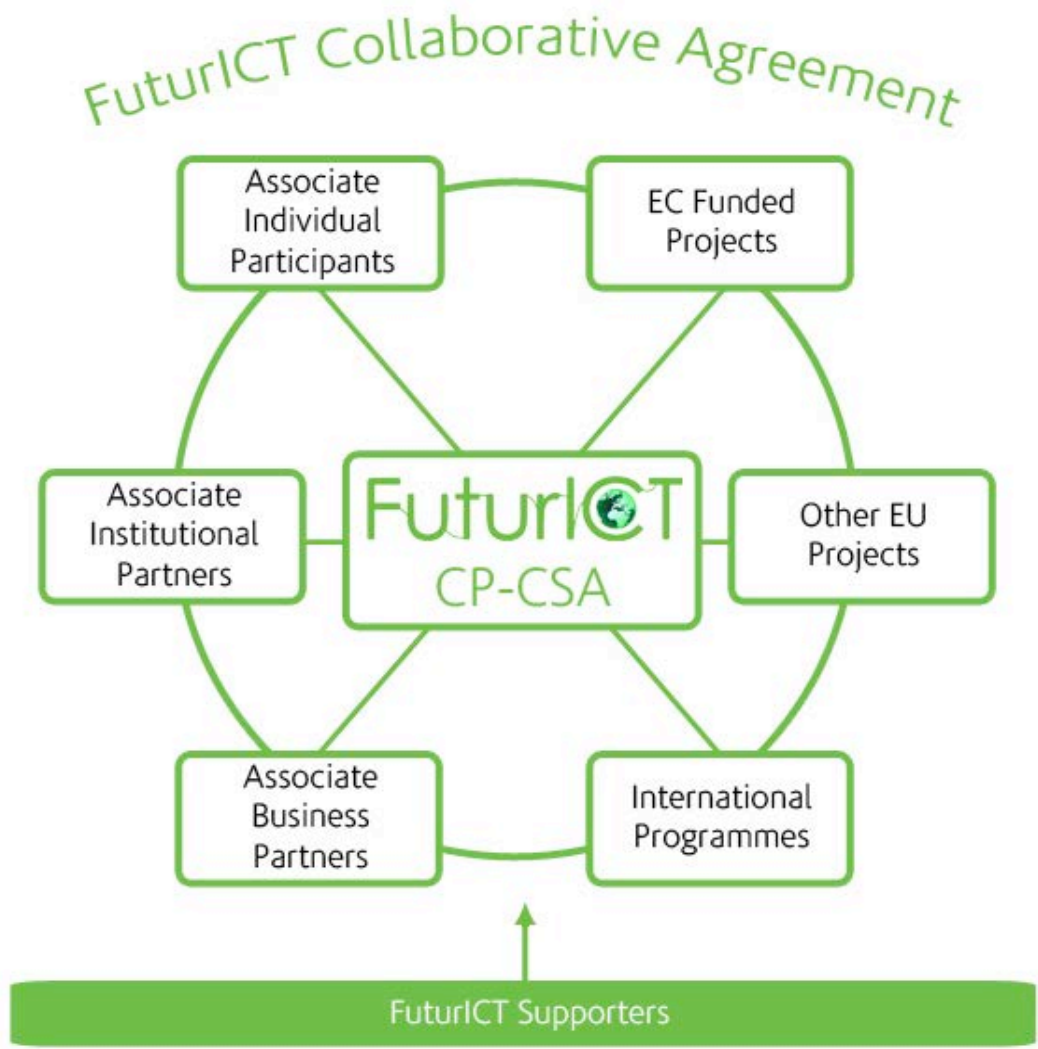

*Figure 2.5* FuturICT will create an open platform with interfaces to Associate Individual Participants, Associate Institutional Partners, Associate Business Partners, EC Funded Projects, other EU projects, and (Inter-)National Research Programs.

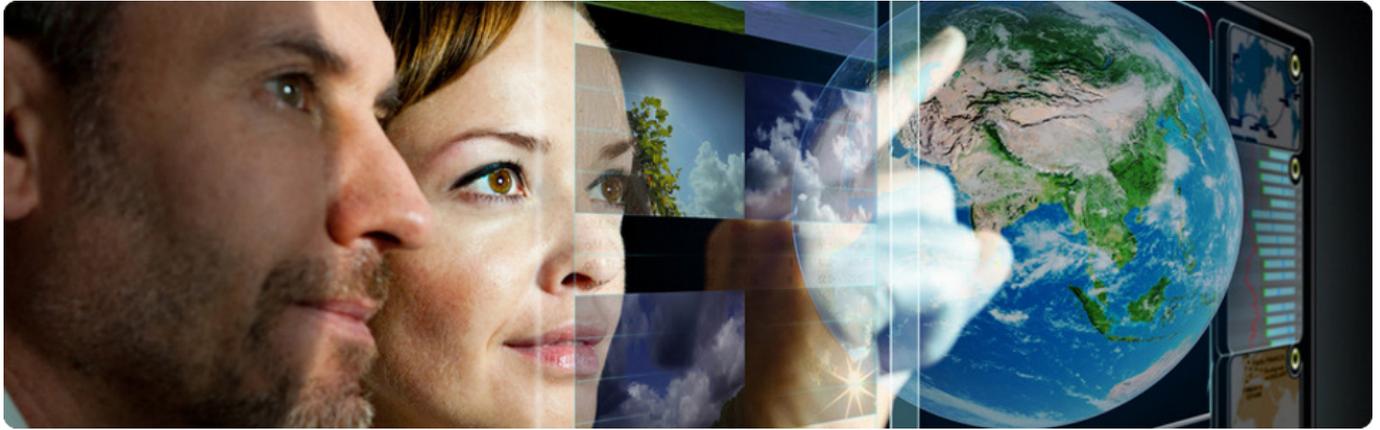

# 3. Governance & Management Structure

The objectives of the governance, management and legal structures of the FuturICT Flagship project are:

1. to provide strong scientific leadership;
2. to ensure fruitful integration of the three scientific disciplines on which FuturICT is built, and to deliver scientific excellence and the highest quality research;
3. to support the scientific endeavours of FuturICT and ensure that its goals are realised, during the CP-CSA phase initially, and onwards throughout the 10-year extended project;
4. to enable opportunities for openness, innovation and collaboration within the FuturICT community;
5. to combine basic science, applied science and innovation to deliver the most effective possible contribution to the creation of wealth and social benefit within Europe,
6. to build human capital across Europe, and
7. to do so in a way that is demonstrably ethical, democratically accountable, legal and fair.

## 3.1 FET Flagship as a Whole

The focus of the FuturICT governance structure is the strong scientific leadership of the Flagship as a whole and the successful delivery of the outputs of the CP-CSA Focus Areas during the FP7 phase. The two principal committees are the Flagship Strategy Board and the CP-CSA Executive Board, with the voting members of both committees being principally scientists. Meetings of FuturICT committees and boards will be convened, attended and minuted by members of the professionally-staffed Project Office. EU citizens, businesspeople and distinguished scientists will contribute to FuturICT through a number of advisory and representative panels. The governance structure is already in place and has been active during the pilot phase of the project; it is robust and well-practised; it includes members of the broad FuturICT community beyond the original group of pilot project partners; it will continue to remain active until the FP7 phase of the project begins, and it will provide the flexibility to enable further transition, openness and growth through the Horizon 2020 phase. The boards and committees of the FuturICT Flagship are shown in Fugure 4.1 below.

# 3. Governence & management Structure

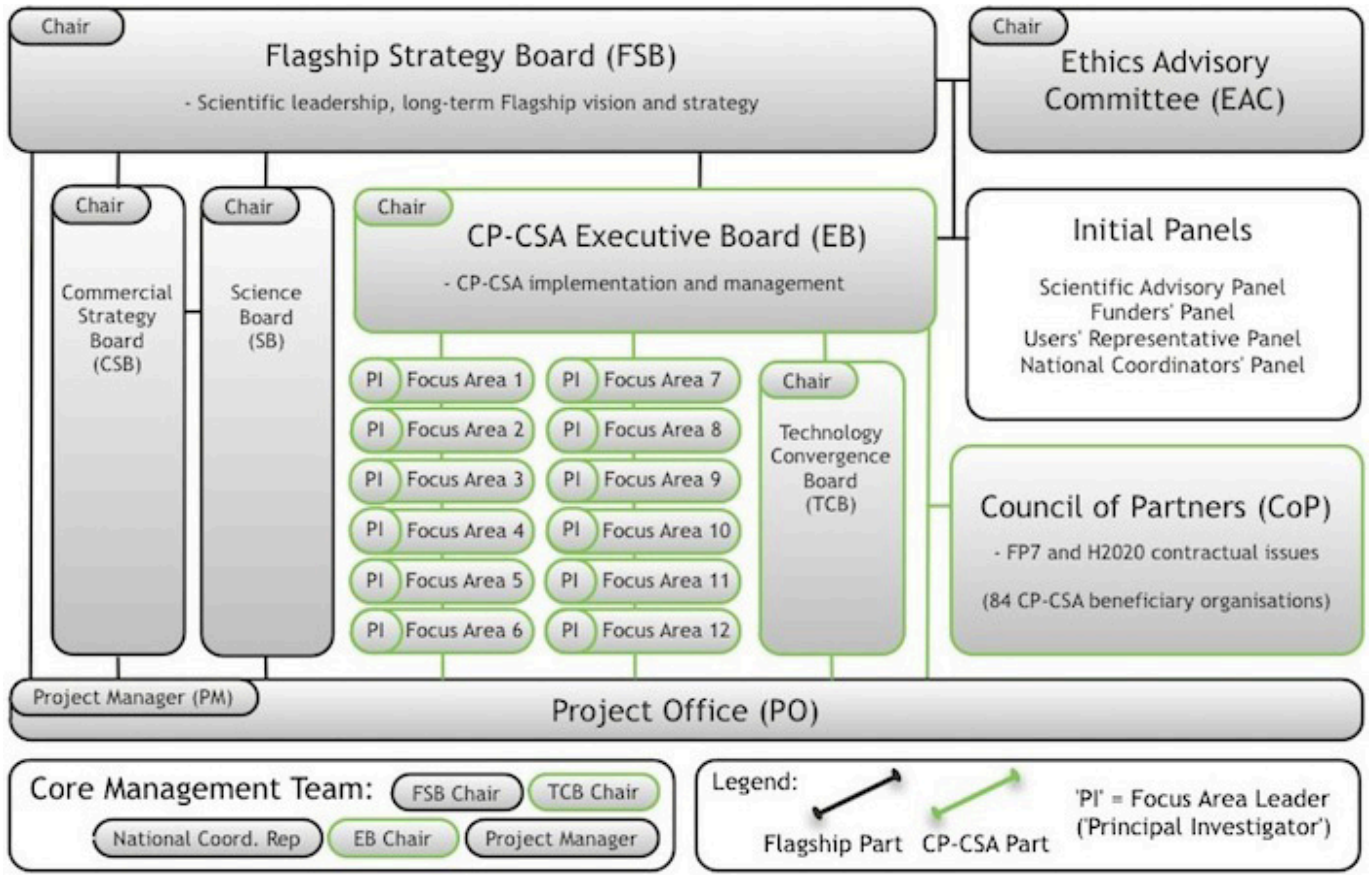

**Figure 3.1** *FuturICT Governance diagram*

The distinction between strategy and operations will be enacted by:

- a Flagship Strategy Board that sets policy throughout the entire lifetime and scope of the FuturICT Flagship,
- a CP-CSA Executive Board that oversees the implementation and operation of the CP-CSA, its Focus Areas and its Work Packages,
- a Science Board that conducts internal reviews of FuturICT operations, identifies topics and areas for Open Calls and ensures integration of the three core FuturICT scientific disciplines.
- a Council of Partners that will be democratically responsible for the core CP-CSA contract and any subsequent EU contracts, and will ensure that operations are compliant with the FP7 rules and regulations.

### FuturICT Funders' Panel

Every year at the FuturICT Week, a representative of each of the organisations (be they national, international or commercial) funding projects under the FuturICT Flagship Initiative will be invited to sit with key members of the FuturICT team to discuss the progress of the project, opportunities for further alignment of funding programmes and identify new areas of interest. The input of the Funders' Panel will be ensured via the participation of at least three members of the Flagship Strategy Board in Funders' Panel meetings, and the results of these meetings will be fed back into this panel via the circulation of minutes amongst the Flagship Strategy Board. In addition, as stated above, a European Commission representative and Coordinator of the aligned ERANET will sit as full members of the Flagship Strategy Board, thus representing a significant number of the Flagship's major funding bodies. Therefore, the widely inclusive Funders' Panel (who will be present and fully involved at the FuturICT Weeks) combined with the narrower participation of major funders in the Flagship Strategy Board, will ensure that the Flagship retains maximum transparency and accountability towards its various funders.

**FuturICT National Coordinators' Panel**

In order to develop the multi-disciplinary FuturICT community within each nation, whether academic, industrial or the general public, locally appointed FuturICT National Coordinators are already actively engaged, with a representative of each nation that wishes to be involved with FuturICT. The Coordinators are not managed centrally, and are in touch with each other at grassroots level so that joint activities and projects may be widely communicated, websites linked, and events cross-promoted across national boundaries. The National Coordinators have a very important role in strengthening and supporting the development and coordination of research proposals and projects within each national community, helping local researchers to form links within their community and suggesting collaborators elsewhere. National Coordinators and representatives from national and regional bodies will be encouraged to participate in the governance of FuturICT as non-voting members of the various boards and committees as necessary. National Coordinators provide an alternate route for local participants to raise issues that they wish to see addressed, if the hierarchical structure of Focus Areas and CP-CSA Executive Board has not been able to address the problem, since a representative of the National Coordinators will be a member of the Core Management Team. The National Coordinators already have local websites, which are linked to the www.FuturICT.eu site. These are under the control of the local group, and are used to publicise a variety of events, projects and local activities. They will also all be invited to contribute to and participate in the events associated with the FuturICT Weeks. The National Coordinators will also continue to link to business and not-for-profit organisations, such as citizen groups, and global organisations such as the UN and the International Court for Justice, based in The Hague.

## 3.2 Scientific Leadership

The first two objectives of the governance of FuturICT are to provide strong scientific leadership and to ensure the fruitful integration of the three scientific disciplines on which FuturICT is built, i.e. ICT, social sciences and complexity science. The scientific leadership will be provided by ensuring that the scientists who are delivering the project (i.e. the Focus Area and Exploratory Leaders) will have key roles in the main decision-making bodies - the Flagship Strategy Board, the Science Board and the CP-CSA Executive Board. Scientific excellence will therefore be at the heart of FuturICT, with a designated board that will drive forwards the scientific research agenda. The Science Board and the Strategy Board will be able to call upon the support and advice of a number of senior scientific figures who can advise on the scientific and business innovations that will impact FuturICT, as supporting technologies or as linked projects, extending the pool of expertise and collaboration on which the scientists of FuturICT may call.

The second key objective, of fruitful integration of the three foundation scientific disciplines, will be achieved in a number of ways. First, FuturICT is building a 'critical mass' of collaboration and networking to ensure that participants do not become 'silo-ed' in their home discipline; fragment into local or national groups or simply retreat back into their home institutions or disciplines. FuturICT has already designed and implemented FuturICT's governance structure upon the model of the FuturICT community as an active network of collaborators who may be involved in more than one Work Package or Focus Area at any time, so that the Work Packages are delivered by flexible, multi-disciplinary and varied teams of researchers, interacting and directed towards a clear goal. Each Focus Area will be the responsibility of a strong leader, collaborating with a team of scientists drawn from across Europe and involving all three of the contributing scientific disciplines. The FuturICT governance structure will also ensure that each scientific discipline is fairly represented in the decision-making arrangements for the Flagship and the CP-CSA part.

Second, the Science Board will act as a senior, cross-cutting body that will provide oversight; integrate and promote the scientific outputs of FuturICT; conduct internal reviews of operations; and manage the calls for new partners and projects. Scientific excellence will be at the heart of the FuturICT project, but with strong links to the innovation ecosystem, of business people, policy makers and the general public, through the network of Advisory Panels and the Commercial Strategy Board.

Third, building on the need for a planned and well-resourced approach to achieving the long-term scientific, and social, integration of the FuturICT team, there will be a series of intensive annual one-week meetings, the FuturICT Weeks, which will bring together the extended FuturICT community to present findings, form collaborations across disciplines and meet with sponsors and commercial representatives. These week-long events will host international special-interest conferences and workshops; hold public demonstrations and showcase events suitable members of the public; present ideas festivals to funding and investment organisations and business developers; provide short courses and exchange events for scientists from different disciplines, and hold training and networking events for particular topic interest groups within the scientific community, such as minorities, women and scientists from developing economies.

The management of the CP-CSA part is built upon the extremely successful implementation of the FuturICT pilot initiative (rated Excellent by its independent EU review), the strong collaborations developed amongst the key project partners and the vast experience brought by the key coordinating organisations (UCL and ETHZ) in managing large-scale projects funded under the Seventh Framework Programme. This includes the use of existing and proven methodologies developed by both institutions' dedicated European Offices, thus reducing organisational risk and minimising the need to devise new procedures and structures for carrying out the basic aspects of managing the FuturICT initiative. Finally, the Project Office and management structures detailed below have been based around the industry-standard PRINCE2 and Managing Successful Programmes methodologies, combined with experiences drawn from other projects (such as the ATLAS detector) which have been proven to be useful and applicable to the management of such large international initiatives as FuturICT.

The FuturICT scientific community has been established, developed and strengthened through community-building meetings that were held during the summer of 2011 in Zurich, London and Baveno. These meetings had the purpose of developing the community and clarifying and agreeing the goals of FuturICT's long-term vision. A special issue of the European Journal of Physics (Vol. 214, 2012) is also in press, with 22 multi-authored papers written by scientists from across the FuturICT community that explain the unifying vision of the project. This publication is a great achievement in integrating such as disparate community so quickly and so effectively, and all of these activities have enabled the multi-disciplinary team to form the robust and effective working relationships. These strong, collegiate, friendly relationships will enable FuturICT to confront and overcome the challenges that will inevitably arise in such a major, significant project.

**Developing and disseminating FuturICT's Vision and flexibility over the extended duration of the project**

In order to build the project vision throughout the early ramp-up phase, and into the Horizon 2020 phase and beyond, FuturICT's vision is being communicated to the broad community of scientists, funders, businesspeople and EU citizens through its website and a very active social media campaign. The FuturICT vision has been clarified through the Pilot phase, which has developed the shared vision within the community. The process of developing and communicating the vision will continue throughout the project, alongside the community-building work. The FuturICT website publishes openly the outcomes of the project so far, in terms of publications, films, videos, press contacts, scientific innovations etc. It will continue to publish committee minutes and present its findings and outputs openly. Apart from the ongoing outreach by the Project Office, the wider FuturICT community will be able to contribute to the development of FuturICT activities through forums on the website, membership of committees and participation in conferences, workshops and events. They will be kept abreast of developments through the publicity events and conferences that will take place during the FuturICT Weeks, as well as those held by local FuturICT groups, supported by the National Coordinators. As well as a program of conferences, public presentations and exhibitions, funding will be made available for students, artists, members of the public, businesspeople etc. to set up new strands of discussion, exhibition and displays. Competitions, training workshops, multi-media showcases will contribute to the building of community, communication of FuturICT project outcomes, and discussing, developing and sharing the FuturICT vision.

## 3.3 Open Calls for New FuturICT Partners and Proposals

The FuturICT project will maintain flexibility and responsiveness to new technologies and evolving research priorities throughout the lifetime of the project. This will require skilful management to maintain the project vision while adapting to the concurrent evolution in technologies and research outcomes, and maintaining the support and commitment of the various stakeholder groups, such as the funding agencies, business and commercial collaborators and the community of scientists. In order to achieve this, the Flagship Strategy

Board will from time to time issue calls for proposals. The subject and broad parameters of the call will be set by the Science Board, together with details of the acceptance criteria, timeline and selection of reviewers. These tasks will be carried out with the support of the Project Office. The management of the actual processes of publicising the call, receiving proposals, sending proposals for review, receiving reviews, calling the selection panel, issuing decision notices and negotiating contracts will be managed by the Project Office. With the consent of the European Commission, grants may also be awarded to organizations who were not original partners through Third Party agreements. This process will be managed by WP11.1, with the support of the Project Office.

The schedule for issuing calls for proposals and partners within FP7 is shown in Figure 3.5 below.

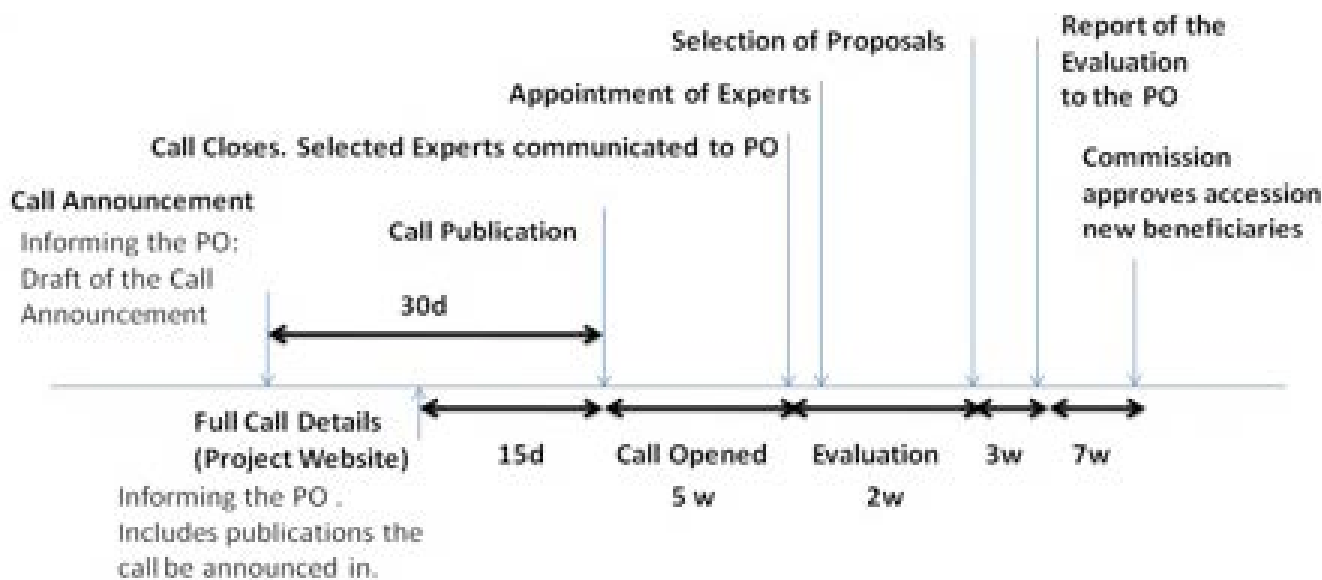

**Figure 3.3** *Estimated timeline for issuing Calls for Proposals and New Partners*

It can be seen from Figure 3.3 that the process from issuing of the call to announcement of the outcome will take no less than about 21 weeks, which is around six months allowing for national holidays. A further 7 weeks may be added to that time for approval by the Commission, with about another 3 – 6 months to make necessary appointments. This amounts to about a year from issuing the call to commencement of the funded activity. Therefore, for a new partner to be able to have a project of 12-18 months duration during the FP7 phase, the FuturICT project will therefore have to be ready to issue its Open Calls for proposals and partners within the first six months of the project. The required legal and administrative structures will also have to be prepared by the Project Office and be ready for deployment from the outset, so that the appropriate criteria and processes may be put in place.

### Additional partners

In view of the requirement to reserve a substantial part of the budget (approaching 20%) for future partners, describe the expected competences, the role of the potential participants and their integration into the running project.

A specific area of the project during the ramp-up phase has been designed for defining and putting in places the strategies for openness and further inclusion. Specifically this activity will be carried out within the Focus Area 11 of FuturICT to provide competitive and transparent mechanisms for new partners to join the FuturICT Flagship project, ensuring dynamic responses to the many challenges of such a large research enterprise, according to the general principles of the EC.

The inclusion of new partners will take place along two main lines of action aimed at:
- integrating expertise and creating opportunities for worldwide collaborations to foster excellence and improvement in research capabilities.

- creating practical opportunities for businesses and start-ups and fostering innovation in a broad circle of application areas.

**The following main paths towards integration have been envisaged to ensure enough flexibility and adaptation of the processes to the main needs of the Project:**

- Projects involving at least 2 new scientific partners that link with other FuturICT activities, or address unforeseen challenges, and are aligned with the FuturICT vision and goals, will be considered for funding subject to budgetary limits, through open competition

- Long term visits of senior or junior researchers who propose a project likely to have a demonstrable impact on FuturICT knowledge transfer and scientific collaboration. The goal is to give the opportunity for excellent researchers to join activities in existing WPs linking to important expertise worldwide and to the non-European FuturICT community.

- Research teams will be integrated into the FuturICT consortium important complementary knowledge and expertise through the instrument of the "Knowledge Integration Awards" to recognise outstanding achievements that might add value to the long-term vision of FuturICT. Along the FuturICT program development, "Hilbert Competitions" could be established to answer the most difficult questions of the research agenda. Awards would also be granted for milestones contributions to these questions from non-FuturICT partners.

**Business Links**

- Proposals from new companies that implement FuturICT activities and create opportunities for short and long-term innovation, in particular within the Focus Areas of the Exploratories PNS, LES and GPP. A competitive call for proposals will be launched during the ramp-up phase calling separately for SME, start-up and large industry partners.

- Proposals from companies or organisations that address policy-maker or regulatory challenges that are aligned with the FuturICT vision and goals.

- Proposals that address the integration/standardization challenges faced by the different areas of the FuturICT project. The accomplishment of high-level of integration and standardization should accelerate the operation of the FuturICT platform and the creation of vital interfaces towards new research and commercial environments (enabling a FuturICT ecosystem).

Competitive calls for all these instruments will be carried out in the light of the principles of EC calls: Excellence, Transparency, Fairness and Impartiality, Confidentiality, Efficiency and Speed. The calls will be adapted to the guidelines and requirements set by EC.

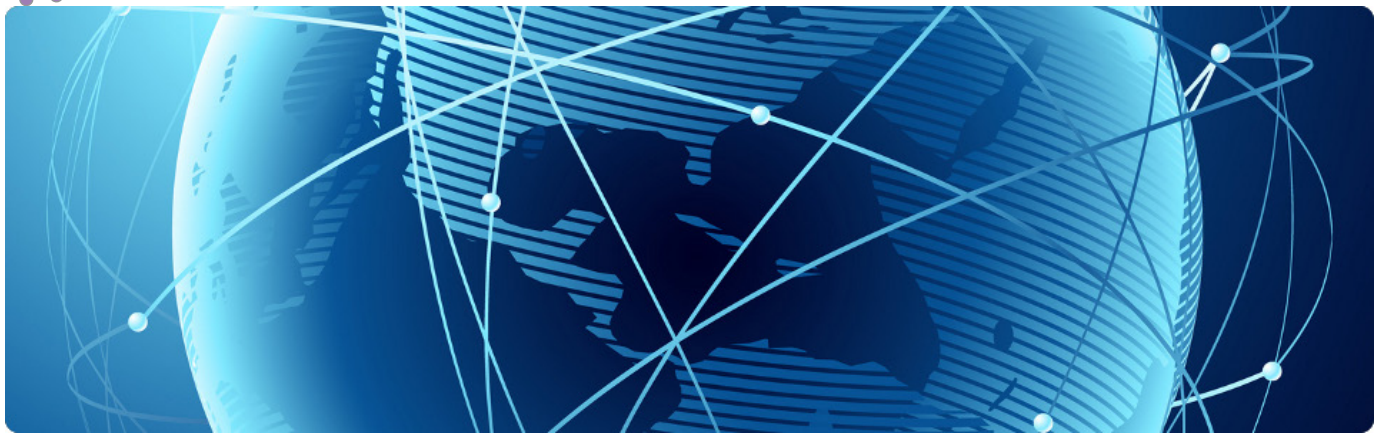

# 4. Impact

## 4.1 Expected Impact

We have built particle accelerators to observe the interaction of sub-atomic particles and understand the forces that make up our physical world. Understanding the principles underlying our strongly connected, techno-socio-economic systems, including the consequences of ever tighter and more complex interaction between human society and modern ICT systems, is a far more challenging objective. Until recently we have had no means systematically to observe the behaviour of such systems *en masse*. Today, the increasing blurring of boundaries between the digital and the physical world has opened up the capacity and at same time the need to do this. The FuturICT vision encompasses the capacity for systematic observation that forms the basis of understanding and explanation of emerging social phenomena. The FuturICT platform will enable observation, provide the tools to understand at systems level, support the design of policy interventions and have the capability to implement new forms of social organisation that achieve policy objectives.

Our techno-socio-economic systems are bringing about unprecedented social change through ubiquitous connectivity and instant, global information access. Yet we do not understand how it impacts our behaviour and the evolution of society. We are dependent on complex ICT for our critical infrastructures and, increasingly, for our daily lives. Yet we are more and more loosing the capability to design and operate large-scale ICT systems in a stable, resilient and trustworthy way. To fill the knowledge gaps and keep up with the fast pace at which our world is changing, the FuturICT flagship project will promote an interdisciplinary integration of natural, social, and engineering sciences with novel information technology paradigms to develop the capacity to explore and manage our future and build more resilient ICT systems. The approach will be based on a fundamental understanding of the institutional and interaction-based principles that make connected systems work well, but powered by opening up and participation of a wide range of knowledgeable workers that go far beyond academics, consultants and politicians. It will be driven by the increasing blurring of the boundaries between the physical and the digital world and the insight that human society and the global ICT network are increasingly becoming a single, tightly interwoven system: **the most complex and dynamic artefact our planet has ever seen.** Today we neither have the knowledge to understand nor the tools to manage this artefact. FuturICT will change that by creating a whole new field of science at the intersection of social science, complexity theory and information technology. This will produce the synergy effects required to address the fundamental societal, economic, and scientific challenges of our 21st century. It will also build the basis for a new paradigm of ICT systems.

In the ramp-up phase over the next thirty months, FuturICT will demonstrate the feasibility of its approach by developing small-scale proof of concept demonstrators of key components of the full vision, together with "best-effort" integration of components and tools.

**Outline of the range of FuturICT Impact**

FuturICT will create considerable benefits in many different domains such as:

- Science and Education:
    - Innovation accelerator
    - Education Accelerator
- Public Sector:
    - Sustainable future cities
    - Healthcare (e.g. epidemics)
    - Exploratories, risk management
- Business and Industry:
    - Financial & insurance risks
    - Managing complexity
    - Transport, traffic, logistics
    - Telecommunication Infrastructures, Supply Chain, Smart grids
- Administration and Governance:
    - Participatory governance
    - Institutional design
    - Consultancy:
    - Scenario analyses for decision-makers

The creation of the societal and economic benefits will happen mainly in three ways:

1. by mitigating systemic risks and techno-socio-economic crises, and the related collateral costs,
2. by creating an information ecosystem and open participatory platforms supporting social and economic collaborative engagement, and
3. by creating a socio-inspired ICT paradigm that will be the basis of new technologies and businesses.

**Reducing Risks and Collateral Damage**

FuturICT will use simulation models and experimental techniques to identify systemic risks, find integrative systems design approaches providing better solutions, and develop new ways of managing complexity. The following enumeration shows that societies are currently paying a high price for an insufficient understanding of socio-economic systems and the related problems to manage them well: (1) Financial crisis: Losses of 2-20 trillion $. (2) Crime and corruption: 2-5% of GDP, about 2 trillion $ annually. (3) Conflict: Global military expenditures of 1.5 trillion $ annually. (4) Terrorism: 90 billion $ lost output of the US economy as a result of 9/11 attacks. (5) Flu: A true influenza pandemic infecting 1% of the world population would cause losses of 1-2 trillion $ per annum. (6) Cyber-crime: 750 billion EUR damage each year in Europe. (7) Traffic congestion: Impact of 7-8 billion £ in the UK alone. If FuturICT would manage to reduce any of those problems just by 1%, the investments into the FuturICT FET flagship project would already pay off multiple times. For comparison: past experiences with applications of socially- and complexity-inspired approaches to traffic control show typical improvements between 10% and 30%.

**Creating Social and Economic Opportunities through Participatory Platforms**

- Prosumers: New potentials through collaborative production and co-production of consumers
- Big Data: The "oil of the 21st century" allows to create innovation and business from information Interactive multi-player online games find serious applications, as experimental platform from marketing to education

- New business models: The functionality of the PNS, LES, and GPP create new opportunities for everyone, for example consultancies of industry and politics
- A better understanding of financial systems and social capital will imply new concepts for value generation
- FuturICT's participatory platforms could create possibilities for bottom-up participation of citizens in democratic decision-making processes and more efficient democratic institutions involving work of volunteers
- The Exploratories of Future Sustainable Cities, of Smart Energy Systems, of Finance and Economics, will be essential for the transitions to a 2000 Watt society, a new energy mix, and more robust financial concepts, and to make use of the economic opportunities related with these. The FuturICT Health Exploratory will accelerate the transformational power increasing the impact of ICT in the public health system.

**Socio-inspired ICT**

Twitter and Facebook are social media that are based on simple social principles such as signalling and social networking. Both companies are worth many billions. How much money could be generated by transferring other principles of social self-organisation, self-regulation, and self-control, such as: social adaptiveness, collective awareness, conflict resolution, trust, norms, values, culture?

**Why this is the right time for the FuturICT revolution?**

Globalisation and technological change have made our world a different place. Moreover, we are currently witnessing a "perfect storm" that brings together a range of developments that have the potential for an even more dramatic, fundamental transformation of our socio-techno-economic system:

1. **Hyperconnectivity** of techno-socio-economic systems, involving billions of subjects and components. This creates new opportunities, but also systemic risks and instabilities, which urge us to come up with new, integrated systems designs.
2. **Growing systemic complexity,** enabling new opportunities for society, novel functionalities and business cases, but also requiring a new approach for understanding emerging threats. These call for a new, decentralised, participatory management approach.
3. **Cybersocial systems**, i.e. ICT systems become tightly connected with the real world, and boundaries between the real and virtual world are becoming increasingly blurred. This calls for a science to grasp and manage the co-evolution of ICT with society and technologies to ensure stability, resilience usability and trustworthiness of critical ICT infrastructure.
4. **Artificial social systems** as a result of ICT systems that are more and more socially interactive. While this can create coordination failures, breakdown of cooperation, conflict, crime, or war, one may also learn from the success principles of society and build socio-inspired technologies, showing self-organisation, adaptiveness, conflict resolution, trust and other forms of social capital and value; norms and values, culture.

The rapidly progressing interweaving of ICT and society has intensified symptoms of the need for systems and institutional renewal, such as: global financial and economic crises, political instabilities and revolutions, the rapid spreading of diseases, disruptions of international supply chains, organised crime, international conflict and increased cyber-risks. This interweaving leaves us dependent on information technology, with consequent complexity, dynamics and increasingly intractable coupling of social processes that threatens the stability, trustworthiness and reliability of socio-techno-economic systems. At the same time established methodologies for designing, managing and securing large scale, interactive ICT systems are increasingly becoming inadequate.

On the other hand, future information and communication technologies also provide the means to achieve the necessary renewal of our systems and institutions.

**Expected Impact**

Building on the age of Information and Communication technology innovations, FuturICT aims to lead Europe into the era of Big Data-driven social science, socio-inspired innovation, and socially interactive ICT. As described below this will create impact in:

- **Science** by promoting the new field of *computational social sciences*, creating the new field of *global systems science*, a new approach to computer science, and an in-depth understanding of the most complex artefact in the history of humanity created by the convergence of society and intelligent, globally networked ICT infrastructure;

- **Technology**, creating new *socially interactive and socially adaptive ICT* that self-organises according to principles inspired by society to achieve new levels of stability, resilience, ease of use, and trust-worthiness;

- **Society** by giving citizens, NGOs and politicians the means to gain a better understanding of the human-Earth system, methods for introspection of complex global processes, tools for assessing the implications of decisions beforehand, and hence to improve our capacity to sustainably manage our system on the basis of well-founded knowledge and inclusive participation;

- **Economy** by creating a whole new ecosystem of Big Data-driven business opportunities, facilitated by the Global Participatory Platform (GPP) and by helping position Europe's companies to collaborate on uncovering new local, national and global whitespace markets.

Many of the scientists who will be working on FuturICT already have experience of working in more than one discipline, so they have a rich and deep scientific understanding, combined with the ability to communicate with colleagues from different disciplines, as well as potential users and other stakeholders of scientific and technical products. The ability to tear down the barriers to interdisciplinary science and barriers which often separate the scientific community from the universe of science's beneficiaries is a great strength of FuturICT, enabling us to quickly move forward on a complex project. This strength will enable the FuturICT team to quickly mobilise the human and financial resources to achieve powerful breakthroughs during the first 30 months of the project, in the FP7 phase.

**Substantial benefits for the European Economy**

Throughout human history, ground breaking scientific and technological paradigm changes have led to great economic opportunities. This was true for the first industrial revolution as it was for the transition from mainframe computers to PCs, from singular PCs to the Internet and for the current shift to pervasive connectivity. Similarly, the FuturICT-led revolution in Big Data-driven, socio-inspired technologies bears unprecedented economic potential. This unprecedented potential arises because FuturICT will provide deep insight into social systems that can be influenced by a wide range of policy measures, incentives, and other framing measures. FuturICT will provide governments with the capacity to predictably build policies that will shape social systems to enable considerable economic benefits rooted in two areas:

- reducing societal costs of systemic problems ranging from epidemics through financial crises, crime and corruption to instability and unreliability of critical ICT infrastructures;

- facilitating a new wave of business opportunities based on the concept of innovation from information, leveraged by a novel way for collaborative, participatory creation and enrichment of Big Data.

To achieve these goals, FuturICT is liaising with governments, businesses, NGOs, SMEs, policy makers and others in the development phase as well as in the technology transfer phase. This will enable a smooth transition of its advances and tools from research into industrial deployment. The scientific and technological developments, skills and knowledge from FuturICT will be shared by the academic research community with a range of organisations and individuals who can further develop and exploit these advances, leading to new technology, products, and services. The generated intellectual property (IP) will help start new spin-off companies, increase collaborations within and among SMEs, spur lucrative licensing agreements with larger businesses, and initiate new ways in which to deploy FuturICT solutions to help implement planetary scale projects. The potential benefits are huge: reducing the impact of major societal and economic problems by only 1 percent would save the European Union billions of Euros every year. Similar to weather forecasts, it is expected that FuturICT can create value that is many times higher than the required investments.

**Economic Potential Resulting from Reducing Damage to Our Techno-Socio-Economic System**

Currently the impact of insufficiently understood health, transportation, economic and social systems on the European and global societies results in the loss of trillions of Euros and causes huge damage to health and wellbeing. For example:

- **Epidemics:** Disruptions associated with SARS led to an immediate economic loss of about 2% of East Asian regional GDP (US$20 billion) in the second quarter of 2003 (WHO). A true influenza pandemic infecting just 0.5-1% of the world population (up to 65 million people) would probably see economic losses of 1 to 2 trillion dollars per annum over a period of 2-3 years, based on current GDP data (Oxford Economic Forecasting Group, 2005).

- **Traffic congestion:** In the UK alone, eliminating existing congestion on the road network would be worth some £7-8 billion of GDP per annum. Every day 7,500 kilometres of European highways are blocked by traffic jams. Congestion on roads and at airports adds 6% to the EU's fuel bill with a corresponding rise in pollution levels.

- **Financial crises**: As the former head of the ECB, Jean-Claude Trichet, pointed out, poor understanding of the financial system contributed to the financial crash, causing trillions of dollars of losses, and damage that is still being felt in the global economy. The global economy needs better models of the socio-economic systems that show how these systems are affected at many scales of time and agent behaviour. For this, it might be necessary to model human traits such as trust, greed or herding behaviour (besides rational motivations). The goal of such models is to learn, how to avoid or diminish the impact of damaging economic situations in the future, through better, more informed analysis of real-time data to detect emerging or possible crises or crashes in financial markets, supply shortages, and social unrest.

- **Corruption:** Several rough estimates of the cost of corruption range from 2 to 5 percent of global GDP, which amounts to somewhere between $800 billion and $2 trillion in current U.S. dollars.

- **Military expenditure:** Global military expenditure annually stands at over $1.5 trillion in annual expenditure at current prices for 2009. This represents a 6% increase in real terms since 2008 and a 49% increase since 2000 (Stockholm International Peace Research Institute). Just a 10% reduction through the mitigation of conflict levels and risks would save $150 billion per year.

- **Cybercrime** is now causing an annual damage of 750 billion EUR in Europe alone, and **cyberwar** has become a worldwide concern.

FuturICT will provide policy makers with a much deeper understanding of the behaviour and interactions between these systems and will provide much better tools to develop and trial policy *in silico*. FuturICT believes this will result in radically new approaches to these systemic problems that could lead to dramatic changes in costs and benefits to society. Even a small percentage improvement in the above losses through a deeper and better systemic understanding of problems, the impact on European economy and society would already be vast and far-reaching.

**Economic Potential Resulting from New Business Opportunities**

The European economy needs new routes to growth and employment. At a policy level, FuturICT will allow policy makers to experiment with new economic models that take account of a much richer picture of the resources available in society and how they can be combined to transform modes of production of economic and social capital. For example, policy makers could consider transforming economic models to take account of the potential of the elderly to make economically significant contributions as they age. This illustrates the synergy between the policy and implementation aspects of FuturICT impact – policy makers can envisage radical transformations in the nature of work because FuturICT provides the implementation potential to realise that transformation. Enabling older workers to contribute effectively in the three to four decades of life beyond current retirement ages offers a massive improvement in the available labour in the European economy.

At an implementation level, the fundamental components of FuturICT, particularly the PNS, LES and GPP, will potentially lead to the development of nimble new businesses and start-ups based on massive open data, mobile technologies, sensors and devices to create new services and deploy them rapidly and efficiently. Delivering a Data and Modelling Commons and an Open Brokerage Platform could lead to high impact innovations in business and services in all sectors. This includes improvements in the efficiency of public services, energy consumption, local awareness and community building. Thus, besides reducing the societal costs of problems such as the above, FuturICT will also facilitate the generation of new economic value based on **innovation from information**. For example:

- **Big Data:** FuturICT will provide the multi-disciplinary and scientific platform to help refine the "oil of the 21$^{st}$ century", in other words: Big Data. The potential value of Big Data for US health care has been enumerated to be $ 300 billion per year (which is more than twice the yearly spending on health care in Spain). The potential annual value for Europe's public administration is 250 billion EUR (which is more than the gross national product of Greece). Furthermore, a surplus of $ 600 billion could be generated by a global use of personal location data. 1.5 million more data experts will be needed in the US alone.

- **Social capital:** Economic value generation would not be possible without many kinds of social capital, such as trust, solidarity, social values, norms, and culture. While absolutely crucial for social wellbeing and the functioning of society, social capital is largely invisible, and poorly understood. We know, however, that social network interactions are a crucial part of it. However, social capital may be damaged or exploited, in a similar way that the environment has been damaged and exploited. Hence, we need to learn how to value and protect social capital. The measurement of the value of social capital is also important for risk insurance. Today, no insurance covers the social capital damaged by disasters and, hence, we are taking higher risks than we should. By understanding and measuring social capital better, FuturICT aims to eliminate this blind spot and hence help bring the maintenance of social capital into policy considerations.

All these aspects will be creating spin-off business opportunities. To foster such entrepreneurial activities, FuturICT will gradually build an Innovation Accelerator platform for the exchange of ideas and experiences among inventors and innovators. Of equal importance is the fact that FuturICT will raise a large pool of experts in complexity science and Big Data, Visual Analytics, etc., to overcome the current bottlenecks in this important part of the job market.

Far more significant in economic terms is the foundational role of FuturICT technologies and approaches in building totally new business *models* that will transcend all industry and government layers. The combination of these new models and of mining Big Data will enable optimised services for the benefit of the end consumer, which could be businesses or citizens. FuturICT will accelerate the design and implementation of these new models and this will confer considerable international competitive advantage both to individual enterprises but also to the states of Europe in providing high-quality public services more efficiently than competitor nations. This effect is the major economic impact of FuturICT and such transformations in the means of production are essential if Europe is to remain internationally competitive. For example, both the Education sector and the Culture and Tourism sectors are highly significant earners of foreign currency for the European economy. FuturICT technologies will help devise radically new business models in these sectors that will see much more individualised delivery and guidance based on data analytics and prosumer-based business models. These models will be implemented on a platform that has notions of trust at its heart and will have European open governance models built in. This will allow enterprises in these sectors to market high-quality, trusted, goods and services with very attractive transparency and governance models that will confer very significant competitive advantage beyond the advantage conferred by the business model.

### Measures to bring about the Economic Impact

A core aspect of FuturICT is the vision of Exploratories that will gather, analyse and model data on various sectors, such as social challenges, health, financial systems etc.. The Exploratories will be inter-linked to eventually form the basis of the Living Earth Simulator and the Global Participatory Platform, which will enable citizens, policy-makers, businesspeople and scientists to access the most up-to-date understanding of complex techno-socio-economic systems.

The key to realising the full economic benefit of FuturICT is the forging of new links and extension of existing links between FuturICT and policy makers at European, National and Regional levels. Policy makers establish the conditions for economic transformation and FuturICT offers unprecedented support for policy making at all scales. In exploring new economic models policy makers will establish the conditions for much wider dissemination of FuturICT and related technologies across Europe.

As policy makers establish the framework that further encourages the adoption of radial new business models analytically driven business will become an area of fierce competition between enterprises. In this context FuturICT will establish a research framework that runs close collaborative projects with a range of enterprises to develop an open platform resource that the public and private sector in Europe can draw on. Enterprises will compete intensely but will be able to focus on in-house innovation where it will yield most benefit. This will create a dynamic competitive environment based on common platform elements.

The concept of democratising Big Data and models through the GPP will also be a major innovation driver. The companies that have gained most from the Internet are in business lines (e.g. social media) which nobody could have imagined when the Internet was conceived. They could emerge because the Internet is an open, participatory platform. Similarly, through the GPP FuturICT will encourage disruptive innovation in areas nobody may be thinking about today.

FuturICT's Innovation Accelerator will drastically reduce the time between scientific inventions and their commercial exploitation. And finally, FuturICT's participatory platforms will generate many opportunities for self-employment and new businesses, particularly along the lines of the prosumers' paradigm (of co-producing consumers).

Today's Data Push and Technology Push are already creating unprecedented opportunities for new business models and innovations. The advent of Big Data from social networking sites, mobile phone apps, and government sources will create novel opportunities for services, and new business and social enterprise models. The growth of new businesses that can exploit the new data resources and ICT technologies of tomorrow will be encouraged and supported by establishing bold, imaginative, far-sighted research projects and a technology transfer team. The technical and research challenges of holding, analysing and understanding this huge new data resource would be within the capability of only a few global corporations, but by putting this capability in a publicly-funded and open research project, this creates a platform for open innovation in this sphere, creating new businesses that fit evolving, innovative, European economic models and social expectations.

The correct policy environment combined with the incentives associated with new business models enabled by FuturICT technologies and other developments will drive companies to adopt FuturICT technologies as a rational choice in the competitive environment. In addition to this aspect of shaping markets the FuturICT project will actively engage with enterprises by involving business collaborators, including SMEs, in the work of FuturICT. Business people will be involved in the management and development of FuturICT through their participation in the Users' Representative Group, which advises the Strategy Board, involving regular consultation on the research outcomes and directions of FuturICT. There will be at least one business representative on the FuturICT Strategy Board. The Strategy Board will also be advised by an Innovation Advisory Board, that will seek out and recommend new ideas, directions and opportunities for realising the benefits of the FuturICT project. A professionally staffed Project Office will help with all aspects of licensing and exploitation of Intellectual Property arising from the project. The scientific leadership of the FuturICT project will be advised by a representative panel of users including businesspeople, to stay informed of the needs of business and their views on the proposed directions for research. Earnings from licensing deals, spin-out companies and royalties will be fed back to support the continuing research work of the FuturICT consortium.

FuturICT Weeks will showcase FuturICT research and development ideas and outcomes to business, including SMEs. Focus Area leaders with experience in setting up their own businesses and policy-makers will be encouraged to suggest new contacts and advise on the presentation of ideas and development of business plans. Small companies will be encouraged to work with young researchers who are interested in bringing new business ideas rapidly to market.

## Social impact of FuturICT

We are in the midst radical social change brought about by the capacity of modern ICT to transform social interaction and the basis for social action. Effective use of these new forms of action and communication with the Internet at their core are credited with bringing about the Obama presidency – a major transformation in 21$^{st}$ century politics. More generally, the capacity of the modern internet to capture and shape individual action into collective action (e.g. Google, Facebook, and others) is in its infancy and we can expect much more radial uses of these technologies to emerge over the next decades.

In this context, the goal of FuturICT is to provide the tools enable policy makers to adapt the top-down social framework in an evidence-based manner and to empower citizens to understand and control their engagement in a rapidly evolving social structure. The social impact of FuturICT is rooted in the core project contributions:

1. Developing new science that will put the processes of identifying, reacting to, and dealing with complex socio-economic problems on a scientific basis. FuturICT will replace former piecemeal approaches with bold and integrated approaches that are adequate to the dynamic, complex and global world we inhabit.

2. The development of a visionary, information platform, integrating insights from social sciences, complexity theory and computer scientists. This system will be able to act as a *'Policy Simulator'* or *'Policy Wind Tunnel'*, allowing everyone: from policymakers to average citizens, to test multiple options about a complex and uncertain world, producing possible outcomes. The platform will analyse data on a massive scale and leverage them with scientific knowledge, thereby giving politicians, businesses, decision-makers and citizens a better understanding upon which to base policy. In the long run, this will enable us to collectively explore the possible or likely consequences – scenarios that would be impossible to otherwise imagine – effectively helping us to see just a little around the corner into the many possible futures that face us.

3. The transition towards socially interactive, participatory, transparent, and democratised ICT that will facilitate new forms of political and economic inclusiveness, information sharing and social dialogue. This will also facilitate the implementation of new business models for social enterprises

FuturICT opens a route to leverage a combination of a new global systems science and the power of next generation socially interactive ICT technology to take policy initiatives that tackle problems and open up new opportunities. These policies can be simulated and tested in the wind tunnel and will be designed to exploit the capacity of FuturICT to enable informed, bottom-up citizen action. This approach will be demonstrable during the ramp-up phase in the first 30 months of the project and can scale up in the later phases of the Flagship.

FuturICT turn the Big Data available at the interface of the digital and the physical world into a European data commons and provide models to interpret and predict from that data. The FuturICT data commons will become central to policy making across Europe helping both **policy makers and citizens** (who will use the data commons through the GPP) (1) understand the dynamics of funded innovation networks (e.g., what are the key components of a successful innovation network?), (2) diagnose problems in existing networks (e.g., is there something missing in European biotechnology innovation networks?) and (3) predict the outcomes of policy decisions (e.g., if the EU increases support for researcher mobility between industry and academia, what impact will result?).

There is a great sensitivity in many cultures to the possibility of wide-reaching government control, and the creation of powerful, exclusive, tools for governments to analyse Big Data might aggravate such fears. That is why participation, openness, trust and transparency have been designed into FuturICT from the start, with a clear focus on empowering democratic participation through future ICT systems. Rather than concentrating power from knowledge, FuturICT aims to distribute knowledge-power in a fair way. The project will open up ICT systems in such a way that everyone has effective access, and a plurality of analyses, models, projections and views can coexist. FuturICT has put in place a transparent, open, flexible and ethical governance procedure in order to ensure that the research directions and outcomes are available to all; are open to scrutiny; are based on ethical principles and reviews; and will lead to the maximum possible social and economic benefits.

To emphasise how FuturICT will benefit a broad cross section of the society and avoid becoming a "Big Brother" tool to govern people, the discussion of the social impact below is explicitly performed from the perspective of decision makers as well as from the perspective of citizens.

**Relevance of the FuturICT flagship for politicians**

Politicians are the democratic link between citizens and state power. As the modern world becomes more complex, the politicians' task becomes increasingly difficult because the consequences of apparently simple, unrelated, decisions can have serious negative consequences. One of the key goals of FuturICT is the development of tools whose potential use includes increasing the effectiveness of politicians in this context. Politicians will be able to make use of tools to evaluate the consequences of the actions of the administration and draw on simulations and visualisations to communicate effectively with their electors. Citizen engagement will engage in better informed debate than ever before and this will change our democratic process in ways we have yet fully to comprehend. This could eventually lead to new political and economic institutions achieving profound impacts on democracy, national governance and prosperity. A leap of imagination is needed to achieve the breakthroughs that will improve our understanding of the world's economic system.

**The FuturICT project and its relevance for EU citizens**

What are the relevant problems for citizens?

Globalisation and technological change have made our world a complex system, which is hard to understand and manage (but not impossible). Although this complexity has created many benefits, this change has also caused strong systemic interdependencies and some serious challenges, such as financial uncertainty, political instability, rapidly spreading pandemics, and insecurities on energy, water and food supplies. Our transportation, communication and financial networks are now generating huge amounts of data, much of which is available from government and public sources. By putting this data together with new models of how communication networks and social systems actually work, we can get a much better understanding of the complex behaviours that exist around us.

Why is it important for citizens that they are addressed?

Conventional models of how economic and political systems work do not consider the interplay of non-linear dynamics, network interactions, heterogeneity and randomness well enough, which may lead to cascading effects such as large-scale financial or economic crises. Consequently, current regulatory regimes may not be sufficient to prevent these problems from occurring. We need an economy that can satisfy the needs of our population, which requires innovations that allow new and old businesses to create wealth and social wellbeing. Politicians, regulators and businesspeople must be equipped with the best possible models and tools to help them understand and make decisions about the interventions, laws and regulations that help to sustain a democratically accountable and effective economy.

How will we address them? Why is the approach worth commitment from citizens and tax-payers?

The FuturICT project will bring together hundreds of scientists, businesspeople, citizens and policy-makers to produce an open and interactive data and modelling commons, i.e. a new public good that will advance our understanding of today's complex world and create a participatory information ecosystem facilitating new collaborative and business opportunities of all kinds. The interactive platform developed by FuturICT will also use wisdom-of-crowds principles beyond Wikipedia and prediction markets.

New **Exploratories** will be created to gather data and to model the forces and behaviours that drive our world. The first tranche of Exploratories will cross disciplinary domains, but be focused on (1) sustainable financial systems, (2) health protection from epidemics, (3) measuring and mitigating social challenges, (4) measuring systemic risks and increasing systemic resilience, (5) finding principles for sustainable future cities, (6) developing smart energy systems, and (7) reaching sustainable environmental systems. Further Exploratories will be added after the ramp-up phase. The Exploratories will develop useful new visual methods to help us understand the consequences of decisions and political interventions. The Exploratories will be linked across Europe to create connected Exploratories, with many elements crossing the four areas of Environment, Economy, Society and Technology. They will bring together data, models, and interactivity in concrete application areas.

**European Leadership in Key Areas**

FuturICT is an essentially European project made possible by its unique collaborative network of excellence that has a history of cutting across disciplines and will give European society, science and business particular advantages in the medium-term. It leverages collaborations across the world under European leadership. FuturICT is unique in the breadth, depth and quality of research directed towards understanding the way complex social systems work, how ICT is changing those systems, and how to build the necessary technical infrastructure to support work at the intersection between ICT, Complexity Science and Social Sciences.

For many years, Europe has had strong research traditions and communities in the areas of complexity science and computational social science. In particular, Europe has unique strengths in agent-based modelling of socio-economic systems, which go beyond the scope of purely qualitative or purely analytical models. FuturICT builds on these achievements, and creates a further step change by combining these theoretical approaches with the mining of Big Data, future supercomputing power, and entirely new opportunities of ICT systems to come. In particular, this has been furthered by four Integrated Projects on techno-social systems funded by the European Commission, namely QLectives, Epiworks, Cyberemotions, and Socionical. The Principal Investigators and further leading researchers of these four projects are on board of FuturICT, as are prominent representatives of the Pervasive Adaptation (PerAda) community for example.

Europe is a centre for world-leading research on pedestrian crowds, traffic, global epidemic tracking, and network theory. The vision of FuturICT is the work of the researchers carrying out this world-leading research. It is the flagship project that targets techno-socio-economic challenges in the most direct way and puts humans in the focus of the project. The social-collective and people-centric perspectives permeate the FuturICT proposal.

This new, social and value-oriented ICT approach differs from the approach of the main ICT competitors and is expected to deliver tailored solutions for Europe's Innovation Union, as well as interesting products and services for the world market.

FuturICT has impressed people all over the world from the very beginning. It was the project featured on the title page of the Scientific American on 10 World-Changing Ideas and by the National Chinese TV (with 1.2 billion spectators). Many people got so excited about the perspectives of the FuturICT flagship project, and its open, participatory approach, that they even offered to engage for the project for free. The USA, China, Japan and Australia are now preparing similar initiatives with both, competitive and cooperative ambitions.

It is worth mentioning that FuturICT has already had societal impact beyond the academic sphere. For example, it has sparked off activities in the areas of arts, music, ethics, business, and politics. Many new ICT calls now have a socio-oriented focus and many social science calls now ask to engage with ICT. FuturICT has become common knowledge, and people with all sorts of backgrounds are talking about the project world wide. European leadership is clearly recognisable. The project ideas will also be featured in a book edited by the world's leading agent of science books, New York-based John Brockman. The influential website http://edge.org has recently created a feature on Computational Social Science (http://edge.org/event/special/-computational-social-science), which opened with a video interview with Dirk Helbing and followed up by other famous scientists such as Craig Venter, Nicholas Christakis, Laszlo Barabasi, and Sandy Pentland. The work of Helbing and Barabasi has also been featured in TV series such as Numb3rs and inspired movies beyond the scope of science and technology channels. Finally, the movies on the GLEAMVIS and FOC projects are noteworthy as well. They can be found in the FuturICT Vimeo channel, see http://vimeo.com/futurict.

**Contributions to the "Europe 2020" Agenda**

FuturICT has the potential to become **the** essential tool to achieve Europe's 2020 priorities and targets.

At a **policy level** it is recognised that sensitivity to sectoral, national, and regional variability is essential to policy success. In Europe, achieving "smart, sustainable and inclusive growth" requires policy tools that are capable of making fine policy distinctions across Europe in order to maximise the growth potential inherent in our diverse, multicultural society. FuturICT will provide the infrastructure to develop and implement a coordinated European Economic Policy that maximises the strengths inherent in that diversity while ensuring open, democratic and trusted participation in policy formation across Europe.

At the **implementation level** where FuturICT will see individual organisations adopting "big data" and modelling approaches to transform their relations with governments, other organisations and customers FuturICT offers an open, trusted, scalable platform that will enable EU companies and organisation to exploit: radical data transparency; the impact of "big data" on competitive strategy, the capacity to offer massive, real-time customisation of goods and services; transformations in management strategies; and the development of new business models. FuturICT will have the capacity to deliver this capacity within a strong, open and transparent EU governance framework that ensures ethical acceptability and privacy preservation.

The policy and implementation levels of Europe 2020 priorities sit within the broader need for a system of **economic governance** that can manage the complexity of the European economy at a policy and regulation level and provide efficient oversight of economic activity across Europe. FuturICT aims to deliver both at a policy level via the pooling of Europe's best economic thought in the Exploratories combined with the capacity of FuturICT tools to monitor economic and social activity in unprecedented detail.

To hit the **employment** targets requires innovation across the spectrum of economic activity involving even very well established sectors. Big data techniques combined with explanatory models have the potential to transform business models across all sectors, including public services. Radical transformations in business models have the potential to develop new areas of demand and bring about sustained employment growth.

Increasingly **R&D** will target complex, multi-disciplinary, problems arising from societal challenges. The vision of the FuturICT Innovation Accelerator is precisely to transform scientific practice to enable it to tackle the challenges of the 21$^{st}$ century.

**Climate Change and Energy** targets will be informed by the FuturICT Exploratories at a policy level but in terms of implementation, FuturICT will provide the platform both for business and social innovation directed towards engaging in "bottom up" lifestyle changes based on open access to better data and models of sustainability that will improve individual and collective decision making on climate and energy.

FuturICT will provide new resources to support **Education** directed to developing citizens with a better understanding of our complex interconnected society. FuturICT will also provide an infrastructure for much more effectively targeted delivery of education this will sustain and develop European Education in the worldwide education market enabling a sustained development of this key sector in the European Economy.

FuturICT will enable new approaches to tackling **Poverty and Social Inclusion** both by contributing to well-informed, evidence-based, debate on the roots of poverty and exclusion and more practically in supporting the evolution of new forms of work organisation (e.g. co-production and co-design) in sectors as diverse as healthcare, creative industries and public service delivery.

# 4.2 Dissemination

**Introduction to dissemination strategy**

FuturICT considers dissemination to be an integral part of its mission and as being pivotal to its overall success. An inclusive and participatory approach has been a hallmark of the FuturICT Flagship from its very inception. A broad range of potential stakeholders have been contacted very early on in the process, and have already rallied behind its long-term roadmap. Yet, dissemination towards society at large as well as the stakeholders will be a longstanding effort throughout and after the project's lifetime. This should result in a widespread awareness of and societal participation in what FuturICT tries to achieve. These goals are indeed worthwhile communicating, as FuturICT is well positioned to support the future creation and sustainability of jobs, growth, and competitiveness targeted by the EC and the different national governments. The project will help propel Europe to the forefront of the next wave of ICT technology and the gathering and analysis of the data by which it will be driven.

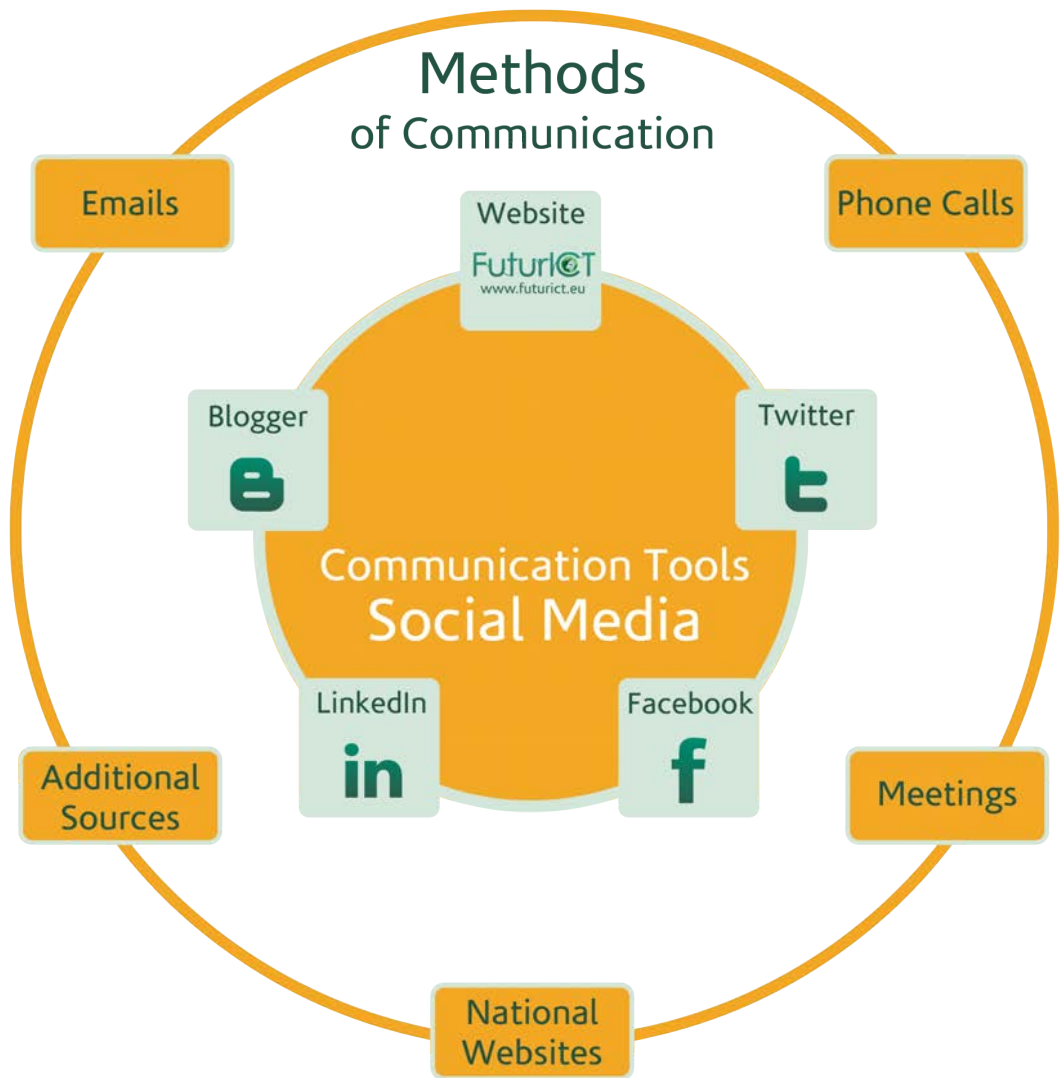

**Figure 4.2.a** *Methods of Communication*

On the one hand, FuturICT will employ a gamut of existing and emerging media to disseminate its work, like the www.futurict.eu web site as well as national level web sites, Facebook, Vimeo and YouTube video channels, a dedicated LinkedIn group, academic conferences, scientific papers, and business and public events. In addition, FuturICT will employ its own Innovation Accelerator and www.livingscience.eu to complement the more familiar dissemination routes. A comprehensive range of dissemination channels and events have been mapped out to target scientists, industry, policy-makers, media, or the general public.

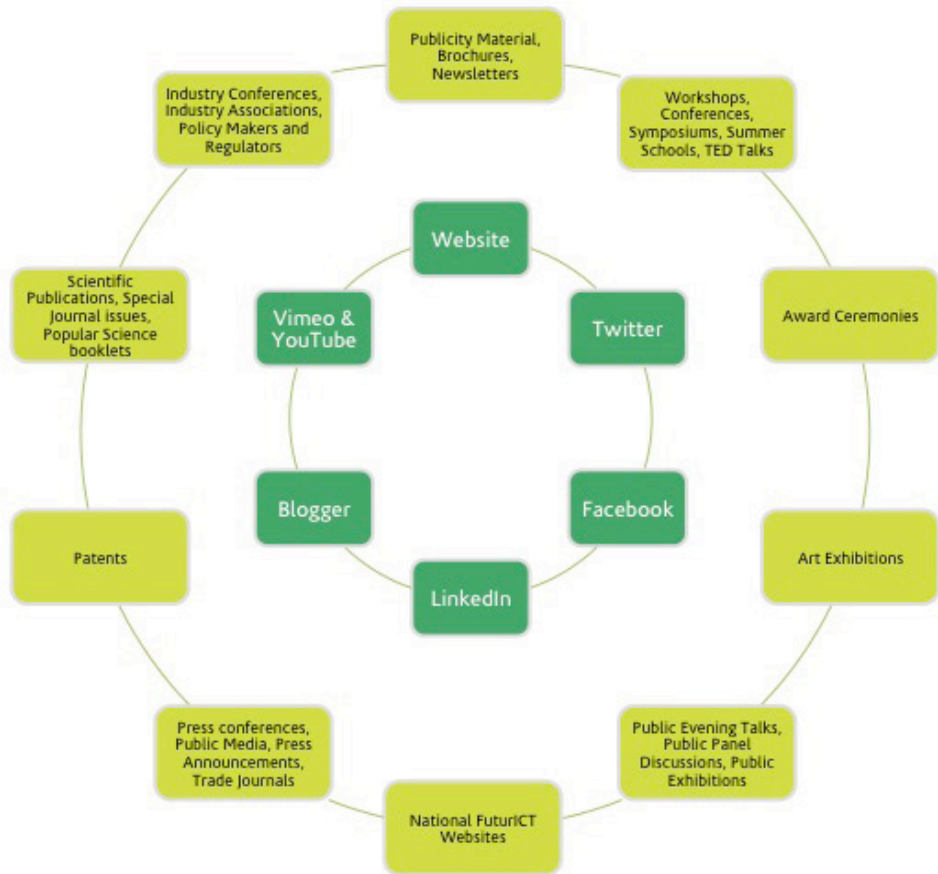

**Figure 4.2.b** *FuturICT has a comprehensive and inclusive dissemination strategy that ensures opportunities for participation of the general public, access to results for businesses, provision of insight to decision makers, and high quality scientific publishing. To achieve such a widespread dissemination to heterogeneous groups a variety of channels, including tradition media, social networks, the WWW, software and data sharing, and scientific publications, are used.*

These events will be supported by top class publicity material hosted on the FuturICT web site (www.futurict.eu) as well as on more than 20 country level web sites and a specialized FuturICT exhibition stand. The material is designed to be accessible and understandable by different target groups and will use different media, including written material, videos, and audio podcasts. Social media channels such as Facebook, LinkedIn and blogs will help add more personal and business dimensions as well as stimulate further interest from relevant groups. The dissemination strategy has been designed to connect with stakeholders in the medium that works best for them. This engagement in a manner that is native and comfortable has meant FuturICT has been able to connect with a very broad range of stakeholders e.g. business people might prefer researchers giving presentations and communication via LinkedIn, some younger people might prefer Facebook while Twitter is favoured by others for its brevity with most people happy to look at web sites intermittently but not every day or even every week. As can be seen from some of our metrics below FuturICT has been very successful in leveraging all these channels and will continue to build on this approach.

## Scientific Dissemination

FuturICT will use all academic publications routes such as peer-reviewed journal and conference publications, special journal issues, books and book series; conference proceedings; and presentations, ranging from invited keynotes at major events to presentations at focused workshops. In particular, the annual FuturICT Week organisers will arrange a number of international inter-disciplinary events each year, including events and demonstrations for business people, journalists, policy-makers and members of the public. These events will be hands-on and interactive, and will be supported by training in the Public Understanding of Science in order to ensure a high quality experience for citizens. FuturICT supports the "Open Access" approach and will endeavour to store all FuturICT derived publications in online publicly accessible repositories.

The FuturICT community is very active in publishing in the best disciplinary and interdisciplinary journals as well as in high impact computer science conferences. This is documented, for example, by the impressive number of more than 50 Nature, Science, and PNAS papers over the past few years. Moreover, new journals such as EPJ Data Science have recently been established which will promote the science FuturICT will be doing along with interdisciplinary journals in complexity science, such as Advances in Complex Systems. FuturICT has furthermore invested in good documentation of the scientific vision and concept behind FuturICT. It is documented by several White Papers of the European Support Action VISIONEER and a special issue with position papers on all major FuturICT activities, with more than 20 contributions, published by EPJ Special Topics. FuturICT's Innovation Accelerator is also planning to establish a new publication and scientific collaboration platform. These initiatives along with FuturICT's liaison of science with arts and the plan to come up with participatory platforms and eventually even a Grammy of Science demonstrate that FuturICT is determined to use all communication channels to maximise information flows and interactive exchange within the scientific community, various stakeholders and with the general public.

**Dissemination to Policy-Makers**

Policy-makers and regulators will be key groups to focus dissemination activities on. FuturICT will enlist the help of supportive policy-makers and regulators in order to further develop use cases, test conditions and success criteria. Those policy-makers and regulators that are interested and capable of engaging in research will be connected to the relevant FuturICT Exploratories. The addition of policy-makers and regulators to the exploratory level industry review and collaboration meeting will ensure a robust exchange of views and approaches for applying FuturICT technology to real world scenarios. The engagement with policy makers and regulators will be further supported by including a section within FA11 Open Call that specifically applies to challenges in these areas. Policy-makers experience two overriding challenges:

- New fields of knowledge keep erupting in their areas of responsibility;
- They have no time to learn.

Policy-makers know they need direct access to knowledgeable people rather than the colourless summaries the bureaucratic briefing machines produce but they're severely time poor. Creating a rich interactive environment, in which they can actively learn with whatever time they can invest is key. This becomes more important if the issue cross-cuts several departmental silos or policy briefs (as so many are these days). So an interactive environment that accelerates learning is vital.

**Dissemination to Interest Groups**

NGOs, industry associations and technology interest groups will be catered for by including them in the exploratory level industry review and collaboration meetings. These meetings will be scheduled at the Focus Area or Exploratory level and could be held on the shoulder of an academic or industry conference. Focus Area 9 and Focus Area 10, which looks after both technological and systems integration as well scientific coordination and other coordination and dissemination measures will facilitate these review and collaboration meetings. Those interest groups that are involved in standard setting initiatives or research will be invited to join the relevant Focus Area or Work Package within FuturICT. Speakers from FuturICT will be made available for presentation of relevant FuturICT Focus Areas at meetings and conferences of relevant interest groups. Two large annual business meetings in London and Zurich are scheduled with national events scheduled within each country. Continuous meeting with business people are scheduled as needed throughout the year.

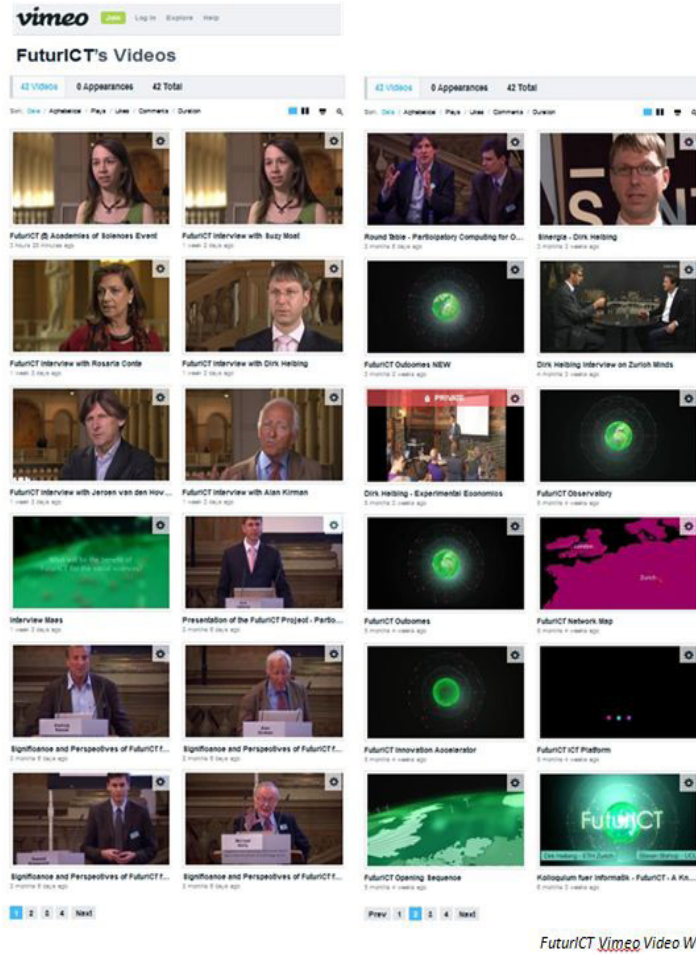

*Figure 4.2.c* FuturICT has released over 50 videos outlining the vision and discussing concrete steps for the future. These videos have been already now seen be more than 18.000 people.

**Dissemination to the Media and the general Public**

FuturICT has its own logo and brand and has set up an innovative website with an exceptionally effective social media strategy. Some of the metrics (see complete table below) that highlight the reach of the project are that it has received 136,156 page views from 1 October 2011 to 1 Oct 2012, representing 48,887 visits by 26,652 unique visitors. Besides the main FuturICT web site www.futurict.eu, there are 26 active websites of the FuturICT communities in Austria, Belgium, Bulgaria, China, Croatia, Czech Republic, France, Germany, Greece, Hungary, Ireland, Israel, Italy, Japan, Latvia, Lithuania, Luxembourg, Netherlands, Poland, Portugal, Romania, Spain, Switzerland, Turkey, and United Kingdom. FuturICT has more than 1,500 registered supporters on its main web site with most of these signed up to receive the monthly FuturICT newsletter. FuturICT has the leading web and social media metrics of all the flagship projects with the key FuturICT statistics displayed in the table below:

| FuturICT Web Site Dissemination Metrics (from 1st of Oct 2011 to 1st Oct 2012) | |
|---|---|
| No. of page views | 136,156 |
| No. of visits | 48,887 |
| No. of unique visits | 26,652 |
| **FuturICT Social Media Channel Metrics (since start date)** | |
| No. of Facebook Likes | 617 |
| No. of Twitter Followers | 767 |
| No. of LinkedIn Followers | 297 |
| No. of FuturICT related Videos (Vimeo) | 50 |
| No. of video views | 18,268 |

*Figure 4.2.d* The table above summaries the web and social media reach of FuturICT and shows that there is already now a great basis for dissemination.

The aim is to grow the central FuturICT web site (see www.FuturICT.eu) while allowing for more integration with the web sites located in each country. This will help stimulate more content for the central web site but also allow more two-way visibility of progress – both at the centre of the project but also at the country partner level. The web site and social media channels will help advertise and stimulate even broader and deeper engagement, both within academia as well as across all sections of industry.

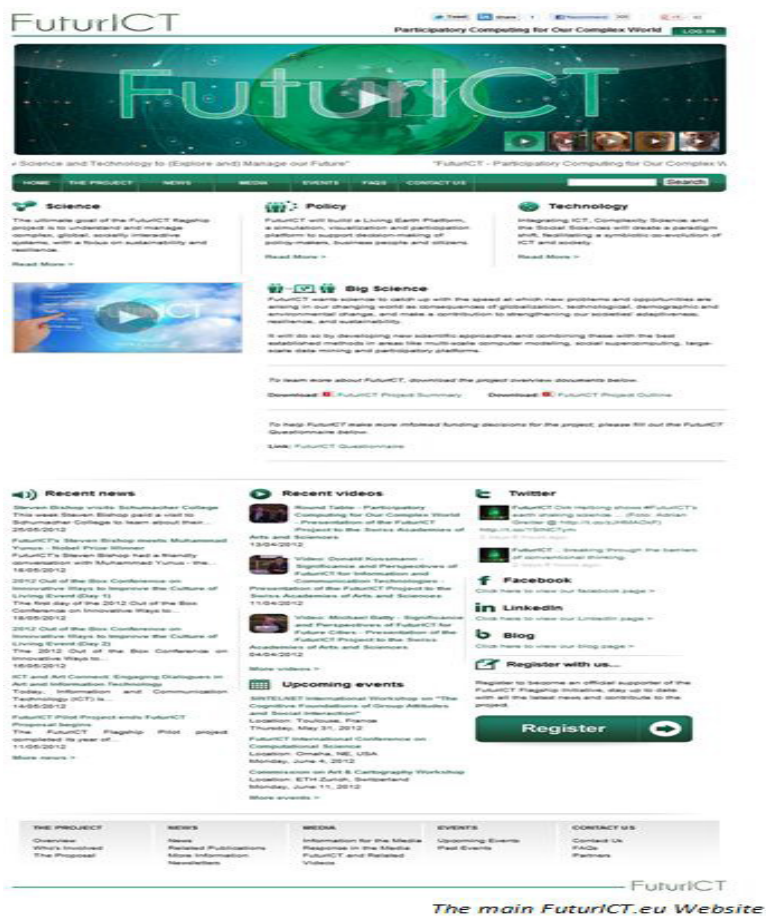

*Figure 4.2.e* The home page of FuturICT is the central starting point for people looking for information on the project and will become the hub for the digital dissemination activities in the project. The aim is to make it as easy as possible to gain access to material provided by FuturICT and offer means to provide feedback and to get involved.

# 4. Impact

FuturICT has very active communities of followers on LinkedIn, Twitter and Facebook. These have delivered a high level of awareness of the work of FuturICT, with updates stimulating hundreds of "likes" on Facebook and nearly 800 "followers" on Twitter made up mostly of people not involved directly in the project. Since many of them are not scientists, this documents a large interest by industry and the general public. In addition, FuturICT has approximately 300 LinkedIn members, who again are a mix of scientists, business and other sectors of society. The FuturICT blog was recently launched and continues to build a strong pool of followers. FuturICT will continue to devote significant resources to social media, as an effective means of promoting awareness of FuturICT's work to our stakeholder communities of interest. The project plans to invest in a roving video wall to support high tech dissemination activities at FuturICT exhibitions that occur during the next phase of the project. Existing content will be re-engineered to work on the video wall displays and exciting new content will be developed along with demonstrations from the research work packages.

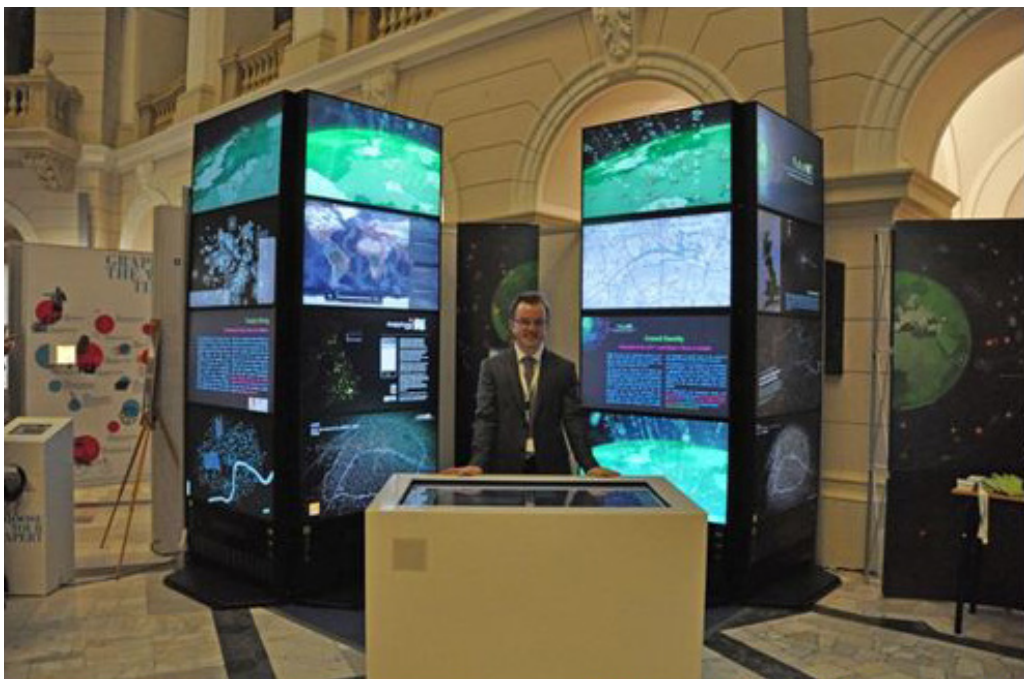

***Figure 4.2.f*** *Tobias Franke pictured with 16 high resolution screens FuturICT created a capturing presence at the Mid Term Meeting in Warsaw.*

Furthermore, FuturICT has been featured by hundreds of newspaper articles, radio and TV contributions in all areas and all major languages of the world (including the Chinese National TV with 1.2 billion spectators). This has made a substantial contribution to the public discussion of and awareness for the huge potentials and risks of Big Data and new Information and Communication Technologies. Scientific American presented the project prominently on the title page of its issue on 10 World Changing Ideas. The USA and China are preparing competitive research programs so it is important that Europe does not get left behind.

## 4.3 Gaining Commercial Value

In FuturICT one of the core aims is in fact not to exploit people but to give them a choice as to whether to join in and donate data or opt out. Therefore although the term exploitation generally means making money from your findings, the preferred route in FuturICT is to talk about gaining commercial benefit, particularly *without* exploiting any one, only the technology.

### Management of IP

FuturICT has the potential to dramatically change the way things are done. It proposes an open source strategy to allow people to enter data and models as a common goods strategy.

### Management of Intellectual Property

FuturICT will employ an innovative yet robust approach to Intellectual Property (IP) management in the Flagship initiative, through a collaborative and strategy-driven approach. The implementation of this strategy will initially be led by the core beneficiaries of the CP-CSA, but with the goal of maximising Intellectual Property exploitation and/or dissemination across the FET Flagship as a whole - by providing clear strategies, understandable frameworks as well as more innovative solutions. The three different phases of implementation will each be formalised: firstly though the terms of the CP-CSA Consortium Agreement, secondly into the terms of the wider Flagship Collaboration Agreement, and finally in a dedicated legal entity that will provide long-term sustainability and ensure long-term impact for the IP generated within the FuturICT.

The FuturICT Flagship also aims at reinforcing the innovation capacity of the European Research Area (ERA) as a whole, creating an entrepreneurship-friendly innovation ecosystem that will encourage the creation of spin-offs and start-ups, but also improving the IPR discovery process and promote wide industrial and SME collaboration. A suitably structured open innovation environment will help reach this goal, strengthen research opportunities and speed up translation to commercial applications in different business contexts. Finally, FuturICT will remain dedicated to the goal of providing open access to publicly funded research publications, and this will be reflected in all of the various contracts, agreements and strategy documents used throughout the Flagship.

The Intellectual Property management strategy for the FET Flagship as a whole will be implemented in four phases:

1. **IP Management Framework**: The first phase will focus on laying the necessary foundations for addressing IP issues within the CP-CSA part (via the project's Consortium Agreement) and the establishment of a flexible, multi-layered approach to managing IP across the Flagship as a whole, led initially by the core beneficiaries of the CP-CSA. This multi-layered approach will provide an immediate workable framework whilst the wider strategies, policies and innovations are developed, and will be underpinned by a series of pre-existing toolkits and document templates that are already in wide use across the sector (see the section on Structure and Toolkits below).

2. **Strategy and Policy Development**: The second phase will focus on developing innovative and effective policies for Intellectual Property management and the appropriate exploitation and/or dissemination of Foreground across the wider Flagship. These policies will be devised by the Commercial Strategy Board and will be developed by the IP & Exploitation Team, with input from the Science Board and subject matter experts across the project. These policies will then be applied to the terms of the Flagship's Collaboration Agreement and disseminated within the project to all relevant parties in a clear and concise format. This phase will also involve interfaces with partners in the Innovation Accelerator Focus Area working on extending the 'VIVO' Linked Open Data platform to include innovative IP management functionality. This extension will allow improved IP discovery, aggregation, tracking and disclosure for participants within the Flagship and lead to the faster deployment of FuturICT's innovations in the market.

3. **Long-term Sustainability**: Long-term sustainability for the use and exploitation of the project's IP will then be ensured by the establishment of a dedicated legal entity to maintain relationships with IP holders, collaborators and licensees and to manage the Flagship's Intellectual Property in conjunction with its owners. Development work for the Foundation will commence during the CP-CSA, and the legal entity will be founded near the end of the ramp-up phase (Month 24), ready to bridge into the Horizon 2020 phase.

4. **Towards a FuturICT Brokerage Platform**: Roadmaps developed within FuturICT's Global Participatory Platform Focus Area will foresee the creation of a 'FuturICT Brokerage Platform', which aims at creating a powerful IP management system for the project forming part of a truly pioneering new innovation ecosystem for Europe. This platform will eventually feature a public 'app store' that will handle FuturICT's IP and promote open and transparent innovation based upon it, including the distribution of revenues to contributors, participating institutions as well as back to the FuturICT Foundation, thus providing financial sustainability for the project as well as a new public good for Europe. FuturICT will also bring together banks and private investors to set up a seed fund, that can help promising spin-offs to take off. London and Zurich being the main FuturICT hubs, but also two of Europe's dominant financial centres, the project is very well positioned in this regard.

The overall goal of the FuturICT Intellectual Property strategy is to ensure a coherent approach to Intellectual Property management across the Flagship and create a true innovation ecosystem, all the while ensuring that the rights of the various IP holders remain protected. This approach will mean that the participants will not simply be using their own IP in a piecemeal and fragmented manner, but that the Flagship will be actively seeking the best and most innovative means to use arising project IP in a joined-up way. This will include various commercial exploitation options as well as approaches such as Open Source licensing, as appropriate. In particular, FuturICT intends to offer particularly good IP conditions to spin-off companies and start-ups, and specific terms to this effect will be set down in the Flagship Collaboration Agreement. These terms will grant such companies the use of IP from across the Flagship in return for a share of revenues to be returned to the project (initially via the beneficiaries and later via the FuturICT Foundation), or in some cases in return for their provision of royalty-free licenses to their own Background.

**Intellectual Property Management: Roles and Responsibilities**

Responsibility for the management, dissemination and/or use of FuturICT IP will initially fall to the CP-CSA beneficiaries generating it, with support from their associated Technology Transfer Offices (TTOs), guidance from the FuturICT IP & Exploitation Team and guidance from the FuturICT Science Board. Responsibility for central coordination of IP issues will then be gradually assumed by the FuturICT Flagship structure, via the development work of the IP & Exploitation Team, Commercial Strategy Board and Science Board, underpinned by the CP-CSA Consortium Agreement, the Flagship Collaboration Agreement and the project's multi-layered IP strategy, as detailed below. The following consortium bodies will have roles and responsibilities relating to Intellectual Property, as part of the wider FuturICT governance structure. The establishment of these consortium bodies is due to be completed by Milestone 12.3.1(month 6):

1. **IP & Exploitation Team**: Constituted within the FuturICT Project Office structure (FA12) but including both dedicated Project Office staff and expert individuals from across the consortium, the IP & Exploitation Team will provide central support and Flagship-level coordination on all issues related to the identification, tracking, management and proper use of FuturICT's Intellectual Property. The team will meet at least every two months, either electronically or in person, and will be led by a full-time FuturICT IP Officer employed at ETHZ, supported by the expertise of 'ETHZ Transfer' (ETHZ's TTO), which has broad expertise in industrial collaborations, negotiating licences and supporting spin-off companies. Prof. Dr. Stefan Bechtold, Professor for Intellectual Property at ETHZ will also ensure that the IP & Exploitation Team remains fully up to date with global IP developments. Further expertise will be provided by both UCL and ETHZ's EU Research offices, which will bring profound experience in IP management (currently administering over 800 running EU and international projects, mainly under FP7). Finally, considering that the Swiss Climate KIC office is also located at ETHZ, its Director Reto Largo will also advise the IP & Exploitation Team, drawing on his experience directing this huge EU initiative.

2. **Commercial Strategy Board:** The Commercial Strategy Board (CSB) will seek out and recommend new ideas, directions and opportunities for achieving the intended commercial and financial impacts of the FuturICT project and ensuring that the Flagship fully realises its innovative potential. The work of the ESB will include analysing Invention Disclosure Forms and the minutes of meetings held between FuturICT beneficiaries and outside organisations, in order to propose Flagship-level strategies for IP management and commercial exploitation. The CSB will also be responsible for developing FuturICT's policies for IP protection, ownership and licensing as well as proper methods for assessing financial and non-financial contributions to shared IP from involved parties; all in line with principles of fairness, ethics and the relevant EU and national laws. These policies will then be fed in to the provisions of the Flagship Collaboration Agreement. The CSB will include nominated experts on Intellectual Property from within the project, senior representatives of key beneficiaries' linked TTOs and members of the FuturICT Executive Board. The CSB will also have access to the project IP database/VIVO system set up by the Project Office, and will help identify commercial opportunities and opportunities for further innovation based on the IP generated across the whole Flagship initiative. The CSB will be coordinated within WP12.3

3. **Science Board:** The FuturICT Science Board will also have a series of responsibilities regarding

the management of FuturICT's IP, with particular regard to the areas of scientific use, academic publication and ensuring that FuturICT's IP management strategy leads to waves of innovation and positive impacts across European society. The Flagship Strategy Board will maintain overall responsibility for setting the terms of reference and determining the balance between the spheres of influence of the Commercial Strategy Board and the Science Board with regard to IP strategy, and a strong daily dialogue will be maintained between these two bodies via the on-going work of the IP & Exploitation Team. Finally, in order to ensure proper dialogue between the Science Board and the Commercial Strategy Board regarding the project's IP and innovation in different areas, a non-voting representative of each will sit on the opposite's board during meetings.

4. **Consortium Partners (beneficiaries)**: The consortium partners (beneficiaries) of the project will, in the short term, own the IP they generate under the project and retain responsibility for its dissemination, use and/or licensing. This role will be facilitated by the work of the IP & Exploitation Team, the Commercial Strategy Board and the Science Board, and by each beneficiary's linked Technology Transfer Office (TTO).

5. **Technology Transfer Offices (TTOs)**: In the ramp-up phase of the project the Technology Transfer Offices (TTOs) of beneficiary organisations will be responsible for practically realising the commercialisation and innovation potential of the Foreground of the FuturICT project, with support and coordination provided by the IP & Exploitation Team and the Commercial Strategy Board. The TTOs will:

    a. Give advice to beneficiaries of the FuturICT project on IP, enterprise and commercialisation strategies;

    b. Manage IP arising from the FuturICT project in collaboration with the Commercial Strategy Board and the Science Board and assess the best means for that innovation to be realised, through commercial or academic means;

    c. Evaluate the commercial potential or innovative impact of IP generated in the project and take suitable action to secure IP protection through patents, copyright, trademarks and design rights where appropriate;

    d. Advise the Commercial Strategy Board and/or Science Board on how best to realise innovation either through commercial or other means.

**Intellectual Property Management: Flagship Structure and Toolkits**

At the commencement of the ramp-up phase the below IP framework for the Flagship (see Figure 6.3.A.1) will be already be finalised and made available to all of the beneficiaries in the form of a set of concise and clear guidelines. To allow quick implementation of IP arrangements, and so as to not delay the innovative and commercial potential of the project, this initial framework is underpinned by existing and proven IP regimes, toolkits and document templates that are already in common usage across the higher education and industrial sectors. Over time this initial IP framework and the underpinning toolkits, shall be adapted and tailored by the IP & Exploitation Team to fit the requirements of Flagship as it evolves, before the commencement of the more innovative FuturICT approaches described below.

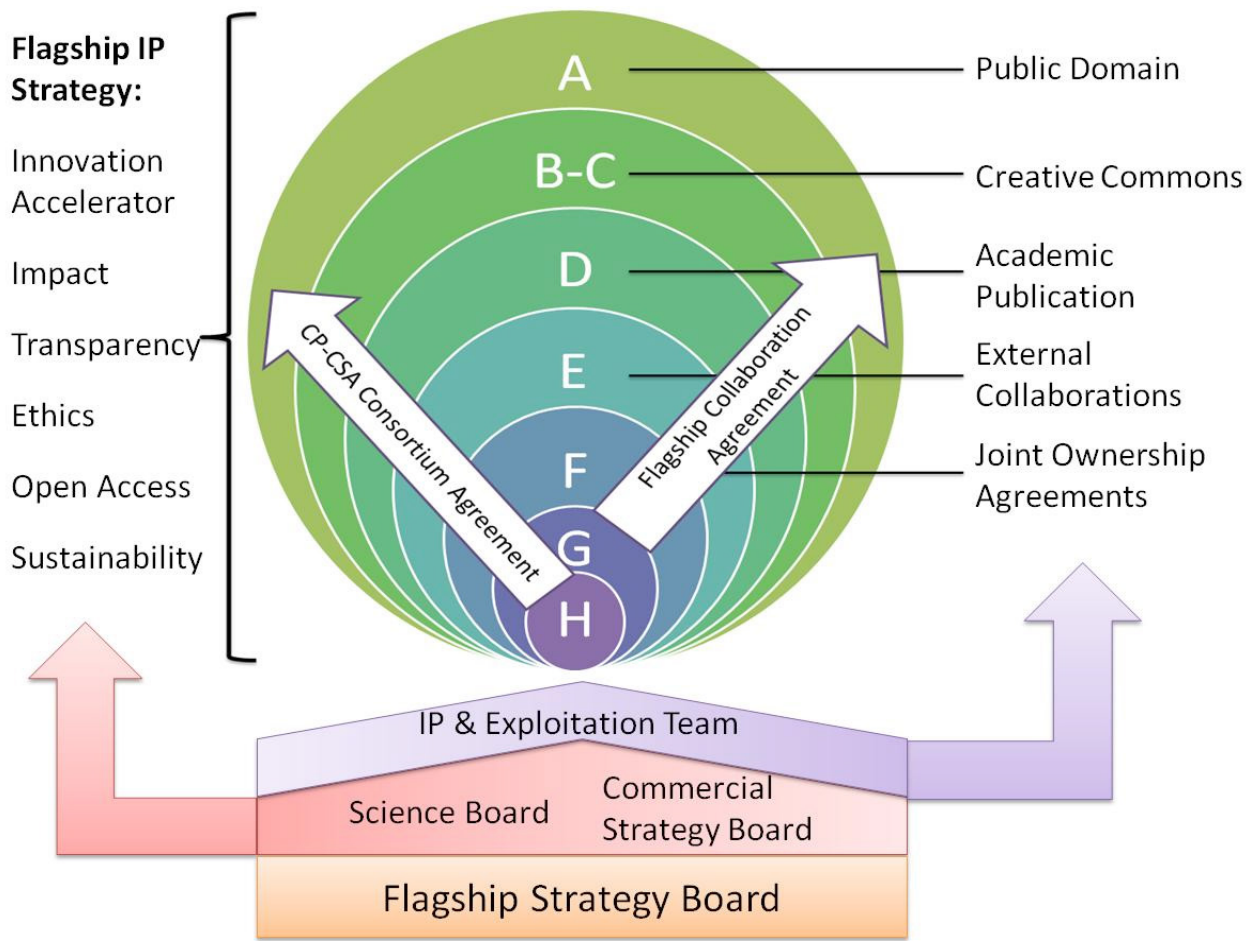

*Figure 4.3.a.1 Illustration of the FuturICT exploitation and IP management structure*

The initial structure shall be based upon eight major IP 'levels', any number of which can be assigned to a piece of IP generated within the FuturICT project ('Foreground'), or to IP arising from a collaboration with an outside organisation. IP shall be assigned these levels in accordance with the wishes of the owning beneficiary(ies) and their TTOs, the advice of the Commercial Strategy Board and, where necessary, under the oversight of the Flagship Strategy Board and the Ethics Advisory Committee. The levels, on a decreasing scale of 'openness' are:

- **Level A (Public Domain)**: Research papers, computer software and data will be released to the public without restriction, and everyone will be able to freely use and reuse material published under this level. The release of FuturICT IP into the Public Domain will only be carried out where appropriate, where ethical (with advice from the Ethics Advisory Committee) and where agreed upon by all parties involved;

- **Level B (inspired by Creative Commons: attribution)**: Research papers, computer software and data will be released under the condition that the author, his or her home institution and FuturICT are mentioned in any use and reuse of the material. Change and commercial use of material is allowed. The specific license used will depend on the type of material and will be decided by the owners with guidance from the Commercial Strategy Board. Whether software and data is licensed with a strong copyleft license (enforcing openness downstream through a viral provision) will depend on the circumstances;

- **Level C (inspired by Creative Commons: attribution & non-commercial)**: Research papers, computer software and data will be released under the condition that the author, his home institution and FuturICT are mentioned in any use and reuse of the material. Change of material and non-commercial use is allowed, but commercial use is not allowed. The specific license used depends on the type of material and will be decided by the owners with guidance from the Commercial Strategy Board. Whether software and data is licensed with a strong copyleft license (enforcing openness downstream through a viral provision) will depend on the circumstances;

- **Level D (Academic Publication)**: Publications will follow the project's commitment to Open

Access and a process for giving proper notification and credit to collaborators, the FuturICT project and the European Commission funding;

- **Level E (External Collaboration Arrangements)**: A series of eight standardised IP regimes and toolkits will be available for use in setting up sub-project specific IP arrangements within the wider Flagship, as well as governing relationships with linked initiatives with outside organisations. All of these sub-levels shall be modelled around the existing 'Lambert' agreement toolkit for collaborative projects published by the UK IP Office, bringing with it a set of eight model agreements. This toolkit shall be made available to all beneficiaries at the beginning of the project, and the key provisions ensuring coherence, fairness and ethical use in such arrangements within the Flagship will be laid out the Flagship Collaboration Agreement and CP-CSA Consortium Agreement. However, the use of the specific Lambert templates will not be mandatory. This approach provides the best balance between a coherent, consistent Flagship structure and flexible implementation under national laws;
- **Level F (Joint Ownership)**: In accordance with the CP-CSA Consortium Agreement and the Flagship Collaboration Agreement Joint Ownership Agreements will be entered into for IPR that is shared between two or more Flagship participants or members of the core CP-CSA consortium;
- **Level G (Flagship Collaboration Agreement)**: The IP provisions of the Flagship Collaboration Agreement will all govern IP and exploitation matters within the wider Flagship, including provisions relating to the joint ownership of IP arising across the Flagship initiative as a whole, and provisions setting out the requirements for collaborations both within the Flagship and without.
- **Level H (CP-CSA Consortium Agreement)**: The IP provisions of the FuturICT Consortium Agreement between the consortium of beneficiaries will govern all IP and exploitation within the core CP-CSA initiative. It will be based on DESCA 3.0 (www.desca-fp7.eu).

**Intellectual Property: Management Procedures**

Following the creation, or the anticipated creation, of a new piece of usable Intellectual Property within the project, the beneficiary generating that IP will be obliged to complete an Invention Disclosure Form (IDF) and submit this to the IP & Exploitation Team for inclusion in the central project IP database/VIVO system. The IDF will be a confidential document, which will be designed to determine the basic facts relating to an invention, design or copyright material and serve an additional purpose of establishing an independent reference point as to when an invention was made. It will provide details on inventorship, ownership, prior art, disclosure, publication schedule and market opportunity. This information will be used to assess the proposed IP so that a decision on next-steps can be made. IDFs will also be included in the schedule to be discussed at the next meeting of the CSB, where the related IP will be discussed within the context of the project's wider exploitation/IP strategies. See figure 3.2.10 for the process.

If this IP is to be developed into a commercial product, or combined with another piece of IP in order to create a new piece of foreground with commercial value, exploitation will be led by the relevant TTO(s) in line with the structures laid out below, with constant support from the IP & Exploitation Team. In the case of jointly owned IP, the lead or best placed institution for that work or the results of a particular Work Package (e.g. the Work Package Leader) will seek assignment of the IP initially so that exploitation of the Foreground IP can be carried out and managed effectively by one institution.

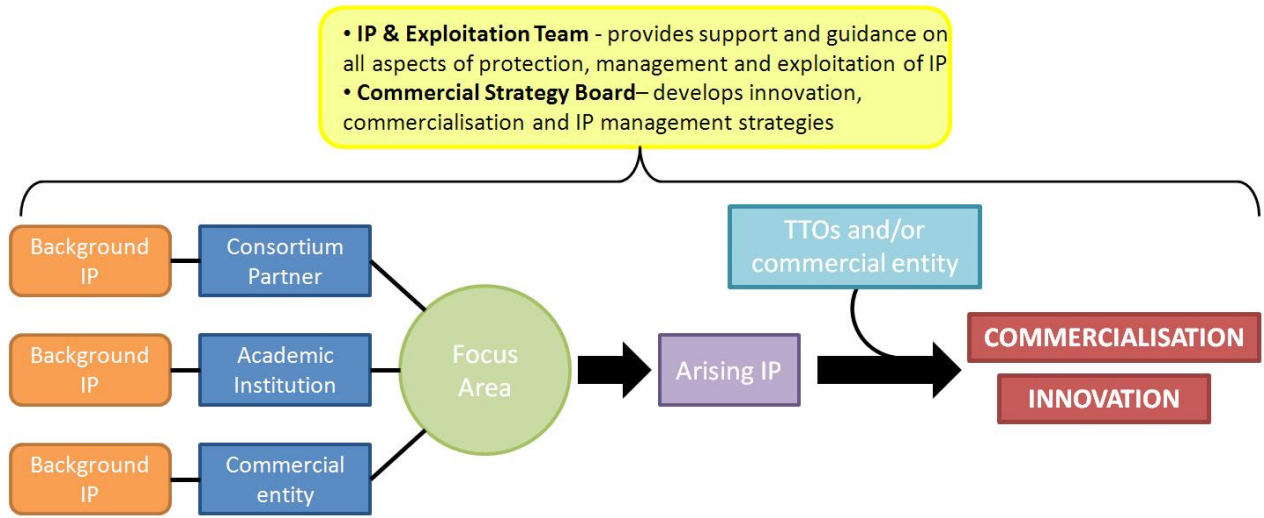

***Figure 6.3.a.2*** *Illustration of the FuturICT exploitation and IP management process*

Project-wide IP policies, supporting documents and the IP toolkits described below will be collated into a clear and concise set of rules and guidelines that shall be circulated to all beneficiaries and linked TTOs electronically once it has been produced. This document shall also provide important information such as the standard formulations for acknowledgements in publications, a summary of the Invention Disclosure process and details of the key IP provisions laid out in the project's Grant Agreement, Consortium Agreement and Flagship Collaboration Agreement explained in clear English, to ensure maximum compliance.

### Intellectual Property: Extensions to the VIVO Platform

Work being carried out in the Innovation Accelerator Focus Area of the CP-CSA phase will include extensions to the 'VIVO' platform, which is an open source semantic web application enabling the sharing and discovery of research and scholarship information between academic institutions. Work on extending this platform, due to be completed in Month 24, will include the addition of functionality to allow FuturICT beneficiary institutions to publish information on IP that they currently hold. This new functionality will have an impact both outside the project (by allowing potential collaborators or licensees globally to discover quickly which VIVO participant holds IP that may be of interest), as well as in the Flagship itself, as appropriate IP information from the FuturICT beneficiaries' Invention Disclosure Forms will be published by the Project Office on the extended VIVO platform. The publication of this data on VIVO will therefore directly improve the management and visibility of the project's IP and facilitate its uptake, exploitation and impact.

### Intellectual Property: FuturICT Foundation

To promote the long-term sustainability of the Flagship initiative, the beneficiaries will also set up a legal entity (the 'FuturICT Foundation'), most probably a 'European Economic Interest Grouping', association or a foundation to sustain relationships with commercial organisations, maintain the Flagship IP and exploitation strategy, regulate the dissemination of licenses and manage the continued use and exploitation of project Intellectual Property. The FuturICT Foundation will be established by Month 24 of the ramp-up phase, and ETHZ will lead the necessary legal and structural work involved in its setup. The legal entity shall be located within the European Research Area and will constitute an important core of the project during the remainder of the FuturICT project lifetime. Each full member of the entity will have equal voting rights, and each project beneficiary (whether a direct member of the legal entity or not) will have the option to license its Foreground IP to the Foundation in order to facilitate the commercial exploitation of its results. A plan for distributing each beneficiary's stake and for providing fair and reasonable compensation will be developed during the CP-CSA by the FuturICT IP Officer. Every beneficiary of the CP-CSA will declare whether they are directly taking part in the legal entity by the end of Month 18. After Month 18, the membership of the Foundation may then be restricted. The Foundation will manage licensing access to Foreground data of the willing partners, and marketing activities regarding possible commercial exploitation of the data will also eventually be taken over by the Foundation.

Taking this into account, the long-term commercial exploitation plan to be carried out by the FuturICT

Foundation contains four key elements:

1. To establish, maintain and use contacts to ensure that business requirements for data structures and functionalities are taken into account throughout the Flagship;
2. To create outputs from the Project that show the demonstrable value of the data pools;
3. To arrange specific meetings to demonstrate the value of the FuturICT results;
4. To set up an outreach strategy and collaborate closely with the Science Board and Dissemination Focus Area stream in order to assure that FuturICT work and results are spread widely within the ERA, the scientific community and the general public.

### Intellectual Property – Contractual Elements

The below section provides details of the three central contracts that will govern the management of FuturICT's Intellectual Property, through the CP-CSA phase and into the post-FP7 phase. As shown in the above Figures, these agreements (particularly the FCA and CA) will define and regulate the IP structure for the Flagship, and will also formalise the relevant roles and responsibilities of the various governing bodies within FuturICT that are involved in the management and use of Intellectual Property.

### Intellectual Property - EC Grant Agreement (EC-GA)

The management of IP within the CP-CSA part will be underpinned by the standard terms of the Grant Agreement that will be concluded between each of the core CP-CSA beneficiaries and the European Commission. The EC-GA sets down a number of basic rules regarding project 'Foreground' (results and IP generated within the CP-CSA), 'Background' (IP, data and tools held by beneficiaries before the start of the CP-CSA) and how both of these are to be dealt with to ensure the project is able to function. The Project Office will include staff that are well versed in IP issues deriving from EC Grant Agreements, and will be responsible for ensuring all beneficiaries are aware of the basic provisions and rules (see WP 12.2). These rules will also be included within the project's set of IP and exploitation policies, which is due to be delivered in Month 15 of the CP-CSA.

### Intellectual Property – Flagship Collaboration Agreement (FCA)

In order to effectively bring the wider Flagship initiative together over the course of the CP-CSA project the Project Office, in conjunction with the key FuturICT governing bodies, will draw up, negotiate and conclude a 'Flagship Collaboration Agreement' (FCA). This agreement will formalise the contractual partnership of the Flagship initiative and prepare the legal framework for further integration, as appropriate. This agreement will include various provisions related to the management of Intellectual Property within the Flagship, and will follow the broad outline of similar agreements used in other such initiatives, including the 'FI-WARE' PPP initiative. These provisions will include:

1. Provisions to ensure that participants across the Flagship will have a level of access to the IP held by other participants to allow the mutually beneficial development and use of IP within the initiative. These will include provisions regarding permissions and the fair distribution of royalties between the owners and the project from any resulting IP or commercial applications;
2. Provisions to allow confidential disclosure of IP (via Invention Disclosure Forms) to the IP & Exploitation Team to allow for mutually beneficial aggregation or exploitation of IP as appropriate;
3. References to the implementation of the multi-layered IPR structure detailed above in dealings and collaborations within the Flagship structure;
4. An outline of the agreed roles of the Commercial Strategy Board and the IP & Exploitation Team within the wider governance and decision-making structures of the Flagship as a whole. These roles will not be given direct authority over a particular organisation's IP without explicit permission;
5. Provisions obliging all beneficiaries to ensure that any use of the FuturICT results by them, or by any outside organisation with which they collaborate, adheres to FuturICT's ethical standards and its guidelines on the responsible use of its results;

6. Provisions foreseeing the establishment, role and eventual revenue-gathering activities of the FuturICT Foundation.

**Intellectual Property - CP-CSA Consortium Agreement (CA)**

The basic IP provisions of the EC-GA will be supplemented by more detailed provisions to be negotiated and agreed within the 'Consortium Agreement' between the core CP-CSA beneficiaries, which will set out the rules and procedures to be followed in the implementation of the CP-CSA. This Consortium Agreement will be based on Version 3.0 of the 'DESCA' standard FP7 CA template that is already accepted and in wide use across the sector. This template will then be adjusted to suit the particular requirements of the FuturICT initiative, and with regard to IP, to include more detailed rules and agreements on specific issues such as:

1. Rules to ensure that CP-CSA beneficiaries will have sufficient access to the IP held by other beneficiaries to allow them to carry out their foreseen work under the project;

2. Provisions to ensure that beneficiaries follow appropriate procedures for obtaining prior permission for, and/or paying royalties following, the use of another beneficiary's IP in the commercialisation and exploitation of the results;

3. Provisions to ensure that all Foreground generated is disclosed (via IDFs) to the IP & Exploitation Team and the other beneficiaries, to ensure that any synergies between Foreground generated within FuturICT can be identified and maximised;;

4. Provisions to ensure that Foreground is made available to the public wherever appropriate (Levels A-E), in line with FuturICT's commitment to openness and the EU's wider drive to promote 'Open Access' to publicly funded research;

5. References to the implementation of the multi-layered IPR structure detailed above in dealings and collaborations with outside parties;

6. An outline of the agreed roles of the Exploitation Strategy Board and the IP & Exploitation Team within the wider governance and decision-making structures of the project;

7. A schedule of the Background to which the beneficiaries agree to grant the other beneficiaries access to, in order that they may carry out their work foreseen within the project and (where appropriate) exploit that work.

8. Provisions obliging all beneficiaries to ensure that any use of the FuturICT results by them, or by any outside organisation with which they collaborate, adheres to FuturICT's ethical standards and its guidelines on the responsible use of its results.

**Fostering Innovation**

Innovation is the process through which value is created and delivered to a community of users in the form of a new solution. FuturICT is a large-scale, disruptive, initiative that will foster innovation at a range of scales from the individual entrepreneur up to the global economic systems and will create different forms of value in the business, civic and academic sphere. FuturICT dissemination strategy targets both citizen engagement that drives innovation bottom-up and engagement with governmental and non-governmental organizations that drives innovation top-down, at international, national, regional and local scales. The stream of new ideas flowing from FuturICT is also heterogeneous and interdisciplinary with a wide range of applicability. Fostering innovation is a key element in FuturICT's dissemination strategy and is built in to many of our activities:

- The primary ICT tools emerging from FuturICT including the LES, PNS, GPP and the Innovation Accelerator are oriented to democratise access to big data analytics and provide a platform for enterprises (particularly SMEs) to develop business analytics solutions that will allow them to compete more effectively in rapidly evolving markets by understanding their customers more deeply and delivering much more customised products and services. This applies equally to public and private sector enterprises and FuturICT anticipates significant innovation in the public sector based on FuturICT tools. In the civic sphere FuturICT anticipates the creation of significant value from innovation in "bottom-up" governance approaches where the availability and promotion of FuturICT tools will drive citizen driven transparency in government and administration and to more active engagement in the democratic process. FuturICT will foster civic sphere innovation by making the tools available to allow significant inversions in traditional information flows. In the academic sphere FuturICT will drive innovation by demonstrating there is a better way to do research in many fields by revolutionise standards of evidence and knowledge dissemination channels.

- At the level of the Exploratories FuturICT envisages significant potential for fostering innovation. The Exploratories provide packaged data and models that allow the exploration of many societal challenge areas. Many FuturICT researchers have close links to governmental and non-governmental policy agencies and a key part of the work of FuturICT will be fostering the incorporation of FuturICT Exploratories and tools into policy making. In the business sphere FuturICT models and tools (including the policy wind tunnel) offer opportunities for regulatory and business environment innovation that will synergise with the move to business models based on deep customisation and business analytics in the market place. For example, businesses will work with much finer market stratification and service delivery mechanisms to achieve this while maintaining trust in privacy will require policy that could impose huge additional costs on enterprises or, with the right policy could be relatively straightforward to implement with low cost overheads. In the civic sphere FuturICT will work closely with non-governmental and international organizations to develop innovative approaches to crises including public health and natural disasters. Working closely at a policy and implementation level FuturICT will foster new approaches to modelling the development and evolution such challenges and on innovation in emergency response via transformations in the development time and responsiveness of crisis observatories and other on-the-ground tools to ensure rapid coordinated responses to unanticipated events.

**Expected Industry Participation**

FuturICT has been closely engaged with industry during the Pilot Phase (see industry letters of support and list of logos below) and plans to translate this support into engagement with the appropriate Focus Area. The FuturICT approach has been to help educate and stimulate the broader ecosystem of stakeholders so they understand the platform and exploratory approach being positioned by the project roadmap. FuturICT's model commons and data commons will help position key foundational components required by many industry solutions built on the FuturICT platform. FuturICT is built on an open, participatory approach, as our ethos has been to stimulate an environment where it is possible and feasible to exploit the results commercially while not exploiting the people engaged in the development. Our open approach includes an "app-store-like" brokerage platform in order to enable and stimulate both free and commercial transaction models.

Large MNCs and SMEs are interested in engaging with the FuturICT research agenda by attending the FuturICT industry meetings and engaging with the Exploratory events as well as reviewing product road maps in the light of the FuturICT research schedule. The differentiation between specialised exploratory and more general FuturICT platform components maps well to the way our industry partners are structured.

The outcomes from the respective Exploratory and platform Focus Areas will stimulate new spin-outs and start-ups since these are topical areas for many industry verticals across the globe. FuturICT has already received several letters of support from SMEs and startups that are interested in collaborating FuturICT. This collaboration with industry partners, NGOs, policy makers and regulators will ensure that FuturICT will focus on large scale pressing issues that could not be solved by established approaches.

Industry associations, policy makers and regulators will be enlisted by the relevant Focus Areas or Work Packages to help stimulate and drive new standards to support the deployment of FuturICT software and services. This will ensure that technical as well as ethical, privacy and confidentiality values are included in the design of any new standard or regulation. Within the Open Call for new partners, there is a specific section for policy makers and regulators.

The FuturICT FET Flagship project is focused on providing innovative new technology, services, processes and business models in order to accelerate the growth and impact of industry and researchers in the EU. FuturICT promises to deliver research excellence in novel areas with a focus on translating these outputs into innovative services that will showcase European start-ups, SMEs and MNCs and their ability to develop and monetise new service-enabled business models. Our FET flagship proposal will stimulate new research centres that are appropriately structured, resourced and focused to meet the demands of a competitive innovation economy and respond to real societal and industry needs. The FuturICT project is focused on leveraging its existing strong linkages with small and large industries all over the EU, the proven base of research centres and staff from the FuturICT consortium of researchers along with its experience in developing new technologies and innovative services. This converged effort will help deliver an enhanced R&D capacity in partnership with European industry, which in turn will help deliver the sustained growth Europe requires of the new knowledge and service intensive economy.

Neelie Kroes (Vice-President of the European Commission responsible for the Digital Agenda) in her speech titled "Building the Connected Continent" in Sofia, 20th September 2012 mentioned "To find its place in the future global economy, Europe needs to become the connected, competitive continent. The e-EU. The right networks, skills and innovation can transform Europe for the better." Her call to arms highlights some of the areas that the EU needs to work on to be better prepared for not just this decade or generation but indeed for the future generations within the EU. The current and future need for additional skills in quantitative environments involving such topical areas as maths, ICT, innovation, and creativity all reinforce the thrust of the FuturICT FET Flagship deliverables which will add tremendous depth to this area from an educational and economic asset perspective. This increased knowledge base will help stimulate activity across third level research centres that act as national or regional hubs while also aligning with the needs of industry by providing the human capital to assist with exploitation efforts around the outcomes of these interactive and interconnected Exploratories.

During the Pilot Phase, FuturICT has ensured that connections between industry and researchers were established early in the research flow, in order that collaborations can be exploited better in the next phase. FuturICT will continue to leverage national and industry associations and other regional support structures successfully not only so that these bodies can help researchers to form partnerships but opportunities should also be sought in applying the outcomes of the FuturICT research. The FuturICT FET Flagship project will deliver on this vision with its novel approach to designing a set of capabilities that can be exploited both within an academic environment, and thus serve as its own case study, as well as exploring and designing technologies and services that will further stimulate industry engagement and exploitation within the European economic context.

The unique groups of academic skill sets and other industrial capabilities required for the FuturICT project will enable this novel, inter-disciplinary research agenda deliver on the pain points highlighted in the same McKinsey Global Institute report as mentioned previously. This report states that the "amount of data in our world has been exploding, and analysing large data sets, so-called Big Data, will become a key basis of competition, underpinning new waves of productivity growth, innovation, and consumer surplus". In order to capitalise on these opportunities, the FuturICT FET Flagship will be a key linchpin to support the transformational change required in the quality and quantity of research undertaken by enterprise directly and in cooperation with third level institutions. Indeed, the FuturICT research initiative has been designed such that it will become the foundation for large-scale academic platform research initiatives in this area such that European industry can build further on in the coming years.

FuturICT will deliver on the promise of accelerated time to engage and deploy Big Data enabled services and technologies for researchers, industry and others by proving the concept with the initial seven Exploratory areas over the first 30 months. The learnings that will accrue from the efforts expended in the first 30 months of the project will enable the development of a blueprint to add further Exploratories in future years. It is well recognised that the potential to use ICT in infrastructural services offers significant competitiveness benefits for the European economy in terms of productivity and reducing costs. By building these prototype capabilities in FuturICT and exploring the means to develop compelling business models with industrial scale solutions, FuturICT ais positioning entrepreneurs in SMEs and MNCs in Europe.

The FuturICT FET Flagship project is confident it can extend its suite of research partners that can provide cash and/or in-kind contributions with additional MNCs and SMEs as well as develop linkages with large quantities of SMEs that are active in the areas of Big Data, Open Data, analytics or services. These linkages will help build and strengthen existing industry R&D centres located in Europe and possibly stimulate the development of new centres focused on themes developed through the FuturICT project. These joint initiatives will stimulate joint publications with industry co-authors since the interdisciplinary nature of this project will enable many different and complementary views to be explored as well as opening new avenues for publications. Since protecting and exploiting the research through patents will be an important facet of the initiative, every opportunity will be leveraged to engage with industry in developing patent applications.

FuturICT is keen to explore all exploitation opportunities and is happy to partner with national agencies to further these goals. In order to provide visibility to the potential of the models, FuturICT will organise presentations and roadshows to publicise the research capability to SMEs in incubation centres, present at industry association events as well as meet directly with MNCs that show interest. Our existing industry partners have stated they will help with introductions to potential research partners as well as services companies that potentially might want to consume the FuturICT IP. Fast moving SMEs have the capability to adopt and integrate these services into their business models quickly so FuturICT is conscious of engaging with this profile of company in order to accelerate licensing opportunities. A comprehensive suite of measurements will be developed in order to track industrial partners, cash contributions, in-kind contributions, NDAs, licenses, patents, papers published, industry conference presentations, academic conferences, social media mentions and website metrics. Many of our industry partners have committed to support the development of case studies around FuturICT Exploratories. These case studies will stimulate additional interest in this area to others by using examples from relevant industry verticals. By working on joint publications and case studies, presentations at industry and academic events, the FuturICT project will be positioned to develop a pipeline of highly trained Data Scientists that can take up employment in the new wave of MNC and SME services companies that will be stimulated by the advent of emerging opportunities based on Open Data and Big Data.

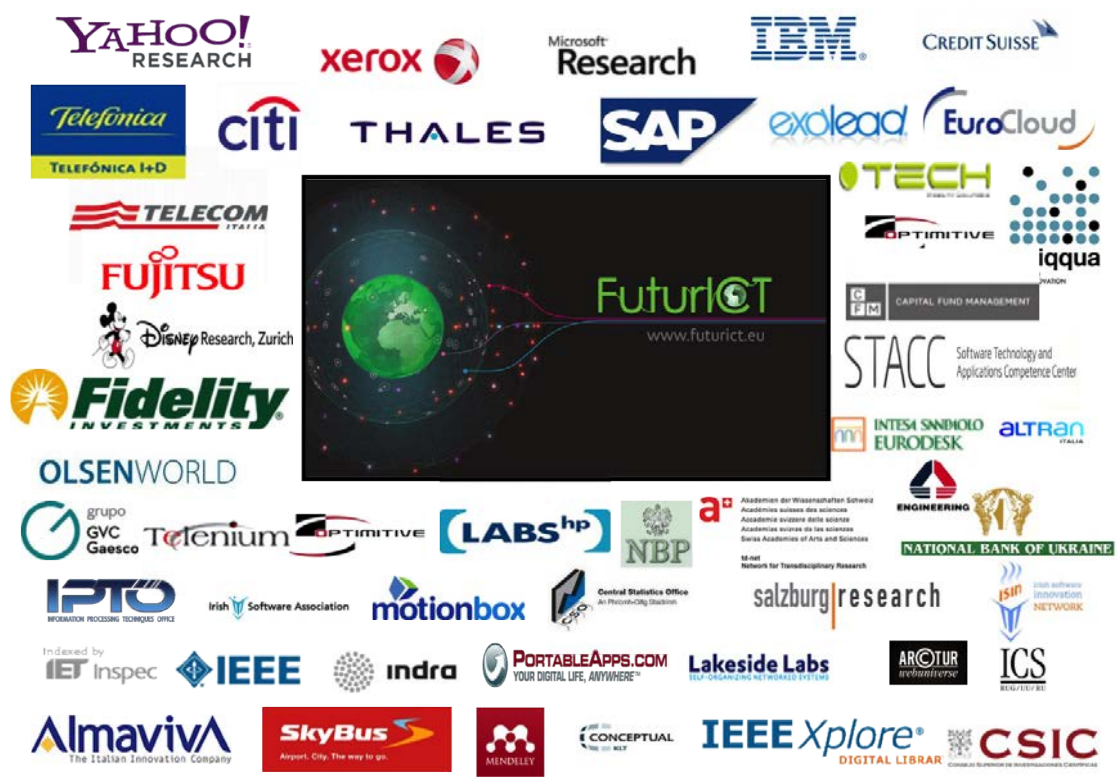

**Figure 4.3.A.3** *FuturICT will collaborate with a wide range of entities to ensure the economic value of the outcome.*

### European research programmes

Within the Commission, FuturICT demonstrator areas are clearly in line with many other research programmes, such as energy, sustainability, health, and governance. Whilst the project could not be carried out within one of these programmes alone, FuturICT will be able to coordinate activities across these disciplines, allowing it to leverage funds accordingly.

FuturICT's plans align perfectly with the efforts of the European Strategy Forum on Research Infrastructures

(ESFRI) towards the scientific integration of Europe through common policy-making on research infrastructures. A range of European projects and initiatives are already working in close collaboration with FuturICT. These include the Climate-KIC, EIT ICT Labs KIC, and the PEER, FOC and VISIONEER projects, to name a few.

**International research programmes**

FuturICT has succeeded in attracting the support of top scientists on an international level. FuturICT has already established collaborations with Sandy Pentland's world leading team at MIT Media Lab. FuturICT will also build on connections to Future City initiatives in Singapore. Support from Japan, Australia and China is also strong.

## 4.4 Social and Legal Implications

**Social Implications**

Globalisation and technological change have made our world a complex system, which is hard to understand and manage (but not impossible). Although this complexity has created many benefits, this change has also caused strong systemic interdependencies and some serious challenges, such as financial uncertainty, political instability, rapidly spreading pandemics, and insecurities on energy, water and food supplies. Our transportation, communication and financial networks are now generating huge amounts of data, much of which is available from government and public sources. By putting this data together with new models of how communication networks and social systems actually work, FuturICT can get a much better understanding of the complex behaviours that exist around us.

Conventional models of how economic and political systems work do not consider the interplay of non-linear dynamics, network interactions, heterogeneity and randomness well enough, which may lead to cascading effects such as large-scale financial or economic crises. Consequently, current regulatory regimes may not be sufficient to prevent these problems from occurring. We need an economy that can satisfy the needs of our population, which requires innovations that allow new and old businesses to create wealth and social wellbeing. Politicians, regulators and businesspeople must be equipped with the best possible models and tools to help them understand and make decisions about the interventions, laws and regulations that help to sustain a democratically accountable and effective economy.

The FuturICT Flagship will bring together hundreds of scientists, businesspeople, citizens and policy-makers to produce an open and interactive data and modelling commons, i.e. a new public good that will advance our understanding of today's complex world and create a participatory information ecosystem facilitating new collaborative and business opportunities of all kinds. The interactive platform developed by FuturICT will also use wisdom-of-crowds principles beyond Wikipedia and prediction markets.

Members of the public will be able to access the outcomes of FuturICT research and to contribute to the setting of research priorities. The FuturICT Strategy Board, which is responsible for the project's leadership and direction, will be advised by a representative panel of members of the public in order to advise on ethical considerations, suggest research priorities and problems, and to comment on the proposed research directions. Every year, there will be at least one "FuturICT Week", which will showcase the research, business and public benefits of the FuturICT projects, in events that will be open for everyone. Each EU member country or region runs its own national or regional hub, which has a website explaining local research. Each hub will host events connecting and demonstrating local research activities, outcomes and benefits. The FuturICT website will carry reports, movies and demonstrations that will explain to members of the public, how they can use FuturICT resources and outcomes.

FuturICT is committed to transparency, openness and ethical behaviour. The project will be under public scrutiny and subject to ethical guidance. The benefits of the project will be available to guide decisions at all levels of society, to stimulate the economy and to deliver social benefits.

**Engagement with Authorities**

Before engineers build a real airplane, they first test it out in a simulated environment – a wind tunnel. With public policies, we usually don't do that: we design and implement the policy, and then hope that it works. FuturICT will develop the methods to build "policy wind tunnels". Through the use of the Planetary Nervous System, the Living Earth Simulator and the Exploratories, FuturICT can simulate, better understand and help to predict the intended and unintended consequences of policy decisions.

The FuturICT project will direct its work to helping policy-makers to address these problems with a pluralistic, value-oriented, democratic approach and within a balanced ethical framework. The changes that technology and Big Data will bring in the next few years could profoundly affect the state of the environment, political stability and economic development. It is important that research on these changes be open and available for public comment and scrutiny, rather than being managed solely by large corporations, which need to pay more attention to profit generation than to individual or public interests. FuturICT will operate within an open environment, making the data and simulations available through a Global Participatory Platform. This will establish a new public good on which all kinds of services can be built, i.e. it will support both commercial and not-for-profit activities. To prevent misuse of the platform and to enable reliable high-quality services, it will be built on the principles of transparency, accountability, and self-regulation, based on reputation, crowd intelligence and alert mechanisms. Moreover, data management will build on principles such as decentralised storage, tailored measurement, aggregation on-the-fly with immediate deletion of sensitive individual-level data, randomisation, encryption, anonymisation, and digital rights management.

FuturICT has been developing these issues in conversation with policy-makers at the highest levels of government, in the EU and nationally.

# 4.5 Human Capital Education and Training

Education is widely considered to be a major force for the economic and social wellbeing of the citizens of Europe and the World. Through education FuturICT aspires to scientific, technological, and societal progress in order to improve the lives of everyone and to achieve social justice, understanding and international peace. Education makes individuals and societies better adapted to thrive in a changing world. Despite these aspirations, millions of people have no access to education at any stage in their lives. Even in Europe many of our citizens are not provided with the education they like or should have, leaving many young people with poor knowledge and skills and poor prospects for work and a satisfying independent life. Education plays a crucial role in the context of FuturICT. On the one hand many individuals must be educated to have the basic knowledge and study skills to participate in the FuturICT endeavour. On the other hand, FuturICT aims to provide new knowledge and understanding of social and technical systems. This opens up completely new methods to accelerate the processes of teaching and learning.

FuturICT identifies various Grand Challenges in Education:

- Enable orders of magnitude more people to learn more effectively than they do today.
- Provide interdisciplinary education that combines depth of knowledge in some disciplines and breadth of knowledge across disciplines.
- Devise new methods of creating education to respond to accelerated acquisition of knowledge and deliver that education to people throughout their lives.
- Provide highly effective interdisciplinary and specialist education to large numbers of people throughout their lives at low cost.
- Provide interdisciplinary education across the traditional scientific domains to create a large cadre of people trained in complex systems science, systemic risk, integrated risk management, integrative systems design and realistic modelling of techno-socio-economic systems.

FuturICT is ideally placed to contribute to a revolution in the means of delivering education, moving away from the traditional approaches of chalk-and-talk lectures, textbooks and printed materials, and moving towards searchable texts, movies, and animations, supported by Peer-to-Peer learning models where students collaborate and use social media to encourage and improve each other's learning. During the life of FuturICT this revolution can be expected in our knowledge about the ways human beings learn, and how ICT can enable a much better understanding of the structure of knowledge and how knowledge networks evolve.

FuturICT aims to combine Complexity Science, Social Science and ICT to create a breakthrough in educational theories, methods and technologies, able to deliver agile adaptable high-quality low-cost personalised education to meet the challenges of the 21$^{st}$ century. The long-term Flagship vision is to meet the grand challenges of enabling people to learn an order of magnitude more effectively, and at an order of magnitude less cost, where order of magnitude means a factor between two and ten.

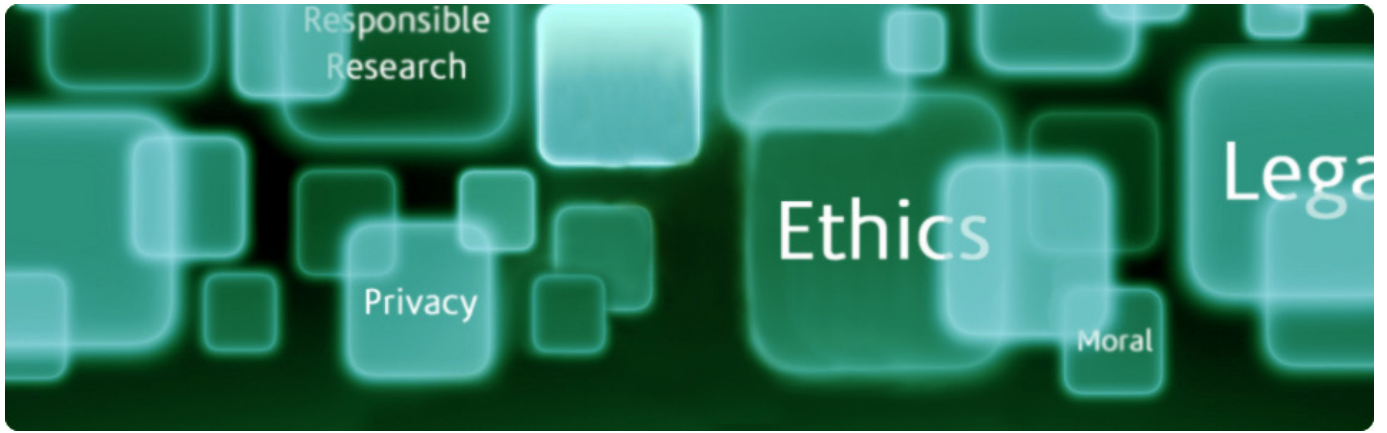

# 5 Potential Ethical & Legal Implications

## 5.1 General Considerations

FuturICT has a strong ethical motivation, which manifests itself in two forms: First, extreme care is taken within the project and among consortium partners to guarantee compliance with relevant laws, regulations, and ethical codes and to prevent unethical outcomes as a result of FuturICT research. Second, the project as a whole exemplifies the aim to promote the development of responsive, responsible, and ethical ICT. As such, FuturICT embodies recent views that the EU is developing on Responsible Research and Innovation and fully appreciates the results of the recent EGE report on the ethical aspects of ICT. Under FuturICT, both individual and institutional decision-making will be enhanced by access to an unprecedented range of data and tools to make sense of them, and decision-makers will be given the ability to explore multiple scenarios. At best, this will lead to better informed decision-making , which in turn will improve individual and collective well-being. The long-term goal is to have a participatory approach, making these capabilities available and understood by all interested individuals.

Over a 10-year time period, FuturICT wants to provide an open data, simulation, exploration, and participatory platform for everyone. The FuturICT proposal is thus fully within the Open Data strategy of the EU and aims at establishing an inclusive platform that guarantees equal access and equal opportunity for all to benefit from the knowledge that is collectively produced in the Flagship Project. This is intended to establish a new public good on which all kinds of services can be built, i.e. it will support both commercial and non-profit activities. To prevent misuse of the platform and enable reliable high-quality services, it will be built on principles of transparency, accountability, reputation, and self-regulation.

FuturICT will collect data of individuals who have consented to sharing their information after having been fully informed of the context in which they do so. Or data will be collected that are, as much as this is technically possible, not easily reducible to individual persons (de-indentified according to standard industry best-practices) and, therefore, do not deal with personal data. FuturICT is not (interested in) tracking individual behavior or gathering data on individual actions. FuturICT's aim is to understand the macroscopic and statistical interdependencies within the highly complex techno-social-economic-environmental system on which we all depend. So the individual as such is not an object of study. Quite to the contrary, FuturICT constitutes an environment for the development of the next generation of advanced privacy-by-design technology and promotes the development of privacy-respecting data mining technologies that give users control over their data. It will strongly engage in preventing and counteracting the misuse of data and the Internet. Researchers will be required to alert public authorities to the ethical and practical implications of the ICT research outcomes, as and when particular issues become apparent within the research process.

Finally, we consider it a moral obligation to push the research directions promoted by the FuturICT project forward as quickly as possible. Given the fragility of the financial and economic systems, the risks that this may ultimately impact the stability of our society and promote crime, corruption, violence, riots, and political extremism, or even endanger our democracies and our cultural heritage, are not negligible.

Quick scientific progress is needed in order to learn how to efficiently stop on-going cascading effects and downward trends. It is similarly important to ensure that social and socio-inspired innovations will best benefit humanity and not end up in the hands of a few stakeholders, as partially happened in genetics (particularly food production).

More broadly, the project will seek public involvement to build and sustain the values of trust, security, fairness, transparency, and ethical behavior. The FuturICT project is also committed to informing and empowering the public *vis a vis* the use of socio-economic data, models, predictions, and policies based upon them.

## 5.2 Privacy and data protection issues

Whilst the goals in themselves may be well-intentioned and morally justified, any technology that collects information about humans implies risks for their privacy. Without privacy-respecting and security technology, ethical processes and procedures, collecting data from devices, services, and data related to users, and following their behaviors, attitudes, and social interactions would potentially enable total surveillance. In the wrong hands or under regimes of differential access, it could give rise to unfair advantages and could lead to social sorting, where some citizens and consumers are not offered certain options and services, and are not even aware that they exist. Effective legal and technical mechanisms need to be identified to prevent and detect discrimination and exclusion, as well as to ensure that legal due process is respected in the potential use of data gathered. FuturICT will apply advanced privacy enhancing technology and advanced computer security expertise that it aims to develop to its own research.

The FuturICT Ethics Board will ensure that the Flagship develops and applies technologies that safeguard against invasions of privacy and manipulation, and will oversee all data collection activities, and develop policies and training for the collection and mining of large-scale datasets in an ethical manner. The Ethics Board will commence its work with a review of state-of-the-art frameworks and mechanisms safeguarding security and privacy - particularly relevant are those developed by existing FET projects in PerAda, such as ATRACO. The proposed Flagship will therefore engage in research into ethics, emergent norms, codes, the development of new data-sharing technologies providing users the control over their data ('new deal on data'), and suitable governance of research and development. Moreover, new policies and procedures to establish future standards and adequate protection of privacy within massive data mining activities shall be elaborated.

## 5.3 Moral Infrastructure of the Consortium and Flagship

The FuturICT Flagship has a Strategy Board which is responsible for the project as a whole. It also has an Ethics Advisory Committee with experts in IT law and IT ethics and external specialist in data protection and research ethics. **Prof Jeroen van den Hoven** (IT Ethics) acts as the Chair of the Ethics Board, and **Prof Stefan Bechtold** (IT Law) is the Vice Chair and also the legal counsel to the Management Team. Their responsibilities are first to guarantee compliance with Article 15 of the FP7 draft rules of participation, which states that any proposal needs to be in compliance with fundamental ethical principles and must fulfill the conditions set out in the specific work programme or in the Call for proposals. Secondly, they will guarantee that FuturICT researchers comply with national legislation, European Union legislation, respect international conventions and declarations and take into account the Opinions of the European Group on Ethics. More specifically, compliance with the Charter of Fundamental Rights of the European Union (concerning human dignity, freedom, equality, solidarity, citizens' rights and justice) and Article 8 of the European Human Rights Convention. FuturICT has a panel of users and representatives of civil society and relevant NGO's. A special group will be appointed which consists of three members, who are recruited from Data protection officers' staff of the EU member countries. It will regularly interact and consult with the article 29 group and take part in conferences of the Data Protection Registrars to report on the development and application of privacy-enhancing technologies and the nature and implications of work as well as give advice on research projects.

Research protocols will be reviewed at the consortium level for both international and EU compliance, the relevant national and international agreements applicable will be used as guideposts for a threshold inquiry, debated and defended by those proposing the research, and either approved as compliant, or suggested for improvements to bring it into compliance where possible. Local legal and ethical requirements, will initially be regarded as the responsibility of principal investigators as being locally compliant. It is expected that any local rules for ethical research conduct will be followed, and certified as such through whatever local mechanisms and certification may exist. In case of doubt, a precautionary approach, by which the consortium must approve of a doubtful study, and specifically inquire into compliance with local rules, will be adopted. In case of absence of local standards, a baseline consortium standard will be used to ensure maximal protection of potential subjects and their privacy

## 5.4 Protocols and mechanisms to ensure responsible research and innovation

Where research in FuturICT involves human subjects, including data gathering, the Ethics Board reviews protocols, safeguards, and consent documents for compliance with the principles enunciated in the Helsinki Declaration. Similarly, where elements of FuturICT involve nexus with human subjects or data-sets gathered from or about human-subjects, the manner by which subjects are informed, and by which data is used and kept must conform with internationally-recognised standards for privacy, consent, respect for persons, and custodial care of data, as well as employ technological risk-reduction strategies such as de-identification and randomization where possible. A sub-committee of the Ethics Board will regularly review proposed protocols for adequate protective measures, and compliance with basic ethical principles (as well as consortia-wide baseline standards, and local compliance documents when possible), especially where data on human subjects, and human subjects themselves are used. While local compliance measures must be respected, an over-arching standard of care based upon well-recognised international legal and ethical principles will serve as a threshold inquiry by the sub-committee. Only where concerns are raised will further elaboration, such as substantive submissions, consent documents, etc., be required, and the ethics subcommittee may raise issues to be addressed for clarification and approval. The subcommittee should be comprised of those with experience on ethics committees, as well as some with legal expertise, and at least one community member. Ongoing compliance will be ensured for long-range studies by periodic review of studies for both basic compliance and success in protections, and the Ethics Board may be notified if a study falls out of compliance and raises concerns which might require changing a study protocol, or stopping it all together. In order to fulfill both the necessity of periodic review of new protocols, and periodic review of ongoing studies, the Ethics Board and its subcommittees shall establish a schedule of regular meetings, as well as a system of reporting and documenting their activities.

Standards for ethical protocol development, as well as some basic background material in ethical research, will be propagated to researchers involved. Thus, researchers will have some preliminary guidance and examples as well as case-studies to draw upon in generating protocols that pass the threshold in the first place, helping to ensure that fewer in-depth reviews by the subcommittee will be necessary in the long-run. The goal is that ethics will be incorporated as an operating principle in all FuturICT activities, and that the culture that surrounds the research will be guided by ethical principles, rather than ethics serving as a trip-wire against which researchers must guard themselves.

Finally as with "legal implications" discussed below, FuturICT offers an opportunity to investigate the interconnection, at a fundamental level, of ethical principles and a certain technology. We may well expect the study to reveal new ethical implications, principles, and mechanisms implicit in the tools and practices in large-scale data collection itself. For instance, trust is a necessary component of numerous online interactions, including commercial ones, and has both technological and ethical implications. Our research ought to open up avenues of examining both in the context of Big Data.

## 5.5 Legal implications

The fundamental nature of FuturICT may give a different sense to the traditional notion of "legal implications" as customarily discussed in EU grant applications. Apart from the legal implications that are discussed above, and specific legal institutional frameworks that are noted in the table below that are within the responsibility of the Ethics Board of FuturICT, the envisaged research may have an impact on our future moral thinking and our future legal frameworks themselves. It may thus have implications for the law, for its revision, and for the direction in which it may need to change. FuturICT will specifically be a deep resource for rethinking and reformulating EU data protection and intellectual property laws in coming years. One of the possible indicators of success after 10 years of FuturICT will be changes to the laws and regulations regarding data, ICT systems, and privacy protection in some or all of the EU countries. These could arise in a number of ways.

Emerging directly from FuturICT's generation and exploitation of new intellectual property may be new ways to organise fair payment for apps that are distributed over many devices and which have complex webs of authorship, due to distributed, cross-platform applications and systems operating over networks.

A new Global Systems Science that can deliver a more accurate and complete analysis of interconnected multi-level systems may have implications for the distributive allocation of responsibilities for designing, building, and operating complex communication networks. If systems have not been amended, after a FuturICT analysis of their risk profile, then new legal responsibilities may be in place. And FuturICT may have to be aware of the consequences of errors in their analysis, which might lead to loss of business or reputational damage for commercial organizations involved. In sum, because the allocation of rights and responsibilities has always been complicated, and has only become more so with the development of complex networks employing ICT, the infrastructure itself is expected to inform and accommodate better distribution of ethical principles, and to ease problems related to both, moral accountability and legal liability. Facing this problem could very well lead to its solution.

FuturICT's intended ability to simulate future scenarios will be valuable to politicians, regulators, policy-makers and citizens, since FuturICT may be able to show that particular courses of action could possibly lead to severe consequences with a higher than acceptable probability. This could lead to changes in regulation, and the establishment of new regulatory regimes. Other possible outcomes could be better models for the balance and level of taxation and welfare benefits, healthcare policies and a nation's energy, water and food supply networks.

The governance and interface with policy making will therefore be monitored and potential conflicts or biases will be signaled and made explicit and transparent.

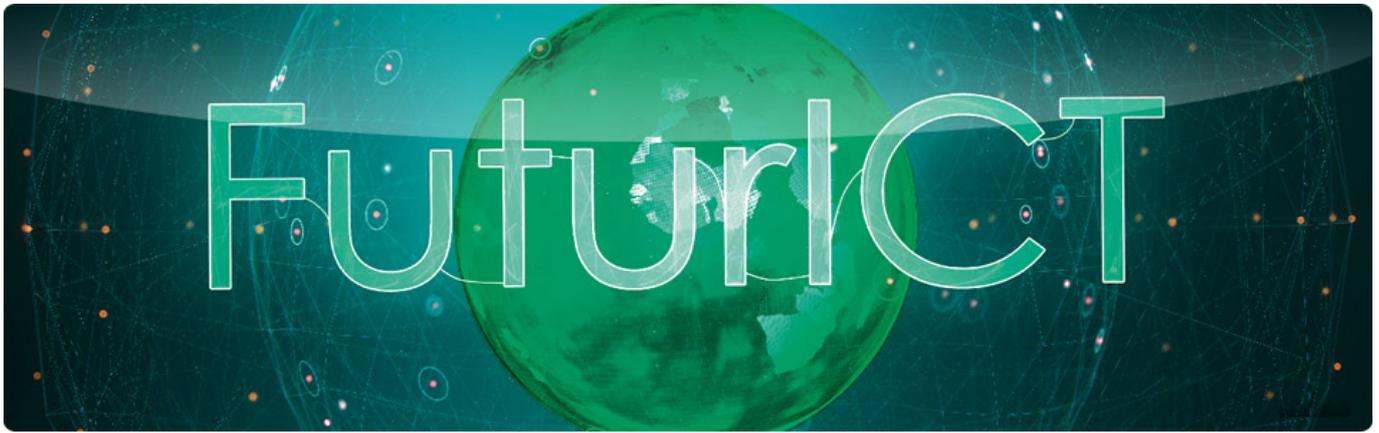

# 6. References

## 6.1 References cited in the proposal

## 6.2 Bibliography

**Selected Recent Key Publications in NATURE, SCIENCE, PNAS
by FuturICT Participants (Participants shown in bold)**